\documentclass[a4,12pt]{article}
\usepackage{comment}
\usepackage{etex}
\usepackage{amsmath,amsthm,amssymb}
\usepackage{graphicx}
\usepackage[usenames,dvipsnames]{xcolor}
\definecolor{MyGreen}{rgb}{0,0.5,0.2}
\colorlet{green}{MyGreen}
\usepackage{wrapfig}
\usepackage{enumerate}
\usepackage{mathrsfs}
\usepackage{geometry}
\usepackage{fancyhdr}
\usepackage{caption}
\usepackage{ascmac}
\usepackage{fancybox}
\usepackage{youngtab}

\usepackage{hyperref}
\usepackage{cite}
\usepackage{bm}
\usepackage{braket}

\allowdisplaybreaks

\usepackage{tikz}
\usetikzlibrary{calc}
\usetikzlibrary{arrows,shapes,positioning}
\usetikzlibrary{decorations.markings}
\usetikzlibrary{decorations.pathreplacing}
\usetikzlibrary{mindmap,trees,shadows}
\usetikzlibrary{patterns}

\usetikzlibrary{external}

\tikzstyle no arrow=[thick]
\tikzstyle end arrow=[thick,postaction={decorate,decoration={markings, mark=at position .999 with {\arrow[scale=1.4]{stealth}}}}]
\tikzstyle mid arrow=[thick,postaction={decorate,decoration={markings, mark=at position .6 with {\arrow[scale=1.4]{stealth}}}}]
\tikzstyle mid-end arrow=[thick,postaction={decorate,decoration={markings, mark=at position .8 with {\arrow[scale=1.4]{stealth}}}}]
\tikzstyle latter arrow=[thick,postaction={decorate,decoration={markings, mark=at position .8 with {\arrow[scale=1.4]{stealth}}}}]
\tikzstyle former arrow=[thick,postaction={decorate,decoration={markings, mark=at position .4 with {\arrow[scale=1.4]{stealth}}}}]
\tikzstyle transarrow=[white,thick,postaction={decorate,decoration={markings, mark=at position .0 with {\arrow[scale=1.4,black]{stealth}}}}]

\newcommand{\circlearrow}[6]{\draw[transarrow] ($#1+(#3:#2)+(#3+90:#4*#2)$) node[black,#5]{#6} --++ (#3+90:#2);}
\newcommand{\anticirclearrow}[6]{\draw[transarrow] ($#1+(#3:#2)+(#3-90:#4*#2)$) node[black,#5]{#6} --++ (#3-90:#2);}

\numberwithin{equation}{section}

\def\SUSYN#1{{\mathcal{N}{=}#1}}

\def\Tr#1{{\rm Tr}\left[#1\right]}
\def\TrR#1#2{{\rm Tr}_{#1}\left[#2\right]}

\def\bbZ{\mathbb{Z}}

\def\bbR{\mathbb{R}}
\def\bbC{\mathbb{C}}

\def\frq{\mathfrak{q}}
\def\prodR#1#2{\prod_{#1}^{#2}\!\!\!\!\!\!\!\!\mbox{\lower1ex\hbox{\rotatebox{39}{$\longrightarrow$}}}}
\def\prodL#1#2{\prod_{#1}^{#2}\!\!\!\!\!\!\!\!\!\!\mbox{\lower0.3ex\hbox{\rotatebox{141}{$\longrightarrow$}}}}

\def\calD{\mathcal{D}}

\def\calF{\mathcal{F}}

\def\calH{\mathcal{H}}

\def\calJ{\mathcal{J}}

\def\calL{\mathcal{L}}

\def\calR{\mathcal{R}}

\def\calW{\mathcal{W}}

\def\calZ{\mathcal{Z}}

\def\ie{{\it i.e.} }

\def\cI{\mathcal{I}}

\def\PE{\mathrm{PE}}
\def\Tr{\mathrm{Tr}}

\def\SU{\mathrm{SU}}
\def\U{\mathrm{U}}
\def\esix#1#2#3#4#5#6{\overset{\scriptscriptstyle #6}{\scriptscriptstyle #1#2#3#4#5}}
\def\Esix#1{\expandafter\esix#1\relax}
\def\inc#1{\includegraphics[scale=.2]{figure/t3/#1}}
\let\bigwedge\wedge

\begin{document}

\begin{titlepage}

\begin{flushright}
IPMU 15-0034 \\
UT-15-08
\end{flushright}
\vskip 2cm

\begin{center}

{\Large \bfseries
On skein relations in class S theories
}

\vskip 1.2cm

Yuji Tachikawa$^{\flat,\sharp}$ and Noriaki Watanabe$^{\sharp}$

\bigskip
\bigskip

\begin{tabular}{ll}
$^\flat$  & Department of Physics, Faculty of Science, \\
& University of Tokyo,  Bunkyo-ku, Tokyo 133-0022, Japan,\\
$^\sharp$  & Kavli Institute for the Physics and Mathematics of the Universe, \\
& University of Tokyo,  Kashiwa, Chiba 277-8583, Japan
\end{tabular}

\vskip 1.5cm

\textbf{Abstract}
\end{center}

\medskip
\noindent
Loop operators of a class S theory arise from networks on the corresponding Riemann surface, and their operator product expansions are given in terms of the skein relations, that we describe in detail in the case of class S theories of type A. 
As two applications, we explicitly determine networks corresponding to dyonic loops of $\SUSYN4$ $\SU(3)$ super Yang-Mills, and compute the superconformal index of a nontrivial network operator of the $T_3$ theory.
 
\bigskip
\vfill
\end{titlepage}
\setcounter{page}{1}

\tableofcontents

\pagebreak

\section{Introduction and summary}\label{sec:intro}

To understand the properties of 4d $\SUSYN{2}$ supersymmetric theories,
it is extremely useful to realize them as  twisted compactifications of the 6d $\SUSYN{(2,0)}$ superconformal field theories (SCFTs) on Riemann surfaces $C$ with  punctures \cite{Gaiotto0904,GMN2}.
These 4d theories are now called class S theories,
and they are closely associated to certain 2d theories defined on $C$.
For example, class S theories on  $S^4$ are related to the Liouville/Toda CFT on $C$ \cite{AGT09,Wyllard09}
and when considered instead on $S^3 \times S^1$ 
they give rise to 2d $q$-deformed Yang-Mills on $C$ \cite{Gadde:2009kb,Gadde:2011ik,Gadde:2011uv}.

How do supersymmetric defects\footnote{In this paper we use the words \emph{defects} and \emph{operators} interchangeably.} appear under these 4d-2d duality relations? 
In this paper, we only consider one-dimensional objects, \emph{loops}, on the 4d side. 
Since 6d $\SUSYN{(2,0)}$ SCFTs have codimension-two defects and codimension-four defects, 
a loop operator on the 4d side should come from a codimension-four defect
on a codimension-one defect on the 2d side.

For class S theories of type $\SU(2)$, or equivalently $A_1$, 
 dyonic charges of loops on the 4d side is in one-to-one correspondence with topologies of non-intersecting loops on the 2d side \cite{DMO09}.
When the 2d theory is the Liouville/Toda theory, these loops on the 2d side have an explicit realization in terms of the Verlinde operators \cite{Verlinde88} as  shown in  \cite{AGGTV09,DGOT09}. 
When the 2d theory is the $q$-deformed Yang-Mills, these loops are instead realized as 2d Wilson loops.\footnote{This fact was not explicitly mentioned in the literature to the authors' knowledge, but apparently it has been known to many practitioners in the field. We spell this out in Sec.~\ref{sec:SCIwith}.}

We only have to consider non-intersecting loops on the Riemann surface for the $A_1$ case, thanks  to the existence of   a  skein relation resolving  each crossing  into a sum of two non-crossing ones : 
\begin{equation}
\raisebox{-0.5\height}{\tikzsetnextfilename{-figure-qYM-A1-R}
\begin{tikzpicture}
\def\l{1.3}
\coordinate (UL) at (0,0);
\coordinate (DL) at ($(UL)+(0,-\l)$);
\coordinate (UR) at ($(UL)+(\l,0)$);
\coordinate (DR) at ($(UR)+(0,-\l)$);

\draw[very thick] (DR) -- (UL);
\draw[line width=5.0pt,white] (DL) -- (UR);
\draw[very thick] (DL) -- (UR);
\end{tikzpicture}
}
=\frq^{-1/2}\;\raisebox{-0.5\height}{\tikzsetnextfilename{-figure-qYM-A1-Id}
\begin{tikzpicture}
\def\w{1.3}
\def\h{1.3}
\coordinate (UL) at (0,0);
\coordinate (DL) at ($(UL)+(0,-\h)$);
\coordinate (UR) at ($(UL)+(\w,0)$);
\coordinate (DR) at ($(UR)+(0,-\h)$);

\draw[very thick] (DR) .. controls ($(DR)+(110:0.3*\h)$) and ($(UR)+(-110:0.3*\h)$) .. (UR);
\draw[very thick] (DL) .. controls ($(DL)+(70:0.3*\h)$) and ($(UL)+(-70:0.3*\h)$) .. (UL);
\end{tikzpicture}
}
+\frq^{1/2}\;\raisebox{-0.5\height}{\tikzsetnextfilename{-figure-qYM-A1-Q}
\begin{tikzpicture}
\def\w{1.3}
\def\h{1.3}
\coordinate (UL) at (0,0);
\coordinate (DL) at ($(UL)+(0,-\h)$);
\coordinate (UR) at ($(UL)+(\w,0)$);
\coordinate (DR) at ($(UR)+(0,-\h)$);

\draw[very thick] (DR) .. controls ($(DR)+(160:0.3*\w)$) and ($(DL)+(20:0.3*\w)$) .. (DL);
\draw[very thick] (UR) .. controls ($(UR)+(-160:0.3*\w)$) and ($(UL)+(-20:0.3*\w)$) .. (UL);
\end{tikzpicture}
},\label{eq:above}
\end{equation}
where $\frq = e^{\pi i b^2}$ in the Liouville theory and $\frq=q^{1/2}$ in the $q$-deformed Yang-Mills. 

However, in the higher rank cases,  namely for $\SU(N)$ with $N>2$ or equivalently for  $A_k$ with $k=N-1>1$,  there is no such simple skein relation, since junctions of lines inevitably appear.
This results in the networks%
\footnote{The same objects are also called as  \emph{webs} or \emph{spiders}. In this paper, we only use the terminology \emph{networks}.}
 of lines on the Riemann surface, as already mentioned in  \cite{DGG10,GomisFloch10}.
Such networks were treated and discussed in \cite{Xie1304,Bullimore13}, putting special emphasis on the $\SU(3)$ case.
In \cite{Xie1304} the analysis was mainly carried out using the approach of the higher Teichm\"uller theory classically ($\frq=1$), and in \cite{Bullimore13} the study was done mainly in the framework of Toda CFTs for general $\frq$. 
These analyses gave rise to the skein relations that  had been discovered in the context of mathematics before \cite{Kuperberg96}.

Our  aim in this paper is to describe the skein relations of the networks in the general $\SU(N)$ case. We have, for example, the relation 
\begin{equation}
\raisebox{-0.5\height}{\tikzsetnextfilename{-figure-qYM-R}
\begin{tikzpicture}
\def\l{1.4}
\coordinate (UL) at (0,0);
\coordinate (DL) at ($(UL)+(0,-\l)$);
\coordinate (UR) at ($(UL)+(\l,0)$);
\coordinate (DR) at ($(UR)+(0,-\l)$);

\draw[end arrow] (DR) node[below]{$\Box$} -- (UL);
\draw[line width=5.0pt,white] (DL) -- (UR);
\draw[end arrow] (DL) node[below]{$\Box$} -- (UR);
\end{tikzpicture}
}
=
\frq^{\frac1N-1} \raisebox{-0.5\height}{\tikzsetnextfilename{-figure-qYM-Id}
\begin{tikzpicture}
\def\l{1.4}
\coordinate (UL) at (0,0);
\coordinate (DL) at ($(UL)+(0,-\l)$);
\coordinate (UR) at ($(UL)+(\l,0)$);
\coordinate (DR) at ($(UR)+(0,-\l)$);

\draw[end arrow] (DR) node[below]{$\Box$} -- (UR);
\draw[end arrow] (DL) node[below]{$\Box$} -- (UL);
\end{tikzpicture}
} 
+ 
\frq^{\frac1N}  \raisebox{-0.4\height}{\tikzsetnextfilename{-figure-qYM-Q}
\begin{tikzpicture}
\def\l{1.4}
\coordinate (UL) at (0,0);
\coordinate (DL) at ($(UL)+(0,-\l)$);
\coordinate (UR) at ($(UL)+(\l,0)$);
\coordinate (DR) at ($(UR)+(0,-\l)$);
\coordinate (UM) at ($(UL)+(\l/2,-\l/4)$);
\coordinate (DM) at ($(DL)+(\l/2,\l/4)$);

\draw[mid arrow] (DR) node[below]{$\Box$} -- (DM);
\draw[mid arrow] (DL) node[below]{$\Box$} -- (DM);
\draw[mid arrow] (DM) -- node[right]{$\bigwedge^{\!2}\Box$} (UM);
\draw[mid arrow] (UM) -- (UL) node[above]{$\Box$};
\draw[mid arrow] (UM) -- (UR) node[above]{$\Box$};
\end{tikzpicture}
} 
\end{equation}
that reduces to the equation \eqref{eq:above} when $N=2$. These skein relations were first found in \cite{MOY2000} in the context of knot invariants.

The guiding principle for us is that the representation theory of the quantum group $\SU_\frq(N)$ underlies these networks and their skein relations. 
The relation of the loop operators of 2d CFTs and  the quantum group has been known for quite some time, mainly in the context when $\frq$ is a root of unity, see e.g.~\cite{book}.
In the case of 2d $q$-deformed Yang-Mills, the relation of their loops and the quantum group is very direct, because the $q$-deformed Yang-Mills is a gauge theory whose gauge group is the quantum group \cite{Alekseev:1994pa,Buffenoir:1994fh}. 

We have two applications of the skein relations.
The first concerns the dyonic loop operators of $\SUSYN{4}$ $\SU(N)$ Yang-Mills theory. 
From the 4d gauge theory perspective, their classification was performed in \cite{Kapustin05}. 
When the electric charge and the magnetic charge of a dyonic loop are parallel in the weight system, 
there is an obvious realization of such dyonic loop in the class S language as a loop wrapping the torus. 
When they are not parallel, it was expected that they are represented by networks on the torus. 
We will give a complete description in the case of $\SU(3)$. 

The second application is the study of a particular loop operator of the $T_3$ theory. 
The $T_N$ theory is a strongly-coupled class S theory that does not have a useful Lagrangian description,
and  there is no direct method to describe a loop operator using the path integral language.
Still, using the class S language, it is easy to see that there is a loop operator described by a nontrivial network on a three-punctured sphere. 
We compute the superconformal index of the $T_3$ theory in the presence of this loop operator,
and confirm that it indeed has an enhanced $E_6$ symmetry, as expected from the fact \cite{Gaiotto0904} that the $T_3$ theory is in fact the $E_6$ theory of Minahan and Nemeschansky \cite{Minahan:1996fg}. 

\bigskip

The rest of the paper is organized as follows:
in Sec.~\ref{sec:review_skein}, we start by reviewing the relation of the operator product expansion of the loop operators on the 4d side and the skein relation on the 2d side. 
We emphasize the universality of the skein relation, independent of the choice of the 4d spacetime.
Next, in Sec.~\ref{sec:qYM}, we describe the skein relations explicitly for general $\SU(N)$.
We also explain how they can be understood in terms of the representation theory of quantum groups.
In Sec.~\ref{sec:gauge}, we turn to an application on 4d $\SUSYN{4}$ gauge theory. 
We discuss the operator product expansion of a pure Wilson loop and a pure 't Hooft loop in  the general $\SU(N)$ theory. 
We also describe all the dyonic loops in the $\SU(3)$ theory in terms of  networks, and point out a relation to the brane tiling. 
In Sec.~\ref{sec:T3junction}, we first explain how the 4d Wilson loop is mapped to the 2d Wilson loop in the context of the $q$-deformed Yang-Mills. 
We then use one of the skein relations  to compute  the superconformal index of the $T_3$ theory 
in the presence of a loop operator corresponding to the simplest nontrivial network. 
In the Appendix~\ref{app:Examples}, we have more examples of the realization of dyonic loops by networks in the general $\SU(N)$ theory.

\section{Loop operators in 4d  and skein relations in 2d }\label{sec:review_skein}

\begin{figure}[h]
\[
\raisebox{-0.4\height}{\includegraphics[width=30mm]{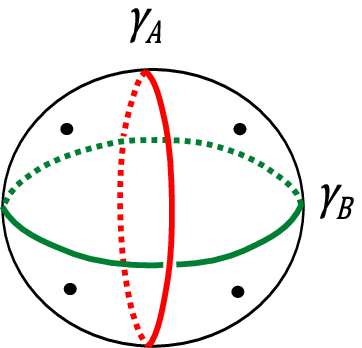}}
\; \leftrightarrow \;
{\calL}_{\gamma_A}{\calL}_{\gamma_B} \neq {\calL}_{\gamma_B} {\calL}_{\gamma_A}
\; \leftrightarrow \;
\raisebox{-0.4\height}{\includegraphics[width=30mm]{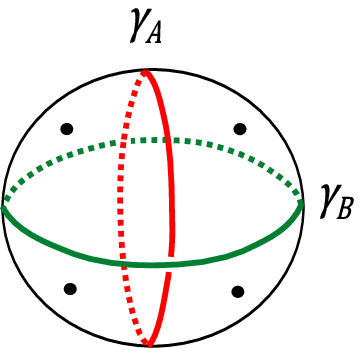}}
\]
\caption{Non-commutativity of actions of loop operators. \label{fig:4punc}}
\end{figure}

In the Liouville theory, a Verlinde loop operator is defined in terms of monodromy actions on the conformal block $\calF$ along a loop $\gamma$. 
It is possible to insert more than one Verlinde operator and ${\calL}_{\gamma_A}{\calL}_{\gamma_B}\calF\neq {\calL}_{\gamma_B}{\calL}_{\gamma_A}\calF$ in general when $\gamma_A$ and $\gamma_B$ intersect each other, as we see from the concrete calculations. See Fig.~\ref{fig:4punc} for an illustration.

Under the 2d-4d correspondence, the Verlinde operators map to loop operators of the 4d theory. 
Therefore, there should be a concept of ordering of loops on the 4d side, such that the product becomes non-commutative. 
In this section we review how this ordering arises, following \cite{Gukov:2006jk,GMN3,Dimofte:2011py,IOT11}. See also the recent reviews \cite{Okuda:2014fja,Gukov:2014gja}.

\subsection{Sums and products of loops}\label{subsec:DPE}
We consider the 4d setups where some kind of localization computations is possible. 
Typically there is a supercharge preserved in the background, 
whose square involves a linear combination of two isometries $k_{1,2}$. 
Supersymmetric loops wrap along the direction of $k_1$ and sit at the fixed point of $k_2$.

In the neighborhood of the loop,  we can approximate the geometry as $S^1\times \bbC\times \bbR$,
where $k_1$ shifts the coordinate along $S^1$ and $k_2$ is the phase rotation of $\bbC$. 
The loop now wraps $S^1$ and sits at the origin of $\bbC$, and the position $x\in \bbR$ is arbitrary. 
Therefore, we can place multiple loops $L_{1,2,\ldots}$ on $x_{1,2,\ldots}$ preserving the same supercharge.

This gives an intrinsic ordering of loops on the 4d side, and furthermore, 
the expectation values are unchanged under infinitesimal changes of the positions $x_i$.
The choice of the local supersymmetric background  at a loop can be characterized by a single parameter which we denote by $\frq$.
These statements can be explicitly checked in the case of the localizations on $S^1\times S^3$ \cite{DGG10},  $S^1\times \bbR^3$ \cite{IOT11} and $S^4_b$ \cite{Hama:2012bg}.

Given two types of loop defects $L_1$ and $L_2$,
we denote loops placed at $x\in \bbR$ by $L_i(x)$. 
We define a formal sum $L_1+L_2$ of two loops by \begin{equation}
\langle \cdots  (L_1+L_2)(x)\rangle :=
\langle \cdots L_1(x)\rangle 
+\langle \cdots L_2(x)\rangle 
\end{equation} where the ellipses stand for other operator insertions. 
The product $L_1\cdot L_2$ of two loops are now defined by 
\begin{equation}
\langle \cdots (L_1 \cdot L_2)(x)\rangle := 
\langle \cdots L_1(x_1) L_2(x_2) \rangle 
\end{equation} where we demand $x_1 > x >  x_2$ so that
$L_1$ and $L_2$ are the loops closest to $x$ from the left and from the right.
Since the expectation values depend only on the order but independent of the relative distance, 
this gives a consistent definition.

At this stage, we can make sense of the operator product expansion for defects. 
Suppose that there is a set $L_{1,2,\ldots}$ of  loops which cannot be decomposed into any sum of other simpler ones.   
We then have the following expansion of correlation functions
\begin{eqnarray}
\langle \cdots {L_i \cdot L_j}(x)\rangle = \sum_k c^{k}_{ij}(\frq) \langle \cdots L_k(x)\rangle
\end{eqnarray}
that can be written succinctly as 
\begin{eqnarray}
L_i \cdot L_j = \sum_k c^{k}_{ij}(\frq) L_k.
\label{eq:DPE}
\end{eqnarray}
The OPE coefficients $c^k_{ij}(\frq)$ are asymmetric under the exchange $i\leftrightarrow j$
due to the intrinsic ordering along $\bbR$.
However, we can simultaneously flip the $\bbR$ direction and $S^1$ direction in the local $S^1\times \bbC\times \bbR$ geometry to obtain another supersymmetric background.
The background is parameterized by $\frq$ as remarked before, 
and we use the parameterization such that this flip is represented by $\frq \mapsto \frq^{-1}$. 
Then we should have the relation 
\begin{equation}
c^{k}_{ij}(\frq)=c^{k}_{ji}(\frq^{-1}).
\end{equation}

\begin{figure}[h]
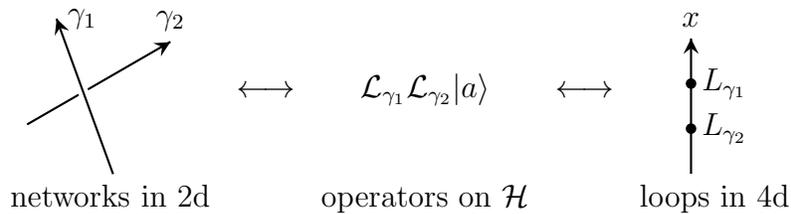

\centering
\[
\begin{array}{ccccc}
\vcenter{\hbox{\input{figure/review/order_on_2d_side.tikz}}} &
\longleftrightarrow &
\calL_{\gamma_1} \calL_{\gamma_2} | a \rangle &
\longleftrightarrow &
\vcenter{\hbox{\input{figure/review/order_on_4d_side.tikz}}} \\
\text{networks in 2d} & & 
\text{operators on $\calH$} & &
\text{loops in 4d}
\end{array}
\]
\caption{
Left: the part of networks $\gamma_1$ and $\gamma_2$. For any other crossing of $\gamma_{1,2}$, $\gamma_1$ is always above $\gamma_2$. 
Middle: the expressions acting on $\calH=\calH_\text{conf}=\calH_\text{hemi}$.
Right: the ordering on the 4d side, along  $\bbR$. 
\label{fig:relation among three orders}
}
\end{figure}

There is also an order for the defect networks on the two dimensional geometry side.
In order to make the relation between two orders, let us consider the case when the 4d side is $S^4_b$
and the 2d side is the Liouville/Toda theory \cite{AGT09,Wyllard09}.
In this case the Verlinde operators associated to the networks on the 2d side act on the space $\calH_\text{conf}$ of conformal blocks,
and the loop operators on the 4d side act on the space $\calH_\text{hemi}$ of holomorphic Nekrasov partition functions defined on a hemisphere \cite{AGGTV09,DGOT09,GOP11,Bullimore13}.
Since $\calH_\text{conf}$ and $\calH_\text{hemi}$ are naturally isomorphic,
we can see the relations among three orderings as shown in Fig.~\ref{fig:relation among three orders}.

Corresponding to \eqref{eq:DPE}, there should be the operator product expansions of defect networks which are indeed skein relations as resolutions of the crossings.
In Sec.~\ref{sec:gauge}, we will see that we can calculate many OPE coefficients in terms of defect networks.

\subsection{Non-commutativity and the angular momentum}
\label{subsec:classical}

Let us recall the origin of the non-commutativity when the geometry is globally $S^1\times \bbC \times\bbR$, following the discussions in \cite{GMN3,IOT11}. 
Considering the $S^1$ as the time direction, 
the partition function of $ S^1 \times \bbC \times \bbR $ is given by \begin{equation}
\Omega=\TrR{\calH}{(-1)^{F}e^{2\pi i \lambda \calJ_3}}
\end{equation}
where $\calH$ is the Hilbert space of the system, $\calJ_3=J_3+I_3$ is the sum of the spin along $\bbR$ and the Cartan of the $\SU(2)$ R-symmetry. 
Our parameter   $\frq$ is then given by $\frq = e^{\pi i\lambda}$.

Suppose now that we have a $\U(1)$ gauge theory, that the first loop $L_1$ is purely electrically charged with electric charge $e$
and that  the second loop $L_2$ is purely magnetically charged with magnetic charge $m$.
Then  there appears the Poynting vector carrying the angular momentum $J_{P}=(\hbar/2) em$ along the $\bbR$ direction.
The magnitude of $J_P$ is  independent of the distance between two particles but the sign depends on the ordering, and therefore we have \begin{equation}
L_1 \cdot L_2 = \frq^{-2em} L_2\cdot L_1.
\end{equation}

In the classical limit ($\frq \to 1$), this product becomes commutative.

\section{Networks and skein relations in 2d}\label{sec:qYM}

In this section we discuss possible types of networks in class S theories of type $\SU(N)$ and their skein relations. 
Our guiding principle is that they are described by the structure of the quantum groups underlying both the $q$-deformed Yang-Mills and the Liouville/Toda CFT, 
and that the skein relations are universal under an appropriate parameter identification.
The skein relations exhibited below already appear in the mathematical works \cite{MOY2000,JeongKim05,CKM12},
up to the overall factors and the changes in conventions. 

Skein relations introduce equivalence relations among all possible networks, and it would be extremely useful if we can pick a natural representative element out of a given equivalence class of networks allowing linear combinations of networks.
In this paper we at least give a general method to simplify a given network:
\begin{itemize}
\item In principle, edges in a network can carry arbitrary representations of $\SU(N)$. We will first rewrite them in terms of representations $\wedge^k \Box$  for $k=1,\ldots, N-1$. 
These are the fundamental representations in the mathematical terminology,\footnote{Contrary to the standard physics usage, we do not restrict the fundamental representation to be the defining $N$-dimensional representation in this paper.}
and within the diagrams we just denote them by $1$, $2$, \ldots, $N-1$.
\item Then we rewrite all crossings in terms of linear combinations of junctions that are at most rectangular. The concrete formulas are given in \eqref{eq:conjectural skein relation of general crossing}. 
We call this process \emph{crossing resolutions}.
\end{itemize}

In the $A_1$ case, these procedures eliminate all the crossings and no junctions remain, 
thus reproducing the classification in \cite{DMO09}.
In the $A_2$ case, we will see that all digons and rectangles can be eliminated, and we will find a natural representative for a given equivalence class of networks.
We will detail this process   in  Sec.~\ref{subsec:network classification on T^2}.

Hereafter we use a version of the standard quantum number defined as 
\begin{equation}
\langle n\rangle:=(-1)^{n-1}[n]:=(-1)^{n-1}\dfrac{\frq^n-\frq^{-n}}{\frq-\frq^{-1}}
\end{equation} 
and the factorial defined as 
\begin{equation}
\langle 0 \rangle!=1,\qquad
\langle n \rangle!=\langle n \rangle \langle n-1\rangle!.
\end{equation}

As this section is rather long, let us pause here to explain the organization: in Sec.~\ref{subsec:gen}, we start by recalling that codimension-4 operators of the 6d $\SUSYN{(2,0)}$ theory are labeled by representations and they can have junctions corresponding to the invariant tensors.
In Sec.~\ref{subsec:representation basis expansion}, we describe how an arbitrary representation can be rewritten in terms of just the fundamental representations of the form $\wedge^k \Box$.
In Sec.~\ref{subsec:canonical}, we then describe the trivalent junctions where three edges labeled by fundamental representations meet.
In Sec.~\ref{subsec:crossing}, we show how a crossing of two edges can be rewritten in terms of junctions. We start from the crossing of two edges labeled by $\Box$ and then describe the general case. 
In Sec.~\ref{subsec:reidemeister} we summarize the Reidemeister moves that are fundamental equivalence relations guaranteeing the isotopy invariance.
In Sec.~\ref{subsec:Skein} we note other useful skein relations that can be used to simplify networks.
Finally in Sec.~\ref{subsec:A3examples}, we explicitly display the skein relations for $A_2$ and $A_3$.

In general, it would also be important in the class S theory to study skein relations with  full and other punctures in \cite{Gaiotto0904} or networks ending on other punctures. We do not, however, consider such objects in this paper.

\subsection{Generalities} \label{subsec:gen}
Before proceeding, 
let us first recall the fact that a codimension-4 operator of the 6d $\SUSYN{(2,0)}$ theory of type $\SU(N)$ has a label given by a representation of $\SU(N)$, and how a multiple number of such operators can be joined. 

A cylinder of the $\SUSYN{(2,0)}$ theory gives rise to a 4d $\SUSYN{2}$ vector multiplet with gauge group $\SU(N)$.  
On the 4d side, then, we can consider the Wilson loop operator in a representation $R$ of $\SU(N)$. 
This should come from some codimension-4 operator of the $\SUSYN{(2,0)}$ theory wrapped around the cylinder. 
Then this codimension-4 operator also needs to be labeled by a representation $R$. 

When we multiply two parallel Wilson loops with representations $R_1$ and $R_2$, we get a Wilson loop with representation $R_1\otimes R_2$, and the product is commutative. 
The same should be then true among codimension-4 operators of the $\SUSYN{(2,0)}$ theory.

On the 4d side, three Wilson loops in representations $R_{1,2,3}$ can be joined at a point consistently if $R_1\otimes R_2\otimes R_3$ contains an $\SU(N)$ invariant subspace, or equivalently when there is an invariant tensor in this triple tensor product. 
The number of independent ways to join them is given by the number of linearly independent invariant tensors. 
Then, three codimension-4 operators of the $\SUSYN{(2,0)}$ theory labeled by $R_{1,2,3}$ can be joined along a one-dimensional subspace when there are invariant tensors in $R_1\otimes R_2\otimes R_3$. 
The number of distinct ways to connect is given by the number of linearly independent invariant tensors.

Since this should be an intrinsic property of codimension-4 operators of the 6d theory, we can join three codimension-4 operators along a one-dimensional loop on the 4d side. 
This gives a junction of three edges labeled by $R_1$, $R_2$, $R_3$ on the 2d side. 
Using many such junctions, we end up having networks on the 2d side, which are our main interest in this paper. 

\subsection{Restriction of labels}\label{subsec:representation basis expansion}

First, note that Wilson loop on the 4d side in a representation $R$ in a direction can be thought of as a Wilson loop in the representation $\bar R$ in the opposite direction.
This feature should also be carried over to the codimension-4 defects of the $\SUSYN{(2,0)}$ theory, and to the networks on the 2d side. This can be represented diagrammatically as 
\begin{eqnarray}
\raisebox{-0.3\height}{\tikzsetnextfilename{-figure-qYM-AN-right_arrow}
\begin{tikzpicture}
\def\l{1.6}
\coordinate (s) at (0,0);
\coordinate (t) at ($(s)+(\l,0)$);

\draw[end arrow] (s) -- (t) node[right]{$R$};
\end{tikzpicture}
}
=\raisebox{-0.3\height}{\tikzsetnextfilename{-figure-qYM-AN-left_arrow}
\begin{tikzpicture}
\def\l{1.6}
\coordinate (s) at (0,0);
\coordinate (t) at ($(s)+(-\l,0)$);

\draw[end arrow] (s) -- (t) node[left]{$R^\ast$};
\end{tikzpicture}
}.
\end{eqnarray}

As it is cumbersome to use arbitrary representations $R$ as labels,  we next rewrite them in terms of fundamental representations $\wedge^k \Box$, $k=1,\ldots, N-1$. 
An irreducible representation $R$ can be specified by a Young diagram $(\ell_i)$
where $\ell_i$ is the number of boxes in the $i$-th row, so that $\sum \ell_i=N$.
For example, $\wedge^k \Box$ is  represented by $(\underbrace{1\;1\;1 \ldots 1}_{k})=(1^k)$.

Note that any symmetric polynomial of $N$ variables $x_1,x_2,\ldots,x_N$, under a constraint $x_1x_2\ldots x_N=1$, can be written as a polynomial of the elementary symmetric polynomials. 
As a character $\chi_{R}(\textrm{diag}(x_1,\ldots,x_N))$ of $\SU(N)$ in a representation $R$ is such a symmetric polynomial,
and $\chi_{(1^k)}(\textrm{diag}(x_1,\ldots,x_N))$ for $k=0,1,2,\ldots,N-1$ are exactly elementary symmetric polynomials,
this means that any representation $R$ can be decomposed as the direct sum (allowing negative integral coefficients) of the tensor products of $\wedge^k \Box$. 

For example, we have the equalities
\begin{align}
\chi_{(2)} & =\chi_{(1)}^2-\chi_{(1^2)}, &
 \chi_{(21)} &=\chi_{(1^2)}\chi_{(1)}-\chi_{(1^3)},\\
\chi_{(3)} &=\chi_{(1)}^3-2\chi_{(1^2)}\chi_{(1)}+\chi_{(1^3)}, & 
\chi_{(2^2)}&=\chi_{(1^2)}\chi_{(1^2)}-\chi_{(1)}\chi_{(1^3)} 
\end{align}
which we can diagrammatically depict, in the case of closed loops, as
\begin{align}
 \raisebox{-0.25\height}{\tikzsetnextfilename{-figure-qYM-rep_basis_exp-2}
\begin{tikzpicture}
\def\R{0.5}

\coordinate (O) at (0,0);

\circlearrow{(O)}{\R}{90}{0.3}{above}{\tiny $\yng(2)$}
\draw[black,thick] (O) circle (\R);
\end{tikzpicture}
}
&=
\raisebox{-0.35\height}{\tikzsetnextfilename{-figure-qYM-rep_basis_exp-1_1}
\begin{tikzpicture}
\def\RI{0.4}
\def\RO{0.6}

\coordinate (O) at (0,0);

\circlearrow{(O)}{\RI}{90}{0.3}{below}{\tiny $1$}
\draw[black,thick] (O) circle (\RI);
\circlearrow{(O)}{\RO}{90}{0.3}{above}{\tiny $1$}
\draw[black,thick] (O) circle (\RO);
\end{tikzpicture}
}
-
\raisebox{-0.35\height}{\tikzsetnextfilename{-figure-qYM-rep_basis_exp-11}
\begin{tikzpicture}
\def\R{0.5}

\coordinate (O) at (0,0);

\circlearrow{(O)}{\R}{90}{0.3}{above}{\tiny $2$}
\draw[black,thick] (O) circle (\R);
\end{tikzpicture}
}, &
\raisebox{-0.2\height}{\tikzsetnextfilename{-figure-qYM-rep_basis_exp-21}
\begin{tikzpicture}
\def\R{0.5}

\coordinate (O) at (0,0);

\circlearrow{(O)}{\R}{90}{0.3}{above}{\tiny $\yng(2,1)$}
\draw[black,thick] (O) circle (\R);
\end{tikzpicture}
}
&=
\raisebox{-0.35\height}{\tikzsetnextfilename{-figure-qYM-rep_basis_exp-11_1}
\begin{tikzpicture}
\def\RI{0.4}
\def\RO{0.6}

\coordinate (O) at (0,0);

\circlearrow{(O)}{\RI}{90}{0.3}{below}{\tiny $1$}
\draw[black,thick] (O) circle (\RI);
\circlearrow{(O)}{\RO}{90}{0.3}{above}{\tiny $2$}
\draw[black,thick] (O) circle (\RO);
\end{tikzpicture}
}
-
\raisebox{-0.35\height}{\tikzsetnextfilename{-figure-qYM-rep_basis_exp-111}
\begin{tikzpicture}
\def\R{0.5}

\coordinate (O) at (0,0);

\circlearrow{(O)}{\R}{90}{0.3}{above}{\tiny $3$}
\draw[black,thick] (O) circle (\R);
\end{tikzpicture}
},
\\
 \raisebox{-0.2\height}{\tikzsetnextfilename{-figure-qYM-rep_basis_exp-22}
\begin{tikzpicture}
\def\R{0.5}

\coordinate (O) at (0,0);

\circlearrow{(O)}{\R}{90}{0.3}{above}{\tiny $\yng(2,2)$}
\draw[black,thick] (O) circle (\R);
\end{tikzpicture}
} 
&=
\raisebox{-0.35\height}{\tikzsetnextfilename{-figure-qYM-rep_basis_exp-11_11}
\begin{tikzpicture}
\def\RI{0.4}
\def\RO{0.6}

\coordinate (O) at (0,0);

\circlearrow{(O)}{\RI}{90}{0.3}{below}{\tiny $2$}
\draw[black,thick] (O) circle (\RI);
\circlearrow{(O)}{\RO}{90}{0.3}{above}{\tiny $2$}
\draw[black,thick] (O) circle (\RO);
\end{tikzpicture}
}
-
\raisebox{-0.35\height}{\tikzsetnextfilename{-figure-qYM-rep_basis_exp-1_111}
\begin{tikzpicture}
\def\RI{0.4}
\def\RO{0.6}

\coordinate (O) at (0,0);

\circlearrow{(O)}{\RI}{90}{0.3}{below}{\tiny $1$}
\draw[black,thick] (O) circle (\RI);
\circlearrow{(O)}{\RO}{90}{0.3}{above}{\tiny $3$}
\draw[black,thick] (O) circle (\RO);
\end{tikzpicture}
},
& \raisebox{-0.25\height}{\tikzsetnextfilename{-figure-qYM-rep_basis_exp-3}
\begin{tikzpicture}
\def\R{0.5}

\coordinate (O) at (0,0);

\circlearrow{(O)}{\R}{90}{0.3}{above}{\tiny $\yng(3)$}
\draw[black,thick] (O) circle (\R);
\end{tikzpicture}
}
&=
\raisebox{-0.35\height}{\tikzsetnextfilename{-figure-qYM-rep_basis_exp-1_1_1}
\begin{tikzpicture}
\def\RI{0.4}
\def\RM{0.6}
\def\RO{0.8}

\coordinate (O) at (0,0);

\circlearrow{(O)}{\RI}{90}{0.3}{below}{\tiny $1$}
\draw[black,thick] (O) circle (\RI);
\circlearrow{(O)}{\RM}{90}{0.3}{left}{\tiny $1$}
\draw[black,thick] (O) circle (\RM);
\circlearrow{(O)}{\RO}{90}{0.3}{above}{\tiny $1$}
\draw[black,thick] (O) circle (\RO);
\end{tikzpicture}
}
-
2\raisebox{-0.35\height}{\tikzsetnextfilename{-figure-qYM-rep_basis_exp-11_1}
\begin{tikzpicture}
\def\RI{0.4}
\def\RO{0.6}

\coordinate (O) at (0,0);

\circlearrow{(O)}{\RI}{90}{0.3}{below}{\tiny $1$}
\draw[black,thick] (O) circle (\RI);
\circlearrow{(O)}{\RO}{90}{0.3}{above}{\tiny $2$}
\draw[black,thick] (O) circle (\RO);
\end{tikzpicture}
}
+
\raisebox{-0.35\height}{\tikzsetnextfilename{-figure-qYM-rep_basis_exp-111}
\begin{tikzpicture}
\def\R{0.5}

\coordinate (O) at (0,0);

\circlearrow{(O)}{\R}{90}{0.3}{above}{\tiny $3$}
\draw[black,thick] (O) circle (\R);
\end{tikzpicture}
}.
\end{align}
These relations are locally applicable on parallel edges. 
Therefore, we can  insert some punctures or networks inside the circle, for example.

\subsection{Canonical junctions and removal of digons}\label{subsec:canonical}

\subsubsection{Canonical junctions}
Now our  edges are labeled  by the fundamental representations $\wedge^k \Box$, $k=0,1,\ldots, N-1$.  
We can just use the integer $k$ to label the edge, and an edge labeled by $0$ can be removed. 
Reversing the orientation now corresponds to replacing the label $k$ by $N-k$.
An edge labeled by $k$ has charge $k$ under the  center  of $\SU(N)$, 
and therefore we call these integer labels as the \emph{charge}.
\!\footnote{Note that it is a special property of $A_k$ that there is the one-to-one correspondence of the set of the fundamental representations $\wedge^k\Box$ including the trivial one and the charge under the center.}

For each trivalent junction, the sum of three inflowing charges must equal to zero modulo $N$.
Say we have three edges labeled by $a$, $b$ and $c=a+b$. 
There is only a single invariant tensor in $\wedge^a \Box \otimes \wedge^b \Box \otimes \wedge^{N-c} \Box$, and this corresponds to the projection from $\wedge^a \Box\otimes \wedge^b \Box$ to $\wedge^{c=a+b} \Box$. 
Therefore, there is no need to place a label on a junction to distinguish the possible invariant tensors.

\begin{figure}[h]
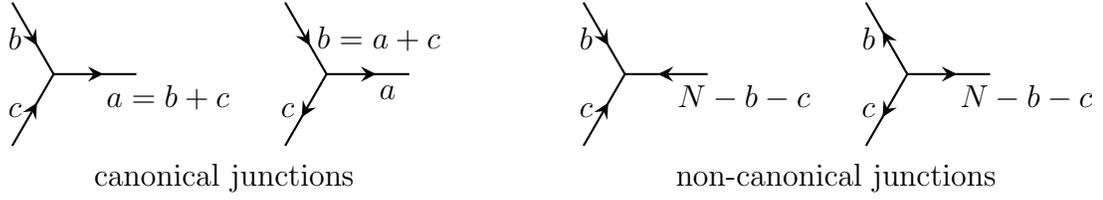

\centering
\begin{tabular}{cc}
\begin{minipage}{0.5\hsize}
\centering
\input{figure/qYM/canonical1.tikz}
\input{figure/qYM/canonical2.tikz}\\
canonical junctions
\end{minipage}
\begin{minipage}{0.5\hsize}
\centering
\input{figure/qYM/noncanonical1.tikz}
\input{figure/qYM/noncanonical2.tikz}\\
non-canonical junctions
\end{minipage}
\end{tabular}
\caption{Canonical and non-canonical junctions\label{fig:canonical}}
\end{figure}

Sometimes these labels $k$ are then taken to be defined modulo $N$ as in \cite{Xie1304,Bullimore13},
but it is useful to consider them just as integers between 0 and $N-1$.
This is because we can  write down the invariant tensor 
rather explicitly using the quantum group representation theory
when the net inflowing charge to a junction vanishes in $\bbZ$.
We call such a junction  \emph{canonical}.
We call a junction non-canonical if the net inflowing charge vanishes only in $\bbZ_N$.
 See Fig.~\ref{fig:canonical} for examples. 

Let us now describe the invariant tensors associated to the canonical junctions.
Let $\Box$ be spanned by the vectors $e_1$, \ldots, $e_N$. 
We define the $\frq$-deformed wedge product by the rule \begin{equation}
e_i \wedge e_j = -\frq  e_j \wedge e_i, \qquad (i \le j)
\end{equation} where in particular $e_i \wedge e_i=0$ where we do not sum over $i$.
This defines the projection $\pi$ from $\Box\otimes \Box $ to $\wedge^2\Box$ by \begin{equation}
\pi_{1,1\to 2}:  e_i \otimes e_j \mapsto e_i \wedge e_j.
\end{equation}
Furthermore, $\wedge^2 \Box$ can be naturally embedded within $\Box\otimes \Box$ by the rule
\begin{equation}
\iota_{2\to 1,1}: e_i \wedge e_j \mapsto -\frq e_i \otimes e_j  + e_j \otimes e_i, \qquad (i<j).
\end{equation}

More generally, we associate to any canonical junction that combines labels $a$, $b$ to $a+b$ the projection \begin{equation}
\pi_{a,b\to a+b}:   e_{i_1}\wedge \cdots \wedge e_{i_a} \otimes e_{j_1}\wedge \cdots \wedge e_{j_b} \mapsto e_{i_1}\wedge \cdots \wedge e_{i_a} \wedge e_{j_1}\wedge \cdots \wedge e_{j_b}
\end{equation} and to any canonical junction that splits the label $a+b$ to two labels $a$, $b$ 
the map $ \iota_{a+b\to a,b}$ where
\begin{multline}
\iota_{a+b\to a,b}: 
e_{k_1}\wedge \cdots \wedge e_{k_{a+b}} \\
\mapsto (-\frq)^{ab} \sum_{i_1< \cdots <i_a,\ j_1< \cdots <j_b}
(-\frq^{-1})^{n(i,j;k)}  e_{i_1}\wedge \cdots \wedge e_{i_a} \otimes e_{j_1}\wedge \cdots \wedge e_{j_b} 
\end{multline}
where we assume $k_1<\cdots<k_{a+b}$,
the sum is over the disjoint split of indices \begin{equation}
\{k_1,\ldots,k_{a+b}\} = \{i_1,\ldots,i_a\} \sqcup \{j_1,\ldots,j_b\},
\end{equation}
and $n(i,j;k)$ is the minimal number of adjacent transpositions to bring the sequence 
$i_1,\ldots,i_a,j_1,\ldots,j_b$ to $k_1,\ldots,k_{a+b}$. 
These maps are described in more detail mathematically in \cite{CT14}.

Then these maps $\pi_{a,b\to a+b}$ and $\iota_{a+b\to a,b}$ naturally combine according to the following diagrams together with the ones with reversed arrows:
\begin{eqnarray}
\raisebox{-0.5\height}{\tikzsetnextfilename{-figure-qYM-flip-composition1}
\begin{tikzpicture}
\def\l{0.6}
\coordinate (O) at (0,0);
\coordinate (U) at ($(O)+(90:\l)$);
\coordinate (M) at ($(O)+(-120:\l)$);
\coordinate (D1) at ($(M)+(-120:\l)$);
\coordinate (D2) at ($(M)+(-60:\l)$);
\coordinate (D3) at ($(O)+(-60:2*\l)$);

\draw[mid arrow] (D1) node[below]{$a$} -- (M);
\draw[mid arrow] (D2) node[below]{$b$} -- (M);
\draw[mid arrow] (D3) node[below]{$c$} -- (O);
\draw[mid arrow] (M) -- node[above left]{$x$} (O);
\draw[mid arrow] (O) -- (U) node[above]{$d$};
\end{tikzpicture}
}
=
\raisebox{-0.5\height}{\tikzsetnextfilename{-figure-qYM-flip-composition2}
\begin{tikzpicture}
\def\l{0.6}
\coordinate (O) at (0,0);
\coordinate (U) at ($(O)+(90:\l)$);
\coordinate (M) at ($(O)+(-60:\l)$);
\coordinate (D1) at ($(O)+(-120:2*\l)$);
\coordinate (D2) at ($(M)+(-120:\l)$);
\coordinate (D3) at ($(M)+(-60:\l)$);

\draw[mid arrow] (D1) node[below]{$a$} -- (O);
\draw[mid arrow] (D2) node[below]{$b$} -- (M);
\draw[mid arrow] (D3) node[below]{$c$} -- (M);
\draw[mid arrow] (M) -- node[above right]{$y$} (O);
\draw[mid arrow] (O) -- (U) node[above]{$d$};
\end{tikzpicture}
}\label{eq:flip}
\end{eqnarray}
or equivalently 
\begin{eqnarray}
\raisebox{-0.5\height}{\tikzsetnextfilename{-figure-qYM-flip-s-channel}
\begin{tikzpicture}
\def\l{1.5}
\coordinate (UL) at (0,0);
\coordinate (DL) at ($(UL)+(0,-\l)$);
\coordinate (UR) at ($(UL)+(\l,0)$);
\coordinate (DR) at ($(UR)+(0,-\l)$);
\coordinate (LM) at ($(UL)+(\l/4,-\l/2)$);
\coordinate (RM) at ($(UR)+(-\l/4,-\l/2)$);

\draw[mid arrow] (DR) node[below]{$c$} -- (RM);
\draw[mid arrow] (DL) node[below]{$b$} -- (LM);
\draw[mid arrow] (LM) -- node[above]{$x$} (RM);
\draw[mid arrow] (UL) node[above]{$a$} -- (LM);
\draw[mid arrow] (RM) -- (UR) node[above]{$d$};
\end{tikzpicture}
}
=
\raisebox{-0.5\height}{\tikzsetnextfilename{-figure-qYM-flip-t-channel}
\begin{tikzpicture}
\def\l{1.5}
\coordinate (UL) at (0,0);
\coordinate (DL) at ($(UL)+(0,-\l)$);
\coordinate (UR) at ($(UL)+(\l,0)$);
\coordinate (DR) at ($(UR)+(0,-\l)$);
\coordinate (UM) at ($(UL)+(\l/2,-\l/4)$);
\coordinate (DM) at ($(DL)+(\l/2,\l/4)$);

\draw[mid arrow] (DR) node[below]{$c$} -- (DM);
\draw[mid arrow] (DL) node[below]{$b$} -- (DM);
\draw[mid arrow] (DM) -- node[right]{$y$} (UM);
\draw[mid arrow] (UL) node[above]{$a$} -- (UM);
\draw[mid arrow] (UM) -- (UR) node[above]{$d$};
\end{tikzpicture}
}\label{eq:flap}.
\end{eqnarray}
Here $x=a+b=d-c$, $y=b+c=d-a$ and $d=a+b+c$.

\subsubsection{Removal of digons}\label{subsec:removal}

Now, we can check that any digons can be removed as
\begin{eqnarray}
\raisebox{-0.45\height}{\tikzsetnextfilename{-figure-qYM-dimension_of_multi_juntion}
\begin{tikzpicture}
\def\w{2.8}
\def\h{1.4}
\coordinate (S) at (0,0);
\coordinate (T) at ($(S)+(0,\h)$);

\fill[black] (S) circle (0.08);
\fill[black] (T) circle (0.08);

\draw[mid arrow] (S) .. controls ($(S)+(-0.9*\w,0.3*\h)$) and ($(T)+(-0.9*\w,-0.3*\h)$) .. node[left]{$i_1$} (T);
\draw[mid arrow] (S) .. controls ($(S)+(-0.6*\w,0.3*\h)$) and ($(T)+(-0.6*\w,-0.3*\h)$) .. node[left]{$i_2$} (T);
\draw[mid arrow] (S) .. controls ($(S)+(-0.3*\w,0.3*\h)$) and ($(T)+(-0.3*\w,-0.3*\h)$) .. node[left]{$i_3$} (T);
\foreach \n in {0,...,3}{\fill[black] ($(S)+(-0.1*\w+\n *0.2*\w,0.5*\h)$) circle (0.05);}
\draw[mid arrow] (S) .. controls ($(S)+(0.9*\w,0.3*\h)$) and ($(T)+(0.9*\w,-0.3*\h)$) .. node[right]{$i_\ell$} (T);
\draw[no arrow] (S) -- ($(S)+(0,0.2*\h)$);
\draw[no arrow] (S) .. controls ($(S)+(0.08*\w,0.1*\h)$) .. ($(S)+(0.1*\w,0.2*\h)$);
\draw[no arrow] (S) .. controls ($(S)+(0.16*\w,0.1*\h)$) .. ($(S)+(0.2*\w,0.2*\h)$);
\draw[no arrow] (T) -- ($(T)-(0,0.2*\h)$);
\draw[no arrow] (T) .. controls ($(T)+(0.08*\w,-0.1*\h)$) .. ($(T)+(0.1*\w,-0.2*\h)$);
\draw[no arrow] (T) .. controls ($(T)+(0.16*\w,-0.1*\h)$) .. ($(T)+(0.2*\w,-0.2*\h)$);
\draw[mid arrow] ($(S)-(0,0.3*\h)$) -- node[right]{$k$} (S);
\draw[latter arrow] (T) -- node[right]{$k$} ($(T)+(0,0.3*\h)$);
\end{tikzpicture}
}
=\dfrac{\langle k\rangle !}{\langle i_1\rangle !\langle i_2\rangle ! \ldots \langle i_l\rangle !} \quad
\raisebox{-0.45\height}{\tikzsetnextfilename{-figure-qYM-dimension_of_multi_juntion_contracted}
\begin{tikzpicture}
\def\w{2.8}
\def\h{1.4}
\coordinate (S) at (0,0);
\coordinate (T) at ($(S)+(0,\h)$);

\draw[mid arrow] ($(S)-(0,0.3*\h)$) -- node[right]{$k$} ($(T)+(0,0.3*\h)$);
\end{tikzpicture}
}
\label{eq:slimming}
\end{eqnarray}
where $\displaystyle \sum_{a=1}^{\ell}i_{a}=k$.

Note that in the classical limit $\frq\to 1$ the prefactor becomes \begin{equation}
(-1)^{\sum_{a<b} i_a i_b} \frac{k!}{i_1 ! i_2 ! \cdots i_l! }
\end{equation} due to the fact that the classical limit of $\iota_{a+b\to a,b}$ 
is $(-1)^{ab}$ times the standard map that follows from the classical epsilon symbol.

This somewhat unusual sign is however necessary to match with the known skein relations in the Liouville/Toda theory,  and it also simplifies the signs appearing in the general crossing resolutions \eqref{eq:conjectural skein relation of general crossing}.
In the $q$-deformed Yang-Mills theory it would be more conventional to drop this sign
and the canonical junctions would be defined to be $\pi_{a,b\to a+b}$ and $(-1)^{ab}\iota_{a+b\to a,b}$. 
We will stick to the Liouville/Toda convention in this paper, except in Sec.~\ref{sec:T3junction}.

When the sum of $i_l$ is $N$, we can use the rule to evaluate  a network with two trivalent junctions, 
since edges labeled by $N$ can be removed. For example, when $i+j+k=N$, we have
\begin{eqnarray}
\raisebox{-0.45\height}{\tikzsetnextfilename{-figure-qYM-dim_for_two_trivalents}
\begin{tikzpicture}
\def\w{0.7}
\def\h{1.2}
\coordinate (S) at (0,0);
\coordinate (T) at ($(S)+(0,\h)$);

\fill[black] (S) circle (0.08);
\fill[black] (T) circle (0.08);

\tikzstyle instant arrow=[thick,postaction={decorate,decoration={markings, mark=at position .53 with {\arrow[scale=1.4]{stealth}}}}]
\draw[instant arrow] (S) .. controls ($(S)+(-\w,0.3*\h)$) and ($(T)+(-\w,-0.3*\h)$) .. node[left]{$i$} (T);
\draw[mid arrow] (S) .. controls ($(S)+(0,0.3*\h)$) and ($(T)+(0,-0.3*\h)$) .. node[left]{$j$} (T);
\draw[instant arrow] (S) .. controls ($(S)+(\w,0.3*\h)$) and ($(T)+(\w,-0.3*\h)$) .. node[left]{$k$} (T);
\end{tikzpicture}
}
=\dfrac{\langle N\rangle !}{\langle i\rangle !\langle j\rangle !\langle k\rangle !}.
\end{eqnarray}
More simply, we can evaluate a closed loop with label $k$ by considering it as a digon with edges labeled by $k$ and $N-k$: \begin{multline}
\tikzsetnextfilename{-figure-qYM-dimension}
\begin{tikzpicture}[baseline=(O.center)]
\def\R{0.5}

\coordinate (O) at (0,0);

\circlearrow{(O)}{\R}{90}{0.3}{above}{$k$}
\draw[black,thick] (O) circle (\R);
\end{tikzpicture}

= \frac{\langle N\rangle!}{\langle k \rangle! \langle N-k \rangle!} \\
= (-1)^{k(N-k)} \chi_{\wedge^k \Box} (\textrm{diag}(\frq^{N-1},\frq^{N-3},\ldots,\frq^{1-N}))
=(-1)^{k(N-k)} \dim_\frq \wedge^k \Box.
\end{multline}

Again, this shows that our convention is different by a factor of $(-1)^{k(N-k)}$ from the convention in the $q$-deformed Yang-Mills. 
We also see at this point  that, to compare with the skein relation of the Toda theory or the $q$-deformed Yang-Mills theory, we need to use the relation \begin{equation}
\frq = e^{\pi i b^2} = q_\text{SCI}^{1/2}.
\end{equation} 

\subsection{Crossing resolutions}
\label{subsec:crossing}

\subsubsection{The $\calR$ matrix}

Let us first discuss the best-known case: the $\calR$-matrix for $\Box \otimes \Box$ of $\SU(N)$,
which is given by 
\begin{eqnarray}
\calR=A ( Q  + \frq^{-1}  I_{\Box \otimes\Box} ).
\end{eqnarray}
Here,  $I_{V}$ is the identity operator on a vector space $V$,
$Q$ is an operator 
\begin{eqnarray}
Q=\sum_{i\neq j} e_{ij} \otimes e_{ji} - \frq \sum_{i<j} e_{ii} \otimes e_{jj} - \frq^{-1} \sum_{i>j} e_{ii} \otimes e_{jj}
\label{eq:Q}
\end{eqnarray} where $e_{ij}$ is a matrix whose only non-zero entry is 1 at the $i$-th row and $j$-th column
and $A$ is the overall normalization which we will be fixed later. 

The action of $Q$ on the base $e_a \otimes e_b$  of $\Box \otimes \Box$ is
\begin{equation}
Q(e_a\otimes e_b) = \begin{cases}
e_b\otimes e_a-\frq e_a \otimes e_b , & (a<b) \\
0, & (a=b) \\
e_b\otimes e_a- \frq^{-1} e_a \otimes e_b , & (a>b) 
\end{cases}.
\end{equation}
The resulting entries are the basis vectors of the second rank antisymmetric representation of $\SU_\frq(N)$. 
Indeed,  the operator $Q$ is the composition of the projection  $\pi_{1,1\to2} :\Box \otimes \Box \to\wedge^{2}_\frq \Box$ and the natural embedding  $\iota_{2\to,1,1}: \wedge^2_\frq \Box\to \Box \otimes \Box$: \begin{equation}
Q= \iota_{2\to 1,1}\pi_{1,1\to 2}.
\end{equation} 
Note also that $Q$ satisfies 
\begin{equation}
Q^2=-(\frq + \frq^{-1}) Q=\langle 2 \rangle Q.
\end{equation}
This is a special case of the digon elimination.

We can now represent the R-matrix $\calR$  diagrammatically as 
\begin{eqnarray}
\raisebox{-0.5\height}{\tikzsetnextfilename{-figure-qYM-R}
\begin{tikzpicture}
\def\l{1.4}
\coordinate (UL) at (0,0);
\coordinate (DL) at ($(UL)+(0,-\l)$);
\coordinate (UR) at ($(UL)+(\l,0)$);
\coordinate (DR) at ($(UR)+(0,-\l)$);

\draw[end arrow] (DR) node[below]{$\Box$} -- (UL);
\draw[line width=5.0pt,white] (DL) -- (UR);
\draw[end arrow] (DL) node[below]{$\Box$} -- (UR);
\end{tikzpicture}
}
=A \left( \raisebox{-0.4\height}{\tikzsetnextfilename{-figure-qYM-Q}
\begin{tikzpicture}
\def\l{1.4}
\coordinate (UL) at (0,0);
\coordinate (DL) at ($(UL)+(0,-\l)$);
\coordinate (UR) at ($(UL)+(\l,0)$);
\coordinate (DR) at ($(UR)+(0,-\l)$);
\coordinate (UM) at ($(UL)+(\l/2,-\l/4)$);
\coordinate (DM) at ($(DL)+(\l/2,\l/4)$);

\draw[mid arrow] (DR) node[below]{$\Box$} -- (DM);
\draw[mid arrow] (DL) node[below]{$\Box$} -- (DM);
\draw[mid arrow] (DM) -- node[right]{$\bigwedge^{\!2}\Box$} (UM);
\draw[mid arrow] (UM) -- (UL) node[above]{$\Box$};
\draw[mid arrow] (UM) -- (UR) node[above]{$\Box$};
\end{tikzpicture}
} + \frq^{-1} \raisebox{-0.5\height}{\tikzsetnextfilename{-figure-qYM-Id}
\begin{tikzpicture}
\def\l{1.4}
\coordinate (UL) at (0,0);
\coordinate (DL) at ($(UL)+(0,-\l)$);
\coordinate (UR) at ($(UL)+(\l,0)$);
\coordinate (DR) at ($(UR)+(0,-\l)$);

\draw[end arrow] (DR) node[below]{$\Box$} -- (UR);
\draw[end arrow] (DL) node[below]{$\Box$} -- (UL);
\end{tikzpicture}
} \right).
\label{eq:skein of fundamental R}
\end{eqnarray}

The inverse  of the R-matrix $\calR$ is 
\begin{eqnarray}
\calR^{-1}=A^{-1}(Q+ \frq I_{\Box \otimes\Box})
\end{eqnarray} that we represent as 
\begin{eqnarray}
\raisebox{-0.5\height}{\tikzsetnextfilename{-figure-qYM-inverseR}
\begin{tikzpicture}
\def\l{1.4}
\coordinate (UL) at (0,0);
\coordinate (DL) at ($(UL)+(0,-\l)$);
\coordinate (UR) at ($(UL)+(\l,0)$);
\coordinate (DR) at ($(UR)+(0,-\l)$);

\draw[end arrow] (DL) node[below]{$\Box$} -- (UR);
\draw[line width=5.0pt,white] (DR) -- (UL);
\draw[end arrow] (DR) node[below]{$\Box$} -- (UL);
\end{tikzpicture}
}
=A^{-1} \left( \raisebox{-0.4\height}{\tikzsetnextfilename{-figure-qYM-Q}
\begin{tikzpicture}
\def\l{1.4}
\coordinate (UL) at (0,0);
\coordinate (DL) at ($(UL)+(0,-\l)$);
\coordinate (UR) at ($(UL)+(\l,0)$);
\coordinate (DR) at ($(UR)+(0,-\l)$);
\coordinate (UM) at ($(UL)+(\l/2,-\l/4)$);
\coordinate (DM) at ($(DL)+(\l/2,\l/4)$);

\draw[mid arrow] (DR) node[below]{$\Box$} -- (DM);
\draw[mid arrow] (DL) node[below]{$\Box$} -- (DM);
\draw[mid arrow] (DM) -- node[right]{$\bigwedge^{\!2}\Box$} (UM);
\draw[mid arrow] (UM) -- (UL) node[above]{$\Box$};
\draw[mid arrow] (UM) -- (UR) node[above]{$\Box$};
\end{tikzpicture}
}
+ \frq \raisebox{-0.5\height}{\tikzsetnextfilename{-figure-qYM-Id}
\begin{tikzpicture}
\def\l{1.4}
\coordinate (UL) at (0,0);
\coordinate (DL) at ($(UL)+(0,-\l)$);
\coordinate (UR) at ($(UL)+(\l,0)$);
\coordinate (DR) at ($(UR)+(0,-\l)$);

\draw[end arrow] (DR) node[below]{$\Box$} -- (UR);
\draw[end arrow] (DL) node[below]{$\Box$} -- (UL);
\end{tikzpicture}
} \right)
\label{eq:skein of inverse fundamental R}
\end{eqnarray}
Below, we call the crossing \eqref{eq:skein of fundamental R} as positive
and the crossing \eqref{eq:skein of inverse fundamental R} as negative.

\paragraph{The $A_1$-case:}  In this case, $\bigwedge^{\!2} \Box$ is the trivial  one-dimensional representation and there is the pseudo-reality condition $\overline{\Box} \simeq \Box$ which we can diagrammatically write as 
\begin{eqnarray}
\raisebox{-0.3\height}{\tikzsetnextfilename{-figure-qYM-A1-right_arrow}
\begin{tikzpicture}
\def\l{1.6}
\coordinate (s) at (0,0);
\coordinate (t) at ($(s)+(\l,0)$);

\draw[end arrow] (s) -- (t) node[right]{$\Box$};
\end{tikzpicture}
}
=\raisebox{-0.3\height}{\tikzsetnextfilename{-figure-qYM-A1-left_arrow}
\begin{tikzpicture}
\def\l{1.6}
\coordinate (s) at (0,0);
\coordinate (t) at ($(s)+(-\l,0)$);

\draw[end arrow] (s) -- (t) node[left]{$\Box$};
\end{tikzpicture}
}
=:\tikzsetnextfilename{-figure-qYM-A1-no_arrow}
\begin{tikzpicture}
\def\l{1.6}
\coordinate (s) at (0,0);
\coordinate (t) at ($(s)+(\l,0)$);

\draw[white,thick] ($(s)-(0,0.1)$) -- ($(t)-(0,0.1)$);
\draw[thick] (s) -- (t);
\end{tikzpicture}
.
\end{eqnarray}

Then the general equation (\ref{eq:skein of fundamental R}) reduces to 
\begin{eqnarray}
\raisebox{-0.5\height}{\tikzsetnextfilename{-figure-qYM-A1-R}
\begin{tikzpicture}
\def\l{1.3}
\coordinate (UL) at (0,0);
\coordinate (DL) at ($(UL)+(0,-\l)$);
\coordinate (UR) at ($(UL)+(\l,0)$);
\coordinate (DR) at ($(UR)+(0,-\l)$);

\draw[very thick] (DR) -- (UL);
\draw[line width=5.0pt,white] (DL) -- (UR);
\draw[very thick] (DL) -- (UR);
\end{tikzpicture}
}
=\frq^{-1/2}\;\raisebox{-0.5\height}{\tikzsetnextfilename{-figure-qYM-A1-Id}
\begin{tikzpicture}
\def\w{1.3}
\def\h{1.3}
\coordinate (UL) at (0,0);
\coordinate (DL) at ($(UL)+(0,-\h)$);
\coordinate (UR) at ($(UL)+(\w,0)$);
\coordinate (DR) at ($(UR)+(0,-\h)$);

\draw[very thick] (DR) .. controls ($(DR)+(110:0.3*\h)$) and ($(UR)+(-110:0.3*\h)$) .. (UR);
\draw[very thick] (DL) .. controls ($(DL)+(70:0.3*\h)$) and ($(UL)+(-70:0.3*\h)$) .. (UL);
\end{tikzpicture}
}
+\frq^{1/2}\;\raisebox{-0.5\height}{\tikzsetnextfilename{-figure-qYM-A1-Q}
\begin{tikzpicture}
\def\w{1.3}
\def\h{1.3}
\coordinate (UL) at (0,0);
\coordinate (DL) at ($(UL)+(0,-\h)$);
\coordinate (UR) at ($(UL)+(\w,0)$);
\coordinate (DR) at ($(UR)+(0,-\h)$);

\draw[very thick] (DR) .. controls ($(DR)+(160:0.3*\w)$) and ($(DL)+(20:0.3*\w)$) .. (DL);
\draw[very thick] (UR) .. controls ($(UR)+(-160:0.3*\w)$) and ($(UL)+(-20:0.3*\w)$) .. (UL);
\end{tikzpicture}
}.
\end{eqnarray}
where we have set  $A=\frq^{1/2}$.
We then have 
\begin{eqnarray}
\frq^{-1/2}\;\raisebox{-0.5\height}{\tikzsetnextfilename{-figure-qYM-A1-R}
\begin{tikzpicture}
\def\l{1.3}
\coordinate (UL) at (0,0);
\coordinate (DL) at ($(UL)+(0,-\l)$);
\coordinate (UR) at ($(UL)+(\l,0)$);
\coordinate (DR) at ($(UR)+(0,-\l)$);

\draw[very thick] (DR) -- (UL);
\draw[line width=5.0pt,white] (DL) -- (UR);
\draw[very thick] (DL) -- (UR);
\end{tikzpicture}
}
-\frq^{1/2}\;\raisebox{-0.5\height}{\tikzsetnextfilename{-figure-qYM-A1-Rflipped}
\begin{tikzpicture}
\def\l{1.3}
\coordinate (UL) at (0,0);
\coordinate (DL) at ($(UL)+(0,-\l)$);
\coordinate (UR) at ($(UL)+(\l,0)$);
\coordinate (DR) at ($(UR)+(0,-\l)$);

\draw[very thick] (DL) -- (UR);
\draw[line width=5.0pt,white] (DR) -- (UL);
\draw[very thick] (DR) -- (UL);
\end{tikzpicture}
}
=(\frq^{-1}-\frq)\;\raisebox{-0.5\height}{\tikzsetnextfilename{-figure-qYM-A1-Id}
\begin{tikzpicture}
\def\w{1.3}
\def\h{1.3}
\coordinate (UL) at (0,0);
\coordinate (DL) at ($(UL)+(0,-\h)$);
\coordinate (UR) at ($(UL)+(\w,0)$);
\coordinate (DR) at ($(UR)+(0,-\h)$);

\draw[very thick] (DR) .. controls ($(DR)+(110:0.3*\h)$) and ($(UR)+(-110:0.3*\h)$) .. (UR);
\draw[very thick] (DL) .. controls ($(DL)+(70:0.3*\h)$) and ($(UL)+(-70:0.3*\h)$) .. (UL);
\end{tikzpicture}
}
\end{eqnarray}
These reproduce the standard skein relations of the Liouville theory found in \cite{AGGTV09,DGOT09} under the identification $\frq = e^{i\pi b^2}$.

The relation $Q^2=\langle 2\rangle  Q$ shows
\begin{eqnarray}
\raisebox{-0.5\height}{\tikzsetnextfilename{-figure-qYM-A1-circle}
\begin{tikzpicture}
\def\l{0.6}

\draw[very thick] (0,0) circle (\l);
\end{tikzpicture}
} = \langle 2\rangle = -\chi_\Box (\textrm{diag}(\frq,\frq^{-1})).
\end{eqnarray}
From this we see  that $\frq=q^{1/2}_\text{SCI}$ where $q_\text{SCI}$ is the parameter used commonly in the literature on the superconformal index. 
The minus sign here is a convention common in the Liouville/Toda literature,
i.e.~the definition of a loop in the representation $\Box$ differs by an overall minus sign
between the Liouville theory and the $q$-deformed Yang-Mills.

\paragraph{The $A_2$-case:}  Here we have $\overline\Box=\bigwedge^{\!2} \Box$, and therefore we have
\begin{eqnarray}
\raisebox{-0.3\height}{\tikzsetnextfilename{-figure-qYM-A2-right_arrow}
\begin{tikzpicture}
\def\l{1.6}
\coordinate (s) at (0,0);
\coordinate (t) at ($(s)+(\l,0)$);

\draw[end arrow] (s) -- (t) node[right]{$\Box$};
\end{tikzpicture}
}
=\raisebox{-0.3\height}{\tikzsetnextfilename{-figure-qYM-A2-left_arrow}
\begin{tikzpicture}
\def\l{1.6}
\coordinate (s) at (0,0);
\coordinate (t) at ($(s)+(-\l,0)$);

\draw[end arrow] (s) -- (t) node[left]{$\bigwedge^2\Box$};
\end{tikzpicture}
}.
\end{eqnarray}
As we now only have one type of the label $\Box$, we can drop it altogether. 
The general R-matrix (\ref{eq:skein of fundamental R}) then becomes
\begin{eqnarray}
\raisebox{-0.5\height}{\tikzsetnextfilename{-figure-qYM-A2-R}
\begin{tikzpicture}
\def\l{1.4}
\coordinate (UL) at (0,0);
\coordinate (DL) at ($(UL)+(0,-\l)$);
\coordinate (UR) at ($(UL)+(\l,0)$);
\coordinate (DR) at ($(UR)+(0,-\l)$);

\draw[end arrow] (DR) -- (UL);
\draw[line width=5.0pt,white] (DL) -- (UR);
\draw[end arrow] (DL)  -- (UR);
\end{tikzpicture}
}
= \frq^{1/3} \raisebox{-0.4\height}{\tikzsetnextfilename{-figure-qYM-A2-Q}
\begin{tikzpicture}
\def\l{1.4}
\coordinate (UL) at (0,0);
\coordinate (DL) at ($(UL)+(0,-\l)$);
\coordinate (UR) at ($(UL)+(\l,0)$);
\coordinate (DR) at ($(UR)+(0,-\l)$);
\coordinate (UM) at ($(UL)+(\l/2,-\l/4)$);
\coordinate (DM) at ($(DL)+(\l/2,\l/4)$);

\draw[mid arrow] (DR)  -- (DM);
\draw[mid arrow] (DL)  -- (DM);
\draw[mid arrow] (UM) --  (DM);
\draw[mid arrow] (UM) -- (UL);
\draw[mid arrow] (UM) -- (UR);
\end{tikzpicture}
} + \frq^{-2/3} \raisebox{-0.5\height}{\tikzsetnextfilename{-figure-qYM-A2-Id}
\begin{tikzpicture}
\def\l{1.4}
\coordinate (UL) at (0,0);
\coordinate (DL) at ($(UL)+(0,-\l)$);
\coordinate (UR) at ($(UL)+(\l,0)$);
\coordinate (DR) at ($(UR)+(0,-\l)$);

\draw[mid arrow] (DR)  -- (UR);
\draw[mid arrow] (DL)  -- (UL);
\end{tikzpicture}
} \label{eq:A2skein}
\end{eqnarray}
where we have set  $A=\frq^{1/3}$.
This  reproduces the fundamental  skein relations of the $\SU(3)$ Toda  theory found in \cite{Bullimore13}, again under the identification $\frq = e^{i\pi b^2}$.

\paragraph{General case:} The analysis so far suggests that we should take \begin{equation}
A=\frq^{\tfrac{1}{N}}
\end{equation} in general. 
As we will see soon, this is consistent with the  general crossing resolutions \eqref{eq:conjectural skein relation of general crossing}.

\subsubsection{General crossing resolutions}
Now let us move on to the crossing resolutions in the general case.
The expression was found in \cite{MOY2000} up to an overall factor, which we quote here:
\begin{eqnarray}
\raisebox{-0.4\height}{\tikzsetnextfilename{-figure-qYM-general_R}
\begin{tikzpicture}
\def\l{1.6}
\coordinate (UL) at (0,0);
\coordinate (DL) at ($(UL)+(0,-\l)$);
\coordinate (UR) at ($(UL)+(\l,0)$);
\coordinate (DR) at ($(UR)+(0,-\l)$);

\draw[end arrow] (DR) -- (UL) node[above]{$a$};
\draw[line width=5.0pt,white] (DL) -- (UR);
\draw[end arrow] (DL) -- (UR) node[above]{$b$};
\end{tikzpicture}
}
=
\frq^{\frac{ab}{N}}
\sum_{i=0}^{s}
\frq^{-i}
\raisebox{-0.5\height}{\tikzsetnextfilename{-figure-qYM-general_Q}
\begin{tikzpicture}
\def\l{1.1}
\def\ll{0.7}
\coordinate (UL) at (0,0);
\coordinate (DL) at ($(UL)+(0,-\l)$);
\coordinate (UR) at ($(UL)+(\l,0)$);
\coordinate (DR) at ($(UR)+(0,-\l)$);
\coordinate (Lout) at ($(UL)+(135:\ll)$);
\coordinate (Rout) at ($(UR)+(45:\ll)$);
\coordinate (Lin) at ($(DL)+(-135:\ll)$);
\coordinate (Rin) at ($(DR)+(-45:\ll)$);

\draw[mid arrow] (DL) -- node[left]{$i$} (UL);
\draw[mid arrow] (DR) -- node[right]{$a+b-i$} (UR);
\draw[mid arrow] (DL) -- node[below]{$b-i$} (DR);
\draw[mid arrow] (UR) -- node[above]{$a-i$} (UL);
\draw[mid arrow] (UL) -- (Lout) node[above]{$a$};
\draw[mid arrow] (UR) -- (Rout) node[above]{$b$};
\draw[mid arrow] (Lin) node[below]{$b$} -- (DL);
\draw[mid arrow] (Rin) node[below]{$a$} -- (DR);
\end{tikzpicture}
}
\label{eq:conjectural skein relation of general crossing}
\label{eq:final form of conjectural skein relation of general crossing}
\end{eqnarray}
where $s=\min(a,b,N-a,N-b)$ and $a,b=0,1,2,\ldots,N-1$.
With this choice of the overall factor, this equality is invariant with a reversal of an arrow and the rotation of the diagrams by $90^\circ$. 
Note also that when $a=b=1$, this equality reduces to \eqref{eq:skein of fundamental R} we already discussed.

Let us introduce the names to the fundamental objects on the right hand side of \eqref{eq:conjectural skein relation of general crossing}:
\begin{eqnarray}
Q^{ab}_{\;\;(i)}:=
\raisebox{-0.5\height}{\tikzsetnextfilename{-figure-qYM-general_Q}
\begin{tikzpicture}
\def\l{1.1}
\def\ll{0.7}
\coordinate (UL) at (0,0);
\coordinate (DL) at ($(UL)+(0,-\l)$);
\coordinate (UR) at ($(UL)+(\l,0)$);
\coordinate (DR) at ($(UR)+(0,-\l)$);
\coordinate (Lout) at ($(UL)+(135:\ll)$);
\coordinate (Rout) at ($(UR)+(45:\ll)$);
\coordinate (Lin) at ($(DL)+(-135:\ll)$);
\coordinate (Rin) at ($(DR)+(-45:\ll)$);

\draw[mid arrow] (DL) -- node[left]{$i$} (UL);
\draw[mid arrow] (DR) -- node[right]{$a+b-i$} (UR);
\draw[mid arrow] (DL) -- node[below]{$b-i$} (DR);
\draw[mid arrow] (UR) -- node[above]{$a-i$} (UL);
\draw[mid arrow] (UL) -- (Lout) node[above]{$a$};
\draw[mid arrow] (UR) -- (Rout) node[above]{$b$};
\draw[mid arrow] (Lin) node[below]{$b$} -- (DL);
\draw[mid arrow] (Rin) node[below]{$a$} -- (DR);
\end{tikzpicture}
}
=
\raisebox{-0.5\height}{\tikzsetnextfilename{-figure-qYM-general_Q2}
\begin{tikzpicture}
\def\l{1.1}
\def\ll{0.7}
\coordinate (UL) at (0,0);
\coordinate (DL) at ($(UL)+(0,-\l)$);
\coordinate (UR) at ($(UL)+(\l,0)$);
\coordinate (DR) at ($(UR)+(0,-\l)$);
\coordinate (Lout) at ($(UL)+(135:\ll)$);
\coordinate (Rout) at ($(UR)+(45:\ll)$);
\coordinate (Lin) at ($(DL)+(-135:\ll)$);
\coordinate (Rin) at ($(DR)+(-45:\ll)$);

\draw[mid arrow] (DL) -- node[left]{$a+b-i$} (UL);
\draw[mid arrow] (DR) -- node[right]{$i$} (UR);
\draw[mid arrow] (UL) -- node[above]{$b-i$} (UR);
\draw[mid arrow] (DR) -- node[below]{$a-i$} (DL);
\draw[mid arrow] (UL) -- (Lout) node[above]{$a$};
\draw[mid arrow] (UR) -- (Rout) node[above]{$b$};
\draw[mid arrow] (Lin) node[below]{$b$} -- (DL);
\draw[mid arrow] (Rin) node[below]{$a$} -- (DR);
\end{tikzpicture}
}.
\label{eq:equivalent relation between Q's}
\end{eqnarray}
Note that the number of possible choices of $i$ matches with the number of irreducible summands of the decomposition of the tensor product $\wedge^a \Box\otimes \wedge^b \Box$.
We expect  that all these $Q^{ab}_{\;\;(i)}$ cannot be further decomposed as parts of networks. 

\subsubsection{The intersection number and the powers of $\frq$}
Let us briefly discuss the significance of the prefactor $\frq^{\frac{ab}{N}}$
in \eqref{eq:conjectural skein relation of general crossing}.
In general, two loop operators in a class S theory of type $\SU(N)$ can be mutually nonlocal,
and the nonlocality can be measured in terms of the Dirac pairing that takes values in $\bbZ_N$ \cite{Tachikawa:2013hya}.
In terms of the 2d networks realizing the 4d loop operators,  
the Dirac pairing is given by their intersection number. 
We can define it by assigning a local intersection number to a crossing  as follows:
\begin{align}
\vcenter{\hbox{\tikzsetnextfilename{-figure-qYM-general_R}
\begin{tikzpicture}
\def\l{1.6}
\coordinate (UL) at (0,0);
\coordinate (DL) at ($(UL)+(0,-\l)$);
\coordinate (UR) at ($(UL)+(\l,0)$);
\coordinate (DR) at ($(UR)+(0,-\l)$);

\draw[end arrow] (DR) -- (UL) node[above]{$a$};
\draw[line width=5.0pt,white] (DL) -- (UR);
\draw[end arrow] (DL) -- (UR) node[above]{$b$};
\end{tikzpicture}
}} : &  + ab, &
\vcenter{\hbox{\tikzsetnextfilename{-figure-qYM-general_inverseR}
\begin{tikzpicture}
\def\l{1.6}
\coordinate (UL) at (0,0);
\coordinate (DL) at ($(UL)+(0,-\l)$);
\coordinate (UR) at ($(UL)+(\l,0)$);
\coordinate (DR) at ($(UR)+(0,-\l)$);

\draw[end arrow] (DL) -- (UR) node[above]{$b$};
\draw[line width=5.0pt,white] (DR) -- (UL);
\draw[end arrow] (DR) -- (UL) node[above]{$a$};
\end{tikzpicture}
}} : &  -ab.
\end{align}
This is consistent with the reversal of arrows, since it sends the label $a$ to $N-a$.

The intersection number $I(\Gamma_1,\Gamma_2)$ of two networks $\Gamma_1$ and $\Gamma_2$ is then defined by summing the contributions from all the crossings:
\begin{eqnarray}
\sum_{c \;: \;{\rm crossing}} \textrm{sign}(c) a^{(1)}_c a^{(2)}_c \in \bbZ_N
\end{eqnarray}
where $\rm{sign}(c)$ is the sign of the crossing $c$ and $a^{(i)}_c$ is the charge of $\Gamma_i$ at $c$.

In the Liouville/Toda setup, we expect the expectation value of any network without crossings is a single-valued function of $\frq=e^{\pi i b^2}$ invariant under $\frq \to e^{2\pi i}\frq$.
Similarly, in the superconformal index,  the expectation value of a loop operator on the 4d side is a single valued function of $\frq=q^{1/2}_\text{SCI}$, since the index is a trace $\Tr (-1)^F q^{\Delta-I_3}$ and the scaling dimensions $\Delta$ of a class S theory are integral or half-integral.

Non-invariance of the expectation value under $\frq\to e^{2\pi i}\frq$ then captures the mutual non-locality, 
and we can  think of the prefactor $\frq^{\frac{ab}{N}}$ in \eqref{eq:conjectural skein relation of general crossing} as encoding the difference in the local intersection number between the left hand side and the right hand side, to keep track of this non-locality.
The difference in the powers of $\frq$ among different summands 
in the resolutions of the crossings should be integral, and the relation \eqref{eq:conjectural skein relation of general crossing} indeed satisfies this requirement.

\subsubsection{$SL(2,\bbZ)$ action on the torus}
We can define the operation $I$ which we call the \emph{inversion} by reversing all the arrows simultaneously.
This is an involution that we can identify with the charge conjugation on the 4d side. 

When the network is on a torus, we can also consider the action of $SL(2,\bbZ)$ on the networks.
Two basic actions  are the $T$-action and the $S$-action.
$T$ corresponds to sending $\alpha\mapsto \alpha$, $\beta\mapsto \beta+\alpha$
and $S$ corresponds to  $\alpha\mapsto \beta$, $\beta\to -\alpha$,
where $\alpha$, $\beta$ are two bases of the 1-cycles on the torus.  

The operation $C=S^2$  generates the center of $SL(2,\bbZ)$ and is the charge conjugation action in the 4d $\SUSYN{4}$ SYM theory.
Then we need to have two operations, $C$ and $I$, to be consistent on the torus.
The relation  (\ref{eq:equivalent relation between Q's}) relating two forms of the networks representing the same object $Q^{ab}_{\;\;(i)}$ is exactly the one required to have $C=I$, 
when these networks are put on the torus and open edges are connected to the opposite ones.

\subsection{Reidemeister moves}
\label{subsec:reidemeister}

In knot theory, a projective representation in two dimension of knots and links in a three dimensional space is not unique, 
and any different representations can be mapped to each other by a combination of three so-called  \emph{Reidemeister moves}, see e.g.~\cite{book:Knots_and_Physics}.
The move I straightens a twist in an edge, the move II slides one edge over another edge to two parallel  edges,
and the move III changes the order of three crossings.
In the presence of junctions, we need to add another move, where we move an edge over a junction.
We call this as the move IV.

Since we  expect that the charge of a loop in the 4d theory is determined by the isotropy class of networks,  we would like to require that a network is invariant under these moves. 
This is indeed possible for the moves II, III and IV,
but the move I results in a  $\frq$-dependent factor. 
In the context of 3d Chern-Simons theory, this can be understood from the change in the framing of the link \cite{Witten:1989wf}. 
Let us describe these moves explicitly below. 

\paragraph{R-Move II:}
This relation  says that the the negative crossing is given by the inverse $\calR^{-1}$ of the R-matrix $\calR$ corresponding to the positive crossing.

\begin{eqnarray}
\raisebox{-0.55\height}{\input{figure/qYM/Rmove/RmII_1.tikz}}
=
\raisebox{-0.55\height}{\input{figure/qYM/Rmove/RmII_0.tikz}}
=
\raisebox{-0.55\height}{\input{figure/qYM/Rmove/RmII_2.tikz}}.
\label{eq:R-Move II}
\end{eqnarray}

\paragraph{R-Move III}
This is the Yang-Baxter equation  which the R-matrix $\calR$ should satisfy.

\begin{eqnarray}
\raisebox{-0.55\height}{\input{figure/qYM/Rmove/RmIII_1.tikz}}
=
\raisebox{-0.55\height}{\input{figure/qYM/Rmove/RmIII_2.tikz}}.
\label{eq:R-Move III}
\end{eqnarray}

\paragraph{R-Move IV}
This is the additional move for the networks with junctions.
\begin{eqnarray}
\raisebox{-0.6\height}{\input{figure/qYM/Rmove/RmIV_1.tikz}}
=
\raisebox{-0.5\height}{\input{figure/qYM/Rmove/RmIV_0.tikz}}
=
\raisebox{-0.6\height}{\input{figure/qYM/Rmove/RmIV_2.tikz}}.
\label{eq:exact R-Move IV}
\end{eqnarray}

\paragraph{R-Move I}
Finally, this move involves a nontrivial factor.
\begin{eqnarray}
\raisebox{-0.55\height}{\input{figure/qYM/Rmove/RmI_1.tikz}}
=
C_V(\frq)
\raisebox{-0.55\height}{\tikzsetnextfilename{-figure-qYM-Rmove-RmI_0}
\begin{tikzpicture}
\def\h{1.8}
\coordinate (U) at (0,0);
\coordinate (D) at ($(U)+(0,-\h)$);

\draw[end arrow] (D) node[below]{$V$} -- (U);
\end{tikzpicture}
},
\qquad
\raisebox{-0.55\height}{\input{figure/qYM/Rmove/RmI_2.tikz}}
=
C_V(\frq^{-1})
\raisebox{-0.55\height}{\tikzsetnextfilename{-figure-qYM-Rmove-RmI_0}
\begin{tikzpicture}
\def\h{1.8}
\coordinate (U) at (0,0);
\coordinate (D) at ($(U)+(0,-\h)$);

\draw[end arrow] (D) node[below]{$V$} -- (U);
\end{tikzpicture}
}.
\end{eqnarray}

When the representation $V$ is $\wedge^k\Box$, the coefficient $C_k(\frq)$ can be calculated 
using the general crossing resolution \eqref{eq:conjectural skein relation of general crossing}
and the relation \eqref{eq:slimming} removing digons.
The $i$-th network on the right hand side of  
\eqref{eq:conjectural skein relation of general crossing}
gives a coefficient $\tfrac{\langle N-k+i\rangle!}{\langle i\rangle!\langle k-i\rangle!\langle N+i-2k\rangle!}$ 
thanks to  \eqref{eq:slimming}.
We then find
\begin{eqnarray}
C_k(\frq)=(-1)^{k(N+1)}\frq^{-(1+\tfrac{1}{N})k(N-k)}.
\end{eqnarray}
A direct understanding of this coefficient in 4d or 6d would be an interesting problem.

\subsection{More simplifying relations }
\label{subsec:Skein}

Let us list various other skein relations that can be used to simplify networks.
All relations except \eqref{eq:R-matrix on tube} are known in \cite{CKM12} \cite{JeongKim05} and references therein.

\subsubsection{Triangle contraction relations}

We  have  rules to remove triangles.
In order to express the rules, we first map all the junctions so that they are canonical.
Then, there are four possibilities up to the mirror images: 
\begin{align}
\raisebox{-0.35\height}{\input{figure/qYM/triangle_contraction/no4.tikz}}
\!\! & =\dfrac{\langle a\rangle !}{\langle i\rangle !\langle j\rangle !}\raisebox{-0.45\height}{\input{figure/qYM/triangle_contraction/trivalent4.tikz}}, &
\raisebox{-0.35\height}{\input{figure/qYM/triangle_contraction/no2.tikz}}
\!\! &=\dfrac{\langle N-a\rangle !}{\langle N-i\rangle !\langle j\rangle !}\raisebox{-0.45\height}{\input{figure/qYM/triangle_contraction/trivalent2.tikz}}, \\
\raisebox{-0.35\height}{\input{figure/qYM/triangle_contraction/no3.tikz}}
\!\! &=\dfrac{\langle a\rangle !}{\langle i\rangle !\langle j\rangle !}\raisebox{-0.45\height}{\input{figure/qYM/triangle_contraction/trivalent3.tikz}}, &
\raisebox{-0.35\height}{\input{figure/qYM/triangle_contraction/no1.tikz}}
\!\!&=\dfrac{\langle N-a\rangle !}{\langle i\rangle !\langle N-j\rangle !}\raisebox{-0.45\height}{\input{figure/qYM/triangle_contraction/trivalent1.tikz}}.
\end{align}
Note that whether three vertices are totally ordered by arrows or not changes the look of the factors.

\subsubsection{Rectangle decaying relations}
The rectangles $Q^{(i)}_{ab}$ that we had in (\ref{eq:equivalent relation between Q's}) 
can not be further simplified, 
but  there are many other rectangles that are  equivalent to sums of simpler ones. 
Let us show one class:
\begin{eqnarray}
\raisebox{-0.5\height}{\tikzsetnextfilename{-figure-qYM-unstable_rectangle-unstable}
\begin{tikzpicture}
\def\l{1.1}
\def\ll{0.7}
\coordinate (UL) at (0,0);
\coordinate (DL) at ($(UL)+(0,-\l)$);
\coordinate (UR) at ($(UL)+(\l,0)$);
\coordinate (DR) at ($(UR)+(0,-\l)$);
\coordinate (Lout) at ($(UL)+(135:\ll)$);
\coordinate (Rout) at ($(UR)+(45:\ll)$);
\coordinate (Lin) at ($(DL)+(-135:\ll)$);
\coordinate (Rin) at ($(DR)+(-45:\ll)$);

\draw[mid arrow] (DL) -- node[left]{$k+j$} (UL);
\draw[mid arrow] (DR) -- node[right]{$\ell-j$} (UR);
\draw[mid arrow] (DR) -- node[below]{$j$} (DL);
\draw[mid arrow] (UL) -- node[above]{$j$} (UR);
\draw[mid arrow] (UL) -- (Lout) node[above]{$k$};
\draw[mid arrow] (UR) -- (Rout) node[above]{$\ell$};
\draw[mid arrow] (Lin) node[below]{$k$} -- (DL);
\draw[mid arrow] (Rin) node[below]{$\ell$} -- (DR);
\end{tikzpicture}
}
=
\sum_{s=\max(0,j-\ell+k)}^{\min(j,k)}
\dfrac{\langle \ell- k\rangle !}{\langle \ell-k+s-j\rangle !\langle j-s\rangle !}
\raisebox{-0.5\height}{\tikzsetnextfilename{-figure-qYM-unstable_rectangle-stablers}
\begin{tikzpicture}
\def\l{1.1}
\def\ll{0.7}
\coordinate (UL) at (0,0);
\coordinate (DL) at ($(UL)+(0,-\l)$);
\coordinate (UR) at ($(UL)+(\l,0)$);
\coordinate (DR) at ($(UR)+(0,-\l)$);
\coordinate (Lout) at ($(UL)+(135:\ll)$);
\coordinate (Rout) at ($(UR)+(45:\ll)$);
\coordinate (Lin) at ($(DL)+(-135:\ll)$);
\coordinate (Rin) at ($(DR)+(-45:\ll)$);

\draw[mid arrow] (DL) -- node[left]{$k-s$} (UL);
\draw[mid arrow] (DR) -- node[right]{$\ell+s$} (UR);
\draw[mid arrow] (DL) -- node[below]{$s$} (DR);
\draw[mid arrow] (UR) -- node[above]{$s$} (UL);
\draw[mid arrow] (UL) -- (Lout) node[above]{$k$};
\draw[mid arrow] (UR) -- (Rout) node[above]{$\ell$};
\draw[mid arrow] (Lin) node[below]{$k$} -- (DL);
\draw[mid arrow] (Rin) node[below]{$\ell$} -- (DR);
\end{tikzpicture}
}.
\label{eq:rectangle decaying}
\end{eqnarray}
This is valid for $k \le l$ and $0 \le j \le \min(k,N-\ell)$ or $k \le j \le \ell \le N-k$.

These relations assure that any network constructed from only rectangles  around a tube can always be decomposed into a sum of closed loops around the tube.
For example, using (\ref{eq:rectangle decaying}) recursively, we can see
\begin{eqnarray}
\frq^{-\tfrac{k^2}{N}}
\raisebox{-0.3\height}{\tikzsetnextfilename{-figure-qYM-unstable_rectangle-R_for_kk}
\begin{tikzpicture}
\def\l{1.6}
\def\ll{0.5}
\coordinate (UL) at (0,0);
\coordinate (DL) at ($(UL)+(0,-\l)$);
\coordinate (UR) at ($(UL)+(\l,0)$);
\coordinate (DR) at ($(UR)+(0,-\l)$);
\coordinate (ULL) at ($(UL)-(\ll,0)$);
\coordinate (URR) at ($(UR)+(\ll,0)$);
\coordinate (DLL) at ($(DL)-(\ll,0)$);
\coordinate (DRR) at ($(DR)+(\ll,0)$);

\draw[mid-end arrow] (DR) -- (UL) node[above]{$k$};
\draw[line width=5.0pt,white] (DL) -- (UR);
\draw[mid-end arrow] (DL) -- (UR) node[above]{$k$};
\draw[thin] (ULL) -- (URR);
\draw[thin] (DLL) -- (DRR);
\end{tikzpicture}
}
=
\frq^{-k}\sum_{i=-k}^k (-1)^i\frq^{i^2+i} 
\raisebox{-0.3\height}{\tikzsetnextfilename{-figure-qYM-unstable_rectangle-fund_rep_in_sum_trR}
\begin{tikzpicture}
\def\l{1.6}
\def\w{1}
\def\ll{0.4}
\coordinate (UL) at (0,0);
\coordinate (DL) at ($(UL)+(0,-\l)$);
\coordinate (UR) at ($(UL)+(\w,0)$);
\coordinate (DR) at ($(UR)+(0,-\l)$);
\coordinate (ULL) at ($(UL)-(\ll,0)$);
\coordinate (URR) at ($(UR)+(\ll,0)$);
\coordinate (DLL) at ($(DL)-(\ll,0)$);
\coordinate (DRR) at ($(DR)+(\ll,0)$);

\draw[thin] (ULL) -- (URR);
\draw[thin] (DLL) -- (DRR);

\draw[mid-end arrow] (DR) -- (UR) node[above]{$k-i$};
\draw[mid-end arrow] (DL) -- (UL) node[above]{$k+i$};
\end{tikzpicture}
},
\label{eq:R-matrix on tube}
\end{eqnarray}
where the two horizontal thin parallel lines signify that they are to be identified
so that the network is on a tube.

There are various other relations. Here we just note one example:
\begin{eqnarray}
\raisebox{-0.5\height}{\tikzsetnextfilename{-figure-qYM-double_rectangle-ex1}
\begin{tikzpicture}
\def\l{1.1}
\def\ll{0.7}
\coordinate (UL) at (0,0);
\coordinate (DL) at ($(UL)+(0,-\l)$);
\coordinate (UM) at ($(UL)+(\l,0)$);
\coordinate (DM) at ($(UM)+(0,-\l)$);
\coordinate (UR) at ($(UM)+(\l,0)$);
\coordinate (DR) at ($(UR)+(0,-\l)$);
\coordinate (Lout) at ($(UL)+(135:\ll)$);
\coordinate (Rout) at ($(UR)+(45:\ll)$);
\coordinate (Lin) at ($(DL)+(-135:\ll)$);
\coordinate (Rin) at ($(DR)+(-45:\ll)$);

\draw[mid arrow] (DL) -- node[left]{$1$} (UL);
\draw[mid arrow] (DM) -- node[right]{$k$} (UM);
\draw[mid arrow] (DR) -- node[right]{$2$} (UR);
\draw[mid arrow] (DL) -- node[below]{$k-1$} (DM);
\draw[mid arrow] (DR) -- node[above]{$1$} (DM);
\draw[mid arrow] (UM) -- node[below]{$2$} (UL);
\draw[mid arrow] (UM) -- node[above]{$k-2$} (UR);
\draw[mid arrow] (UL) -- (Lout) node[above]{$3$};
\draw[mid arrow] (UR) -- (Rout) node[above]{$k$};
\draw[mid arrow] (Lin) node[below]{$k$} -- (DL);
\draw[mid arrow] (Rin) node[below]{$3$} -- (DR);
\end{tikzpicture}
}
=
\langle 3\rangle \raisebox{-0.5\height}{\tikzsetnextfilename{-figure-qYM-double_rectangle-ex1_1}
\begin{tikzpicture}
\def\l{1.1}
\def\ll{0.7}
\coordinate (UL) at (0,0);
\coordinate (DL) at ($(UL)+(0,-\l)$);
\coordinate (UR) at ($(UL)+(\l,0)$);
\coordinate (DR) at ($(UR)+(0,-\l)$);
\coordinate (Lout) at ($(UL)+(135:\ll)$);
\coordinate (Rout) at ($(UR)+(45:\ll)$);
\coordinate (Lin) at ($(DL)+(-135:\ll)$);
\coordinate (Rin) at ($(DR)+(-45:\ll)$);

\draw[mid arrow] (DL) -- node[left]{$k+1$} (UL);
\draw[mid arrow] (DR) -- node[right]{$2$} (UR);
\draw[mid arrow] (DR) -- node[below]{$1$} (DL);
\draw[mid arrow] (UL) -- node[above]{$k-2$} (UR);
\draw[mid arrow] (UL) -- (Lout) node[above]{$3$};
\draw[mid arrow] (UR) -- (Rout) node[above]{$k$};
\draw[mid arrow] (Lin) node[below]{$k$} -- (DL);
\draw[mid arrow] (Rin) node[below]{$3$} -- (DR);
\end{tikzpicture}
}
+
\langle 2\rangle \raisebox{-0.5\height}{\tikzsetnextfilename{-figure-qYM-double_rectangle-ex1_2}
\begin{tikzpicture}
\def\l{1.1}
\def\ll{0.7}
\coordinate (UL) at (0,0);
\coordinate (DL) at ($(UL)+(0,-\l)$);
\coordinate (UR) at ($(UL)+(\l,0)$);
\coordinate (DR) at ($(UR)+(0,-\l)$);
\coordinate (Lout) at ($(UL)+(135:\ll)$);
\coordinate (Rout) at ($(UR)+(45:\ll)$);
\coordinate (Lin) at ($(DL)+(-135:\ll)$);
\coordinate (Rin) at ($(DR)+(-45:\ll)$);

\draw[mid arrow] (DL) -- node[left]{$k+2$} (UL);
\draw[mid arrow] (DR) -- node[right]{$1$} (UR);
\draw[mid arrow] (DR) -- node[below]{$2$} (DL);
\draw[mid arrow] (UL) -- node[above]{$k-1$} (UR);
\draw[mid arrow] (UL) -- (Lout) node[above]{$3$};
\draw[mid arrow] (UR) -- (Rout) node[above]{$k$};
\draw[mid arrow] (Lin) node[below]{$k$} -- (DL);
\draw[mid arrow] (Rin) node[below]{$3$} -- (DR);
\end{tikzpicture}
}.
\label{eq:double rectangle contraction example}
\end{eqnarray}

\subsection{Examples: $A_2$ and $A_3$}
\label{subsec:A3examples}
The skein relations of the $A_1$ case and the $A_2$ case have already been described in the literature. 
\subsubsection{$A_2$}
Let us record the $A_2$ case as a summary. We have two types of junctions :
\begin{equation}
\raisebox{-0.45\height}{\tikzsetnextfilename{-figure-gauge-classification-cano_in}
\begin{tikzpicture}
\def\l{0.7}
\coordinate (O) at (0,0);
\coordinate (A) at ($(O)+(0:\l)$);
\coordinate (B) at ($(O)+(120:\l)$);
\coordinate (C) at ($(O)+(240:\l)$);

\draw[mid arrow] (B) -- node[left]{$1$} (O);
\draw[mid arrow] (C) -- node[left]{$1$} (O);
\draw[mid arrow] (O) -- node[below right]{$2$} (A);
\end{tikzpicture}
}
=
\raisebox{-0.45\height}{\tikzsetnextfilename{-figure-gauge-classification-inflowing_junction}
\begin{tikzpicture}
\def\l{0.7}
\coordinate (O) at (0,0);
\coordinate (A) at ($(O)+(0:\l)$);
\coordinate (B) at ($(O)+(120:\l)$);
\coordinate (C) at ($(O)+(240:\l)$);

\draw[mid arrow] (B) -- (O);
\draw[mid arrow] (C) -- (O);
\draw[mid arrow] (A) -- (O);
\end{tikzpicture}
}
,\qquad
\raisebox{-0.45\height}{\tikzsetnextfilename{-figure-gauge-classification-cano_out}
\begin{tikzpicture}
\def\l{0.7}
\coordinate (O) at (0,0);
\coordinate (A) at ($(O)+(0:\l)$);
\coordinate (B) at ($(O)+(120:\l)$);
\coordinate (C) at ($(O)+(240:\l)$);

\draw[mid arrow] (O) -- node[right]{$1$} (B);
\draw[mid arrow] (O) -- node[left]{$1$} (C);
\draw[mid arrow] (A) -- node[below right]{$2$} (O);
\end{tikzpicture}
}
=
\raisebox{-0.45\height}{\tikzsetnextfilename{-figure-gauge-classification-outgoing_junction}
\begin{tikzpicture}
\def\l{0.7}
\coordinate (O) at (0,0);
\coordinate (A) at ($(O)+(0:\l)$);
\coordinate (B) at ($(O)+(120:\l)$);
\coordinate (C) at ($(O)+(240:\l)$);

\draw[mid arrow] (O) -- (B);
\draw[mid arrow] (O) -- (C);
\draw[mid arrow] (O) -- (A);
\end{tikzpicture}
}.
\end{equation}
The basic skein relation was \eqref{eq:A2skein}, which we copy here \cite{Kuperberg96,Xie1304,Bullimore13}:
\begin{equation}
\raisebox{-0.5\height}{\tikzsetnextfilename{-figure-qYM-A2-R}
\begin{tikzpicture}
\def\l{1.4}
\coordinate (UL) at (0,0);
\coordinate (DL) at ($(UL)+(0,-\l)$);
\coordinate (UR) at ($(UL)+(\l,0)$);
\coordinate (DR) at ($(UR)+(0,-\l)$);

\draw[end arrow] (DR) -- (UL);
\draw[line width=5.0pt,white] (DL) -- (UR);
\draw[end arrow] (DL)  -- (UR);
\end{tikzpicture}
}
= \frq^{1/3} \raisebox{-0.4\height}{\tikzsetnextfilename{-figure-qYM-A2-Q}
\begin{tikzpicture}
\def\l{1.4}
\coordinate (UL) at (0,0);
\coordinate (DL) at ($(UL)+(0,-\l)$);
\coordinate (UR) at ($(UL)+(\l,0)$);
\coordinate (DR) at ($(UR)+(0,-\l)$);
\coordinate (UM) at ($(UL)+(\l/2,-\l/4)$);
\coordinate (DM) at ($(DL)+(\l/2,\l/4)$);

\draw[mid arrow] (DR)  -- (DM);
\draw[mid arrow] (DL)  -- (DM);
\draw[mid arrow] (UM) --  (DM);
\draw[mid arrow] (UM) -- (UL);
\draw[mid arrow] (UM) -- (UR);
\end{tikzpicture}
} + \frq^{-2/3} \raisebox{-0.5\height}{\tikzsetnextfilename{-figure-qYM-A2-Id}
\begin{tikzpicture}
\def\l{1.4}
\coordinate (UL) at (0,0);
\coordinate (DL) at ($(UL)+(0,-\l)$);
\coordinate (UR) at ($(UL)+(\l,0)$);
\coordinate (DR) at ($(UR)+(0,-\l)$);

\draw[mid arrow] (DR)  -- (UR);
\draw[mid arrow] (DL)  -- (UL);
\end{tikzpicture}
}.
\end{equation}
The following two relations are useful to simplify the networks:
\begin{equation}
\raisebox{-0.4\height}{\tikzsetnextfilename{-figure-gauge-classification-QQ}
\begin{tikzpicture}
\def\R{0.3}
\def\l{0.7}
\coordinate (O) at (0,0);
\coordinate (L) at ($(O)+(-\R,0)$);
\coordinate (R) at ($(O)+(\R+\l,0)$);

\draw[mid arrow] (L) -- ++(-\l,0);
\draw[mid arrow] (R) -- ++(-\l,0);
\anticirclearrow{(O)}{\R}{90}{0.3}{above}{}
\circlearrow{(O)}{\R}{-90}{0.3}{below}{}
\draw[black,thick] (O) circle (\R);
\end{tikzpicture}
}
= \langle 2\rangle \raisebox{-0.4\height}{\tikzsetnextfilename{-figure-gauge-classification-Q}
\begin{tikzpicture}
\def\l{0.8}
\coordinate (O) at (0,0);
\coordinate (L) at ($(O)+(-\l,0)$);
\coordinate (R) at ($(O)+(\l,0)$);

\draw[white,line width=1cm] (L) -- (R);
\draw[mid arrow] (R) -- (L);
\end{tikzpicture}
},\qquad
\raisebox{-0.45\height}{\tikzsetnextfilename{-figure-gauge-classification-A2rectangle}
\begin{tikzpicture}
\def\l{0.8}
\def\ll{0.5}
\coordinate (UL) at (0,0);
\coordinate (DL) at ($(UL)+(0,-\l)$);
\coordinate (UR) at ($(UL)+(\l,0)$);
\coordinate (DR) at ($(UR)+(0,-\l)$);
\coordinate (Lout) at ($(UL)+(135:\ll)$);
\coordinate (Rout) at ($(UR)+(45:\ll)$);
\coordinate (Lin) at ($(DL)+(-135:\ll)$);
\coordinate (Rin) at ($(DR)+(-45:\ll)$);

\draw[mid arrow] (DL) -- (UL);
\draw[mid arrow] (UR) -- (DR);
\draw[mid arrow] (DL) -- (DR);
\draw[mid arrow] (UR) -- (UL);
\draw[mid arrow] (Lout) -- (UL);
\draw[mid-end arrow] (UR) -- (Rout);
\draw[mid-end arrow] (Rin) -- (DR);
\draw[mid arrow] (DL) -- (Lin);
\end{tikzpicture}
}
=
\raisebox{-0.45\height}{\tikzsetnextfilename{-figure-gauge-classification-II}
\begin{tikzpicture}
\def\w{0.9}
\def\h{1.4}
\coordinate (UL) at (0,0);
\coordinate (DL) at ($(UL)+(0,-\h)$);
\coordinate (UR) at ($(UL)+(\w,0)$);
\coordinate (DR) at ($(UR)+(0,-\h)$);

\draw[end arrow] (DR) .. controls ($(DR)+(100:0.3*\h)$) and ($(UR)+(-100:0.3*\h)$) .. (UR);
\draw[end arrow] (UL) .. controls ($(UL)+(-80:0.3*\h)$) and ($(DL)+(80:0.3*\h)$) .. (DL);
\end{tikzpicture}
}
+
\raisebox{-0.5\height}{\tikzsetnextfilename{-figure-gauge-classification-W}
\begin{tikzpicture}
\def\w{1.2}
\def\h{0.9}
\coordinate (UL) at (0,0);
\coordinate (DL) at ($(UL)+(0,-\h)$);
\coordinate (UR) at ($(UL)+(\w,0)$);
\coordinate (DR) at ($(UR)+(0,-\h)$);

\draw[end arrow] (UL) .. controls ($(UL)+(-10:0.3*\w)$) and ($(UR)+(-170:0.3*\h)$) .. (UR);
\draw[end arrow] (DR) .. controls ($(DR)+(170:0.3*\h)$) and ($(DL)+(10:0.3*\h)$) .. (DL);
\end{tikzpicture}
}.\label{eq:a2simplify}
\end{equation}

\subsubsection{$A_3$}
Let us now discuss the next nontrivial case of $A_3$. 
Note that the label $3$ can be traded with $1$ by reversing the arrow,
and since $2$ is a real representation we do not have to exhibit the direction for edges labeled by $2$.
In this case, there are also two types of junctions as we see below.

There are three types of crossing resolutions :
\begin{align}
\raisebox{-0.3\height}{\tikzsetnextfilename{-figure-qYM-A3-R11}
\begin{tikzpicture}
\def\l{1.3}
\coordinate (UL) at (0,0);
\coordinate (DL) at ($(UL)+(0,-\l)$);
\coordinate (UR) at ($(UL)+(\l,0)$);
\coordinate (DR) at ($(UR)+(0,-\l)$);

\draw[end arrow] (DR) -- (UL) node[above]{$1$};
\draw[line width=5.0pt,white] (DL) -- (UR);
\draw[end arrow] (DL) -- (UR) node[above]{$1$};
\end{tikzpicture}
}
&=
\frq^{1/4}\raisebox{-0.5\height}{\input{figure/qYM/A3/Q11_0.tikz}}
+
\frq^{-3/4}\raisebox{-0.35\height}{\input{figure/qYM/A3/Q11_1.tikz}},\\
\raisebox{-0.3\height}{\tikzsetnextfilename{-figure-qYM-A3-R12}
\begin{tikzpicture}
\def\l{1.3}
\coordinate (UL) at (0,0);
\coordinate (DL) at ($(UL)+(0,-\l)$);
\coordinate (UR) at ($(UL)+(\l,0)$);
\coordinate (DR) at ($(UR)+(0,-\l)$);

\draw[end arrow] (DR) -- (UL) node[above]{$1$};
\draw[line width=5.0pt,white] (DL) -- (UR);
\draw[no arrow] (DL) -- (UR) node[above]{$2$};
\end{tikzpicture}
}
&=
\frq^{1/2}\raisebox{-0.5\height}{\input{figure/qYM/A3/Q12_0.tikz}}
+
\frq^{-1/2}\raisebox{-0.5\height}{\input{figure/qYM/A3/Q12_1.tikz}},\\
\raisebox{-0.3\height}{\tikzsetnextfilename{-figure-qYM-A3-R22}
\begin{tikzpicture}
\def\l{1.3}
\coordinate (UL) at (0,0);
\coordinate (DL) at ($(UL)+(0,-\l)$);
\coordinate (UR) at ($(UL)+(\l,0)$);
\coordinate (DR) at ($(UR)+(0,-\l)$);

\draw[no arrow] (DR) -- (UL) node[above]{$2$};
\draw[line width=5.0pt,white] (DL) -- (UR);
\draw[no arrow] (DL) -- (UR) node[above]{$2$};
\end{tikzpicture}
}
&=
\frq\raisebox{-0.5\height}{\input{figure/qYM/A3/Q22_0.tikz}}
+
\raisebox{-0.5\height}{\input{figure/qYM/A3/Q22_1.tikz}}
+
\frq^{-1}\raisebox{-0.35\height}{\input{figure/qYM/A3/Q22_2.tikz}}.
\end{align}

There are three decaying  relations for one rectangle:
\begin{align}
 \raisebox{-0.45\height}{\input{figure/qYM/A3/S2121_2.tikz}}
&=
\raisebox{-0.35\height}{\input{figure/qYM/A3/II21.tikz}}
+
\raisebox{-0.5\height}{\input{figure/qYM/A3/Y21.tikz}}, \\
 \raisebox{-0.45\height}{\input{figure/qYM/A3/S3131_2.tikz}}
&=
\langle 2\rangle \raisebox{-0.45\height}{\input{figure/qYM/A3/II31.tikz}}
+
\raisebox{-0.5\height}{\input{figure/qYM/A3/W13.tikz}}, \\
 \raisebox{-0.45\height}{\input{figure/qYM/A3/S3113_2.tikz}}
&=
\langle 2\rangle \raisebox{-0.45\height}{\input{figure/qYM/A3/Q31_1.tikz}}.
\end{align}

\section{Networks for $\SUSYN4$ Yang-Mills}\label{sec:gauge}

As the fundamental aspects of defect networks in the two dimensional theories have been discussed, we can move on to the discussion of the correspondence between networks on the 2d side and loops on the 4d side. 
In this section, we restrict ourselves to the most familiar $\SUSYN{4}$ case.
On the gauge theory side, the charges of the loop operators were   classified in \cite{Kapustin05}.
It is not easy to construct the corresponding networks for the general $A_k$ case, 
but we will see that the skein relations allow us to describe and classify the networks for $A_2$ concretely.

Before proceeding, let us quickly recall the possible charges of the loop operators of $\SUSYN4$ $\SU(N)$ Yang-Mills, following \cite{Kapustin05}.
We denote the weight lattice by $\Lambda$. 
We use the notations $\omega_i$ for the fundamental weights
and $h_i$ for the weight vectors in the defining $N$-dimensional representation.
They are explicitly given by 
\begin{align}
 \omega_i&=(1-\tfrac{i}{N},\ldots,\overset{i}{1-\tfrac{i}{N}},\overset{i+1}{-\tfrac{i}{N}},\ldots,-\tfrac{i}{N}), \\
 h_i & =(-\tfrac{1}{N},\ldots,-\tfrac{1}{N},\overset{i}{1-\tfrac{1}{N}},-\tfrac{1}{N},\ldots,-\tfrac{1}{N}). 
\end{align}
Note that $\omega_1=h_1$ and $\omega_{N-1}=-h_N$.

Let us consider a Wilson loop labeled by an irreducible representation $R$.
We can also use its highest weight $\lambda$ as the label,
and possible highest weights are in one-to-one correspondence with $\Lambda/\calW$
where $\calW$ is the Weyl group.
Similarly, a 't Hooft loop can be characterized by a charge vector in $\Lambda$, considered up to the action of the Weyl group.

For a dyonic loop operator, we need to specify a pair of electric and magnetic charges $(\mu,\lambda) \in \Lambda\times \Lambda$ but the charges need to be identified under a simultaneous action of the Weyl group. 
Therefore a dyonic charge corresponds to an element in $(\Lambda\times \Lambda)/\calW$ and represent the element as $[(\mu,\lambda)]$. 
We also call it $[(\mu',\lambda')]$ \emph{lower than} $[(\mu,\lambda)]$ if $\mu'$ and $\lambda'$ are lower than $\mu$ and $\lambda$ respectively when mapped to $\Lambda/\calW$.

\subsection{The product of Wilson loops and 't Hooft loops}\label{subsec:Fundamenal dyonic loop}

It is well known  how pure Wilson loops and pure 't Hooft loops are represented as loops on the torus :
\begin{eqnarray}
W_{\omega_a}
\Longleftrightarrow
\raisebox{-0.4\height}{\input{figure/gauge/Wilson_on_torus.tikz}},
\qquad
T_{\omega_b}
\Longleftrightarrow
\raisebox{-0.4\height}{\input{figure/gauge/tHooft_on_torus.tikz}}
\end{eqnarray}
where we identify each pair of parallel opposite edges to make the parallelogram the torus.
Here $W_R$ is the  Wilson loop  in the representation $R$,
and we identify an irreducible representation and its highest weight vector. 
We use a similar notation for the 't Hooft loop.
We also fix the horizontal one cycle as $\alpha$-cycle and the vertical one as $\beta$-cycle.
Note that the $S$ transformation on the torus is naturally identified with $S$ duality transformation of $\SUSYN{4}$ gauge theory.

We can now decompose the product $W_{\omega_a}\cdot T_{\omega_b}$ using  the crossing resolution (\ref{eq:conjectural skein relation of general crossing}). 
Here we express it  in  a form to make the data of electromagnetic charges manifest:
\begin{eqnarray}
\raisebox{-0.4\height}{\tikzsetnextfilename{-figure-gauge-general_R}
\begin{tikzpicture}
\def\l{1.6}
\coordinate (UL) at (0,0);
\coordinate (DL) at ($(UL)+(0,-\l)$);
\coordinate (UR) at ($(UL)+(\l,0)$);
\coordinate (DR) at ($(UR)+(0,-\l)$);

\draw[end arrow] (DR) -- (UL) node[above]{$b$};
\draw[line width=5.0pt,white] (DL) -- (UR);
\draw[end arrow] (DL) -- (UR) node[above]{$a$};
\end{tikzpicture}
}
=
\sum_{[(\mu^{(i)}_b,\lambda^{(i)}_a)]\in \calD(\omega_b,\omega_a)}
\frq^{-\langle \mu^{(i)}_b,\lambda^{(i)}_a \rangle}
\raisebox{-0.5\height}{\input{figure/gauge/general_Q.tikz}}
\label{eq:OPE of two fundamental chrages}
\end{eqnarray}
where \begin{equation}
\calD(\mu,\lambda):=(\calW(\mu)\times\calW(\lambda))/\calW 
\label{eq:def. bi-charge set}
\end{equation} is the set parameterizing the possible ways to combine 
a magnetic charge $\calW(\mu)$ Weyl-conjugate to $\mu$
and an electric charge $\calW(\nu)$ Weyl-conjugate to $\nu$.

The number of elements in the set $\calD(\omega_b,\omega_a)$ is given by 
 $s=\min(a,b,N-a,N-b)$ and is in a one-to-one correspondence 
with the label $i$ in the summation \eqref{eq:conjectural skein relation of general crossing}.
The label $i$ and a representative $[(\mu^{(i)}_b,\nu^{(i)}_a)]\in \calD(\omega_b,\omega_a)$ 
can be naturally related by the equation \begin{equation}
i=\dfrac{ab}{N}+\langle \mu^{(i)}_b,\lambda^{(i)}_a \rangle.
\end{equation}

Recalling the fact reviewed in Sec.~\ref{subsec:classical} and that $\langle \mu^{(i)}_b,\lambda^{(i)}_a \rangle$ is the $x$-component of the classical angular momentum associated to the Poynting vector under electric charge $\lambda^{(i)}_a$ and magnetic charge $\mu^{(i)}_b$ in the Coulomb phase with the gauge group $\mathrm{U}(1)^{N-1}$, we naturally expect the following correspondence :
\footnote{There are other three equivalent networks connected to each other under \eqref{eq:flip} or \eqref{eq:flap}.}
\begin{eqnarray}
\raisebox{-0.5\height}{\input{figure/gauge/Q_on_torus.tikz}}
\Longleftrightarrow
D_{[(\mu^{(i)}_b,\lambda^{(i)}_a)]}
\end{eqnarray}
where we use the symbol $D_{[(\mu,\nu)]}$ to denote the dyonic loop operator with the charge $[(\mu,\nu)]\in (\Lambda\times\Lambda)/\calW$.
We also use a simple symbol $D_{(i)}^{b,a}$ for $D_{[(\mu^{(i)}_b,\lambda^{(i)}_a)]}$.
Below, we call these dyonic loops $D_{(i)}^{b,a}$ and the corresponding networks \emph{elementary}.

\subsection{Analysis in the Liouville/Toda theory}

Let us now connect our analysis so far with a computation on the Liouville/Toda theory side, using the localization in the gauge theory \cite{Pestun07,GOP11}.

Hereafter we use symbols $a=(a_1,a_2,\ldots,a_N)$ under the constraint $\sum_{i=1}^{N} a_i=0$, $A_i(a):=\exp[2\pi i b \langle a,h_i \rangle]$ and $A(a)=\textrm{diag}(A_1(a),A_2(a),\ldots,A_N(a))$.
We also denote the $\SUSYN{4}$ holomorphic partition function by $\calZ(a)$.

As seen in \cite{Pestun07} and \cite{GOP11}, Wilson loop $W_R$ and 't Hooft loop $T_R$ are written in the form of matrix model :
\begin{eqnarray}
\langle W_R \rangle &=& \int_{i \bbR^{N-1}} [da] \calZ(a)^\ast \chi_R(A(a)) \calZ(a), \\
\langle T_R \rangle &=& \int_{i \bbR^{N-1}} [da] \calZ(a)^\ast \sum_{\lambda \in \Pi(R)} T^{(R)}_{\lambda}(a) \calZ(a-b \lambda)
\end{eqnarray}
where $\chi_R$ is the character of $R$ and $\Pi(R)$ is the set of weights corresponding to the irreducible representation $R$.
$T^{(R)}_{\lambda}(a)$ are some functions of $a$ related to the character $\chi_R$ via a Fourier transformation in $a$ \cite{DGG10,GOP11} but the concrete expressions are unnecessary hereafter.

In general, any loop operator is expected to be represented as
\begin{align}
\langle X \rangle = \int_{i \bbR^{N-1}} [da] \calZ(a)^\ast \sum_{\nu} X_{\nu}(a) \calZ(a - b\nu)
\end{align}
where $\nu$ runs over some finite set in the weight lattice $\Lambda$ and $X_{\nu}(a)$ are some functions the detail of which we do not need either.
The additions of $W_R$ and $T_R$ in the ordering of loops seen in Sec.~\ref{subsec:DPE} are written as follows :
\begin{eqnarray}
\nonumber \langle W_R X \rangle &=&
\int_{i \bbR^{N-1}} [da] \calZ(a)^\ast W_R \sum_{\nu} X_{\nu}(a) \calZ(a - b\nu) \\
&=& \int_{i \bbR^{N-1}} [da] \calZ(a)^\ast \sum_{\nu} X_{\nu}(a) \chi_R(A(a-b\nu)) \calZ(a - b\nu), \\
\nonumber \langle T_R X \rangle &=&
\int_{i \bbR^{N-1}} [da] \calZ(a)^\ast T_R \sum_{\nu} X_{\nu}(a) \calZ(a - b\nu) \\
&=& \int_{i \bbR^{N-1}} [da] \calZ(a)^\ast \sum_{\mu \in \Pi(R)} \sum_{\nu} T^{(R)}_{\mu}(a-b\nu) X_{\nu}(a) \calZ(a - b\nu-b\mu).
\end{eqnarray}

In particular, let us choose $R$ as one of the fundamental representations $\wedge^{n}\Box$ and introduce $W^{(k)}:=W_{\wedge^k\Box}$ and $T^{(\ell)}:=T_{\wedge^\ell\Box}$.
Then consider insertions both of $W^{(k)}$ and of $T^{(\ell)}$.
One way to insert is
\begin{align}
\nonumber \langle \cdots T^{(\ell)} W^{(k)} X \rangle &= 
\int_{i \bbR^{N-1}} [da] \calZ(a)^\ast \cdots \\
& \sum_{\mu \in \Pi(\wedge^{\ell}\Box)} \sum_{\nu} T^{(\ell)}_{\mu}(a-b\nu) X_{\nu}(a) \chi_{\wedge^{k}\Box}(A(a-b\nu)) \calZ(a - b\nu - b\mu)
\end{align}
where the ellipsis represents further insertions of other loops.

Recalling $\displaystyle \chi_R(A(a)) = \sum_{\lambda \in \Pi(R)} \exp[2\pi ib\langle a,\lambda \rangle]$, then define the following operators labeled by $m=1,2,\ldots,\min(k,\ell)$ :
\begin{align}
\nonumber \langle \cdots [TW]^{(\ell,k)}_m X \rangle &:= 
\int_{i \bbR^{N-1}} [da] \calZ(a)^\ast \cdots \\
& \sum_{(\lambda,\mu) \in \Pi(\wedge^{\ell}\Box,\wedge^{k}\Box)_m} \sum_{\nu} T^{(\ell)}_{\mu}(a-b\nu) X_{\nu}(a)  \exp[2\pi ib\langle a-b\nu,\lambda \rangle] \calZ(a - b\nu - b\mu).
\end{align}
where we decompose the set $\Pi(\wedge^{\ell}\Box) \times \Pi(\wedge^{k}\Box)=\calW(\omega_\ell) \times \calW(\omega_k)$ into several sectors defined by
\begin{eqnarray}
\Pi(\wedge^{\ell}\Box,\wedge^{k}\Box)_m:=\{ (\mu,\lambda) \in \Pi(\wedge^{\ell}\Box) \times \Pi(\wedge^{k}\Box) \;|\; \langle \mu,\lambda \rangle=m-\tfrac{k\ell}{N} \}=\calD(\omega_\ell,\omega_k).
\end{eqnarray}
We  then have
\begin{align}
\langle \cdots T^{(\ell)} W^{(k)} X \rangle
=\sum_m \langle \cdots [TW]^{(\ell,k)}_m X \rangle.
\end{align}
Note that the decomposition of $T^{(\ell)} W^{(k)}$ is independent of the ellipsis $\ldots$ and $X$ assuring that this expansion is local and represent the product as $T^{(\ell)} \cdot W^{(k)}$ or $T^{(\ell)} \times W^{(k)}$.

On the other hand, the insertion in the opposite order is
\begin{align}
\nonumber \langle \cdots W^{(k)} T^{(\ell)} X \rangle &= 
\int_{i \bbR^{N-1}} [da] \calZ(a)^\ast \cdots \\
& \sum_{\mu \in \Pi(\wedge^{\ell}\Box)} \sum_{\nu} T^{(\ell)}_{\mu}(a-b\nu) X_{\nu}(a) \chi_{\wedge^{k}\Box}(A(a-b\mu-b\nu)) \calZ(a - b\nu - b\mu)
\end{align}
and we also have
\begin{align}
\langle \cdots T^{(\ell)} W^{(k)} X \rangle
=\sum_m \langle \cdots \frq^{2\left(\tfrac{k\ell}{N}-m\right)} [TW]^{(\ell,k)}_m X \rangle
\end{align}
where we use
\begin{eqnarray}
\exp[2\pi ib\langle a-b\mu-b\nu,\lambda \rangle]=\frq^{2\left(\tfrac{k\ell}{N}-m\right)}\exp[2\pi ib\langle a-b\nu,\lambda \rangle].
\end{eqnarray}

In summary, we have found the relations
\begin{align}
 {T}^{(\ell)} \times {W}^{(k)}& =\sum_{m=0}^{\min(k,\ell)} [TW]^{(\ell,k)}_m ,
& 
 {W}^{(k)} \times {T}^{(\ell)}& =\sum_{m=0}^{\min(k,\ell)} \frq^{2\left(\tfrac{k\ell}{N}-m\right)}[TW]^{(\ell,k)}_m.
\label{eq:crossing resolution as operators}
\end{align}

Comparing the product expansion \eqref{eq:crossing resolution as operators} and the graphical expansion (\ref{eq:OPE of two fundamental chrages}) 
we find the following identification :
\begin{equation}
\begin{array}{ccccccc}
\text{($Q^{\ell k}{}_{(m)}$ on  $T^2$)} 
& \leftrightarrow &
D_{(m)}^{\ell,k} 
& \leftrightarrow &
[(\mu_\ell^{(m)},\lambda_k^{(m)} )] 
& \leftrightarrow &
\frq^{-\langle\mu_\ell^{(m)},\lambda_k^{(m)} \rangle} 
[TW]^{(\ell,k)}_{m} \\
\text{network} && \text{4d loop} && \text{charge} && \text{operator}
\end{array}.
\end{equation}
Here the pair of weights $[(\mu_\ell^{(m)},\lambda_k^{(m)} )]$ was chosen as in
\eqref{eq:OPE of two fundamental chrages}, and therefore we have 
 $\langle\mu_\ell^{(m)},\lambda_k^{(m)}\rangle=m-\tfrac{k\ell}{N}$.

Let us see how $T$ transformation of the $SL(2,\bbZ)$ duality action acts on these loop operators.
The $\theta$ dependence originally comes from the classical part of $\SUSYN{4}$ partition function $\calZ(a)=\exp[-\pi i \tau \langle a,a \rangle]$ where holomorphic gauge coupling $\tau=\tfrac{\theta}{2\pi}+\tfrac{4\pi i}{g^2_{YM}}$ and the monodromy action under  the change $\theta \to \theta+2\pi$ is following :
\begin{eqnarray}
\nonumber \calZ(a)^\ast\calZ(a-b\lambda) &\underset{\tau \to \tau+1}{\longrightarrow}& \exp[-\pi i (\langle a-b\lambda,a-b\lambda \rangle-\langle a,a \rangle)] \calZ(a)^\ast\calZ(a-b\lambda) \\
&=& \frq^{-\langle \lambda,\lambda \rangle}e^{2\pi ib\langle \lambda,a \rangle} \calZ(a)^\ast.\calZ(a-b\lambda)
\end{eqnarray}
The Witten effect on the partition function can be re-expressed in the loop operators which acts on the partition functions. In particular, when $\lambda$ is in $\Pi(\wedge^\ell \Box)=\calW(\omega_\ell)$, $\calZ(a)^\ast\calZ(a-b\lambda)$ is accompanied by
\begin{eqnarray}
\frq^{-\ell+\tfrac{\ell^2}{N}} e^{2\pi ib\langle \lambda,a \rangle}
\end{eqnarray}
as $\theta$ shifts by $2\pi$. Summing it up over $\Pi(\wedge^\ell\Box)$,
\begin{eqnarray}
T^{(\ell)} \longrightarrow \frq^{-\ell+\tfrac{\ell^2}{N}} [TW]_\ell^{(\ell,\ell)}=D_{[(\omega_\ell,\omega_\ell)]}=D^{\ell,\ell}_{(\ell)}
\end{eqnarray}
under $\theta \to \theta+2\pi$.
Since $D_{[\omega_\ell,\omega_\ell]}=D^{\ell,\ell}_{(\ell)}$ is given by \begin{equation}
\raisebox{-0.4\height}{\input{figure/gauge/twist_Tk_on_torus.tikz}}.
\end{equation}
This $\theta \to \theta+2\pi$ action is graphically represented as
\begin{eqnarray}
\begin{array}{cccc}
T : & \raisebox{-0.4\height}{\input{figure/gauge/Tk_on_torus.tikz}}
& \longrightarrow &
\raisebox{-0.4\height}{\input{figure/gauge/twist_Tk_on_torus.tikz}} \\
&T_{\omega_\ell} &  & D_{[(\omega_\ell,\omega_\ell)]}.
\end{array}
\end{eqnarray}
and matches with the $T$ transformation on the torus.

\subsection{Examples of products of loops in $A_2$}

Let us  focus on the $A_2$ case and perform some explicit computations. The examples in the  general $A_k$ case will be given in Appendix \ref{app:Examples}.
We will see the geometric $SL(2,\bbZ)$ action on the torus is nicely mapped to the $SL(2,\bbZ)$ action on the electric and magnetic weight systems.

\begin{figure}[h]
\centering
\input{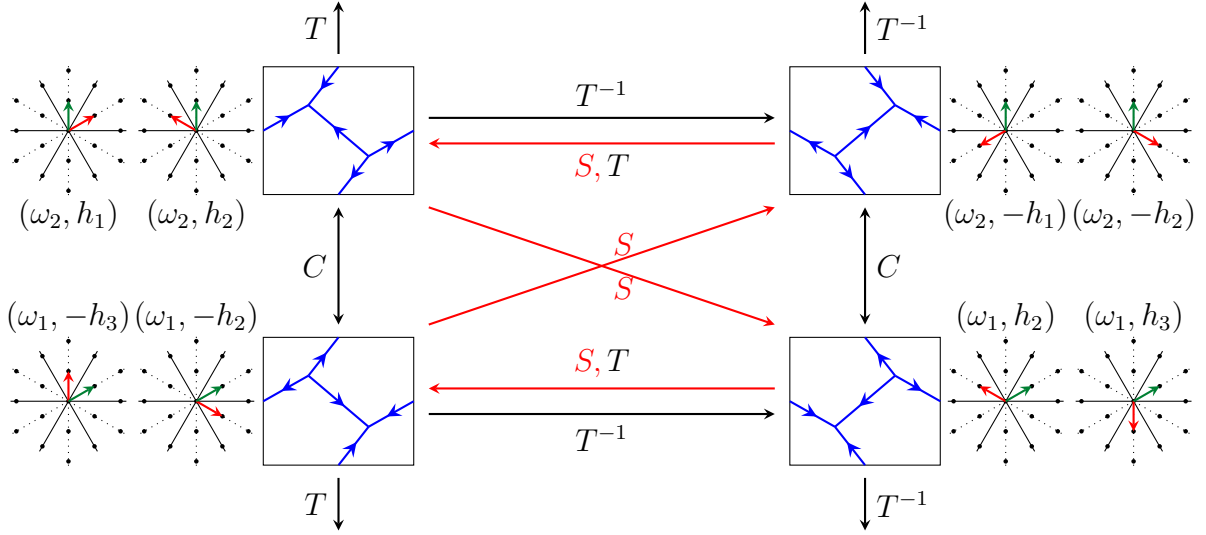}
\caption{The $SL(2,\bbZ)$ duality orbit of $D_{(0)}^{1,1}$ (below right) and their dyonic charges. Two adjacent expressions of a pair of weights are equivalent via some Weyl reflections. Red weights correspond to electric weights and green's to magnetic ones.}
\label{figure:duality orbit of D_0^1,1}
\end{figure}

\paragraph{Example 1.}   The simplest case is $W_{\Box} \times T_{\Box}$, which corresponds to the equation  (\ref{eq:OPE of two fundamental chrages}) with $a=1$ and $b=1$ :
\begin{eqnarray}
W_{\Box} \times T_{\Box}=\frq^{1/3} D_{(0)}^{1,1} +\frq^{-2/3}D_{(1)}^{1,1}
\end{eqnarray}
where $D_{(0)}^{1,1}=D_{[(\omega_1,h_2)]}=D_{[(\omega_1,h_3)]}$ and $D_{(1)}^{1,1}=D_{[(\omega_1,\omega_1)]}$. This was originally found in \cite{Xie1304,Bullimore13} in the context of class S theory.

The dyonic loop $D_{(1)}^{1,1}$ is obtained from the 't Hooft loop $T_{\Box}$  
by an application of the $T$ operation.  In particular this loop can be mapped to a Wilson loop in some duality frame.
The object $D_{(0)}^{1,1}$  cannot be mapped into a network localized on any one cycle by the torus modular transformations. 
In the language of charges,  this means that  the electric weight and the magnetic weight are not parallel.
We can now work out how the $SL(2,\bbZ)$ transformations act on this particular network and the pair of weights, see  Fig.\ref{figure:duality orbit of D_0^1,1}.

\paragraph{Example 2.}  The next example is $W_{\Box\!\Box} \times T_{\Box}$:
\begin{multline}
\raisebox{-0.35\height}{\tikzsetnextfilename{-figure-gauge-W2S_Tf-OPE}
\begin{tikzpicture}
\def\vl{1.8}
\def\hl{2}
\coordinate (H) at (0:\hl);
\coordinate (V) at (90:\vl);
\coordinate (DL) at (0,0);
\coordinate (DR) at ($(DL)+(H)$);
\coordinate (UL) at ($(DL)+(V)$);
\coordinate (UR) at ($(UL)+(H)$);

\coordinate (L1) at ($(DL)!0.5!(UL)$);
\coordinate (R1) at ($(DR)!0.5!(UR)$);
\coordinate (U1) at ($(UL)!0.5!(UR)$);
\coordinate (D1) at ($(DL)!0.5!(DR)$);

\draw (UL) -- (UR) -- (DR) -- (DL) -- (UL);

\draw[green,mid-end arrow,ultra thick] (D1) -- (U1) node[black,above]{$\Box$};
\draw[white,line width=3pt] ($(L1)+0.1*(H)$) -- ($(R1)-0.1*(H)$);
\draw[red,mid-end arrow,ultra thick] (L1) -- (R1) node[black,right]{$\Box\!\Box$};
\end{tikzpicture}
}\;
=\;\frq^{2/3}\; \left[\quad\raisebox{-0.45\height}{\tikzsetnextfilename{-figure-gauge-W2S_Tf-0A}
\begin{tikzpicture}
\def\vl{1.8}
\def\hl{2}
\coordinate (H) at (0:\hl);
\coordinate (V) at (90:\vl);
\coordinate (DL) at (0,0);
\coordinate (DR) at ($(DL)+(H)$);
\coordinate (UL) at ($(DL)+(V)$);
\coordinate (UR) at ($(UL)+(H)$);

\coordinate (L1) at ($(DL)!0.7!(UL)$);
\coordinate (L2) at ($(DL)!0.3!(UL)$);
\coordinate (R1) at ($(DR)!0.7!(UR)$);
\coordinate (R2) at ($(DR)!0.3!(UR)$);
\coordinate (U1) at ($(UL)!0.5!(UR)$);
\coordinate (D1) at ($(DL)!0.5!(DR)$);

\coordinate (M1) at ($(U1)!0.2!(D1)$);
\coordinate (M2) at ($(U1)!0.4!(D1)$);
\coordinate (M3) at ($(D1)!0.4!(U1)$);
\coordinate (M4) at ($(D1)!0.2!(U1)$);

\draw (UL) -- (UR) -- (DR) -- (DL) -- (UL);

\draw[mid arrow] (L1) -- (M2);
\draw[mid arrow] (L2) -- (M4);
\draw[mid arrow] (M1) -- (R1);
\draw[mid arrow] (M3) -- (R2);
\draw[mid arrow] (D1) -- (M4);
\draw[mid arrow] (M1) -- (U1);
\draw[mid arrow] (M3) -- (M2);

\draw[mid arrow] (M3) -- (M4);
\draw[mid arrow] (M1) -- (M2);
\end{tikzpicture}
}\;
-\quad\raisebox{-0.45\height}{\tikzsetnextfilename{-figure-gauge-W2S_Tf-0B}
\begin{tikzpicture}
\def\vl{1.8}
\def\hl{2}
\coordinate (H) at (0:\hl);
\coordinate (V) at (90:\vl);
\coordinate (DL) at (0,0);
\coordinate (DR) at ($(DL)+(H)$);
\coordinate (UL) at ($(DL)+(V)$);
\coordinate (UR) at ($(UL)+(H)$);

\coordinate (L1) at ($(DL)!0.5!(UL)$);
\coordinate (R1) at ($(DR)!0.5!(UR)$);
\coordinate (U1) at ($(UL)!0.5!(UR)$);
\coordinate (D1) at ($(DL)!0.5!(DR)$);

\draw (UL) -- (UR) -- (DR) -- (DL) -- (UL);

\def\a{0.05}
\def\b{0.25}
\draw[mid arrow] (D1) .. controls ($(D1)-\a *(H)+\b *(V)$) and ($(L1)+\b *(H)-\a *(V)$) .. (L1);
\draw[mid arrow] (R1) .. controls ($(R1)-\b *(H)+\a *(V)$) and ($(U1)+\a *(H)-\b *(V)$) .. (U1);
\end{tikzpicture}
} \right. \\
\left. +\; \frq^{-1} \quad\raisebox{-0.45\height}{\tikzsetnextfilename{-figure-gauge-W2S_Tf-1A}
\begin{tikzpicture}
\def\vl{1.8}
\def\hl{2}
\coordinate (H) at (0:\hl);
\coordinate (V) at (90:\vl);
\coordinate (DL) at (0,0);
\coordinate (DR) at ($(DL)+(H)$);
\coordinate (UL) at ($(DL)+(V)$);
\coordinate (UR) at ($(UL)+(H)$);

\coordinate (L1) at ($(DL)!0.5!(UL)$);
\coordinate (R1) at ($(DR)!0.5!(UR)$);
\coordinate (U1) at ($(UL)!0.5!(UR)$);
\coordinate (D1) at ($(DL)!0.5!(DR)$);

\coordinate (M1) at ($(UL)!0.3!(DR)$);
\coordinate (M2) at ($(DR)!0.3!(UL)$);

\draw (UL) -- (UR) -- (DR) -- (DL) -- (UL);

\draw[mid arrow] (M1) -- (L1);
\draw[mid arrow] (D1) -- (M2);
\draw[mid arrow] (M2) -- (R1);
\draw[mid arrow] (M1) -- (U1);
\draw[mid arrow] (M1) -- (M2);
\end{tikzpicture}
}\;
+\; \frq^{-2} \quad\raisebox{-0.45\height}{\tikzsetnextfilename{-figure-gauge-W2S_Tf-2A}
\begin{tikzpicture}
\def\vl{1.8}
\def\hl{2}
\coordinate (H) at (0:\hl);
\coordinate (V) at (90:\vl);
\coordinate (DL) at (0,0);
\coordinate (DR) at ($(DL)+(H)$);
\coordinate (UL) at ($(DL)+(V)$);
\coordinate (UR) at ($(UL)+(H)$);

\coordinate (L1) at ($(DL)!0.7!(UL)$);
\coordinate (L2) at ($(DL)!0.3!(UL)$);
\coordinate (R1) at ($(DR)!0.7!(UR)$);
\coordinate (R2) at ($(DR)!0.3!(UR)$);
\coordinate (U1) at ($(UL)!0.5!(UR)$);
\coordinate (D1) at ($(DL)!0.5!(DR)$);

\draw (UL) -- (UR) -- (DR) -- (DL) -- (UL);

\draw[mid arrow] (L2) -- (R1);
\draw[mid arrow] (D1) -- (R2);
\draw[mid arrow] (L1) -- (U1);
\end{tikzpicture}
}\quad  \right].
\label{eq:2nd ex of OPE for A_2}
\end{multline}

In this example, 
the first term on the right hand side is a network that cannot be mapped by $SL(2,\bbZ)$ to any of the networks we already studied explicitly. 
It is natural to posit the following expansion
\begin{multline}
 W_{\Box\!\Box} \times T_{\Box} 
 =\frq^{2/3} D_{[(\omega_1,2h_{2})]} + \frq^{-1/3} D_{[(\omega_1,h_1+h_{2})]} + \frq^{-4/3} D_{[(\omega_1,2\omega_1)]} \\
 + (\mbox{loops with lower weights} )
\end{multline}
since we expect that the exponent of $\frq$ multiplying $D_{[(\lambda_m,\lambda_e)]}$ equals  $-\langle \lambda_m,\lambda_e \rangle$ to capture the angular momentum. 
Then we can identify  \begin{equation}
D_{[(\omega_1,2h_{2})]} \sim \raisebox{-0.45\height}{\tikzsetnextfilename{-figure-gauge-W2S_Tf-0A}
\begin{tikzpicture}
\def\vl{1.8}
\def\hl{2}
\coordinate (H) at (0:\hl);
\coordinate (V) at (90:\vl);
\coordinate (DL) at (0,0);
\coordinate (DR) at ($(DL)+(H)$);
\coordinate (UL) at ($(DL)+(V)$);
\coordinate (UR) at ($(UL)+(H)$);

\coordinate (L1) at ($(DL)!0.7!(UL)$);
\coordinate (L2) at ($(DL)!0.3!(UL)$);
\coordinate (R1) at ($(DR)!0.7!(UR)$);
\coordinate (R2) at ($(DR)!0.3!(UR)$);
\coordinate (U1) at ($(UL)!0.5!(UR)$);
\coordinate (D1) at ($(DL)!0.5!(DR)$);

\coordinate (M1) at ($(U1)!0.2!(D1)$);
\coordinate (M2) at ($(U1)!0.4!(D1)$);
\coordinate (M3) at ($(D1)!0.4!(U1)$);
\coordinate (M4) at ($(D1)!0.2!(U1)$);

\draw (UL) -- (UR) -- (DR) -- (DL) -- (UL);

\draw[mid arrow] (L1) -- (M2);
\draw[mid arrow] (L2) -- (M4);
\draw[mid arrow] (M1) -- (R1);
\draw[mid arrow] (M3) -- (R2);
\draw[mid arrow] (D1) -- (M4);
\draw[mid arrow] (M1) -- (U1);
\draw[mid arrow] (M3) -- (M2);

\draw[mid arrow] (M3) -- (M4);
\draw[mid arrow] (M1) -- (M2);
\end{tikzpicture}
}
\end{equation}
 up to the lower  contribution $D_{[(\omega_1,h_{i}+h_{j(>i)})]}$ from lower weights.
 Hereafter, we try to map networks and charges of the dyonic loops  up to the contributions from lower weights.\footnote{The complication comes from two sources. One is  common with what we encountered in Sec.~\ref{subsec:representation basis expansion}:   irreducible representations are linear combinations of networks even in the Wilson loop case.  Another is  related to the bubbling effect of the monopole moduli space. See the related works to this subject \cite{KapustinWitten06,GOP11,Saulina:2011qr,MoraruSaulina12}.}

\begin{figure}[ht]
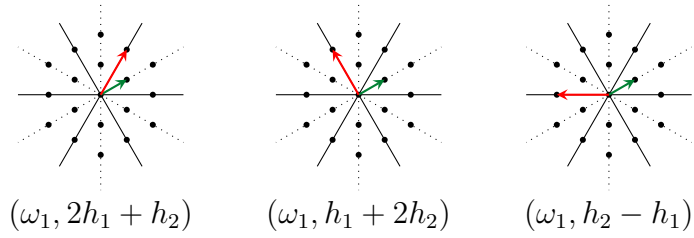

\centering
\begin{tabular}{ccc}
\input{figure/gauge/WAdj_Tf/0A_weight.tikz} &
\input{figure/gauge/WAdj_Tf/1A_weight.tikz} &
\input{figure/gauge/WAdj_Tf/2A_weight.tikz}
\end{tabular}
\caption{Representative weights in $\calD(\omega_1,\omega_1+\omega_2)$\label{fig:www}}
\end{figure}

\paragraph{Example 3.} 
The third example is $W_\text{Adj} \times T_{\Box}$ : The skein relation gives us 
\begin{align}
 \raisebox{-0.35\height}{\input{figure/gauge/WAdj_Tf/OPE.tikz}}\;
=
\;\frq \quad\raisebox{-0.45\height}{\input{figure/gauge/WAdj_Tf/0A.tikz}}\;
+\quad\raisebox{-0.45\height}{\input{figure/gauge/WAdj_Tf/1A.tikz}}\;
+\; \frq^{-1} \quad\raisebox{-0.45\height}{\input{figure/gauge/WAdj_Tf/2A.tikz}}
\end{align}
while from gauge theory we expect 
\begin{align}
\nonumber & W_\text{Adj} \times T_{\Box} \\
\nonumber & \quad =\frq D_{[(\omega_1,h_2-h_1)]} + D_{[(\omega_1,h_1+2h_2)]} +\frq^{-1}D_{[(\omega_1,2h_1+h_2)]} \\
& \quad + (\mbox{loops with  lower weights}).
\end{align} For a graphical representation of weights involved, see Fig.~\ref{fig:www}.

The first term and the third term  can be obtained by $SL(2,\bbZ)$ transformations on $D^{1,1}_{(0)}$.
The second term is a new type: \begin{equation}
D_{[(\omega_1,h_1+2h_2)]} \sim \raisebox{-0.45\height}{\input{figure/gauge/WAdj_Tf/1A.tikz}}.
\end{equation}

\begin{figure}[ht]
\centering
\input{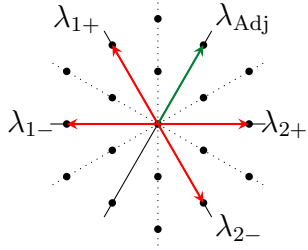}
\caption{Representative weights in $\calD(\omega_1+\omega_2,\omega_1+\omega_2)$ \label{fig:weights_in_adj}
}
\end{figure}

\paragraph{Example 4.} 
Our final example is $W_\text{Adj} \times T_\text{Adj}$. The skein relation gives
\begin{align}
\nonumber &\raisebox{-0.35\height}{\input{figure/gauge/WAdj_TAdj/OPE.tikz}}\;
=
4(2+\frq^{2}+\frq^{-2})\quad\raisebox{-0.45\height}{\tikzsetnextfilename{-figure-gauge-WAdj_TAdj-empty}
\begin{tikzpicture}
\def\vl{1.8}
\def\hl{2}
\coordinate (H) at (0:\hl);
\coordinate (V) at (90:\vl);
\coordinate (DL) at (0,0);
\coordinate (DR) at ($(DL)+(H)$);
\coordinate (UL) at ($(DL)+(V)$);
\coordinate (UR) at ($(UL)+(H)$);

\draw (UL) -- (UR) -- (DR) -- (DL) -- (UL);
\end{tikzpicture}
} +\\
\nonumber & \quad
\;2 \quad\raisebox{-0.45\height}{\input{figure/gauge/WAdj_TAdj/WAdj.tikz}}\;
+\; 2 \quad\raisebox{-0.45\height}{\input{figure/gauge/WAdj_TAdj/TAdj.tikz}}\;
+\; \frq^{-2} \quad\raisebox{-0.45\height}{\input{figure/gauge/WAdj_TAdj/DAdj1.tikz}}\;
+\; \frq^{2} \quad\raisebox{-0.45\height}{\input{figure/gauge/WAdj_TAdj/DAdj2.tikz}} \\
& \quad
+\frq^{-1}\; \left( \raisebox{-0.45\height}{\input{figure/gauge/WAdj_TAdj/3A.tikz}}\;
+\quad\raisebox{-0.45\height}{\input{figure/gauge/WAdj_TAdj/3B.tikz}}\right)
+\frq\; \left( \raisebox{-0.45\height}{\input{figure/gauge/WAdj_TAdj/1A.tikz}}\;
+\quad\raisebox{-0.45\height}{\input{figure/gauge/WAdj_TAdj/1B.tikz}}\right)
\end{align}
while the gauge theory computation yields
\begin{align}
\nonumber & W_\text{Adj} \times T_\text{Adj}
 \;\; =\frq^{-2} D_{[(\lambda_\text{Adj},\lambda_\text{Adj})]}
+ \frq^{2} D_{[(\lambda_\text{Adj},-\lambda_\text{Adj})]} \\
\nonumber & \qquad
+ \frq^{-1}D_{[(\lambda_\text{Adj},\lambda_{1+})]}
+ \frq^{-1}D_{[(\lambda_\text{Adj},\lambda_{2+})]}
+ \frq D_{[(\lambda_\text{Adj},\lambda_{2-})]}
+ \frq D_{[(\lambda_\text{Adj},\lambda_{1-})]} \\
& \qquad + (\mbox{loops with the lower weights})
\end{align}
where $\lambda_\text{Adj}=\omega_1+\omega_2$ is the highest weight of the adjoint representation and see Fig.~\ref{fig:weights_in_adj} for $\lambda_{1,2 \pm}$.
It would be interesting to reproduce the terms with lower weights from a purely 4d gauge theoretic computations.

\subsection{Classification of networks on $T^2$ for $A_2$}
\label{subsec:networks vs bicharge lattice}
\label{subsec:network classification on T^2}

We have seen some basic examples of products of loops  and the identification of the charge and the network.
Here we establish the general mapping between the  networks and the charges in the case of the $A_2$ theory on the torus, or equivalently the $\SUSYN4$ $\SU(3)$ Yang-Mills.
This is a minimal extension of the dictionary of Drukker, Morrison and Okuda \cite{DMO09}.

Let us first classify the possible $A_2$ networks on the torus purely in terms of the skein relation.
First, recall that all networks with crossings are resolved into those with junctions only.
In particular, for the $A_2$ case, there are only two types of junctions,
namely the one where the heads of three arrows meet and 
another one where the tails of three arrows meet. 
Therefore the networks are bipartite \cite{Xie1304} and there appear only polygons with degree-even vertices.

We now use the skein relations we discussed so far.  Recall the basic conventions we discussed in Sec.~\ref{subsec:A3examples}.
All digons can be contracted, and all rectangles are resolved to two pairs of curves, as we discussed in \eqref{eq:a2simplify}.

At this point, the network might contain several disconnected components.  If there are no vertices at all, the network consists of parallel loops wrapping the same one-cycle on $T^2$. 
Assume now there is at least one vertex.
Pick a connected component. It has the topology of either a disk, an annulus or a torus. 

Now, let us  denote the number of edges, or equivalently the number of vertices, of the $i$-th polygon in this connected component by $p_i (=2,4,6,\ldots)$.  
Denote the number of polygons by $f$.
The total number of vertices, edges and faces of the network is then given by 
\begin{equation}
V=\frac{1}{3}\sum_i p_i,\quad
E=\frac{1}{2}\sum_i p_i,\quad
F=f.
\end{equation}
Furthermore, denote the number of boundary edges by $B$ which vanish if the connected component has the topology of torus.
From Euler's theorem we should have \begin{equation}
\chi+\frac{1}{6} B=V-E+F=F-\frac{1}{6}\sum_i p_i  \ge 0
\end{equation}  since the connected component is either a disk ($\chi=1,B>0$), an annulus ($\chi=0,B>0$), or a torus ($\chi=0,B=0$).
Since we removed all digons and rectangles, $p_i\ge 6$, and therefore we have 
\begin{equation}
F-\frac{1}{6}\sum_i p_i  \le 0.
\end{equation}
From this we see that the connected component has the topology of the torus,
and every polygon is a hexagon. 
Therefore, the possible $A_2$ networks on $T^2$ are mapped into the bipartite hexagon tilings with three corner condition at every vertex.

It is interesting to note at this point that bipartite hexagon tilings of the torus appeared in the string theory literature in the context of brane tilings \cite{Hanany:2005ve,Franco:2005rj,Hanany:2010cx}.
In this case the bipartite hexagon tilings corresponded to Abelian orbifolds of $\bbC^3$.

\begin{figure}[htb]
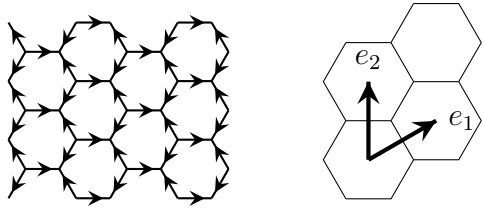

\centering
\input{figure/gauge/hexagon_orientation.tikz}
\qquad 
\input{figure/gauge/basis_of_one_cycle.tikz}
\caption{The infinite bipartite hexagonal tiling and its basis vectors  \label{fig:hexagon_orientation}}
\end{figure}

\begin{figure}[ht]
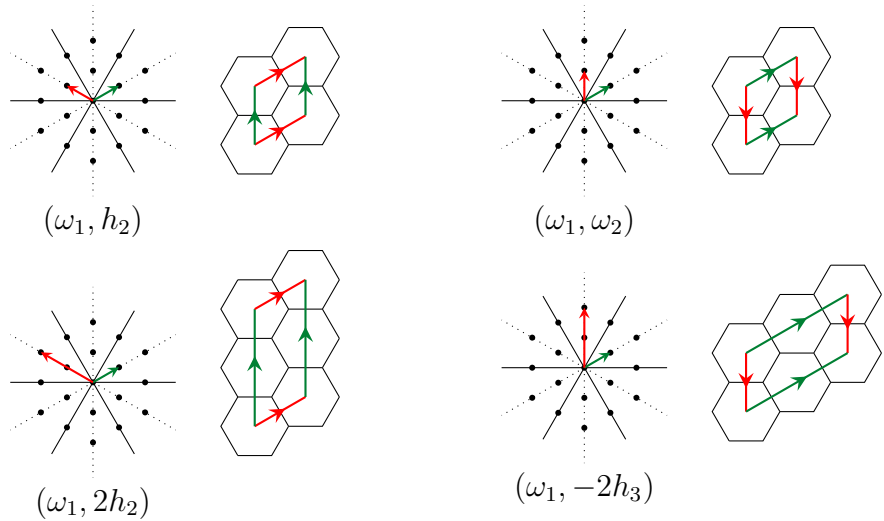

\centering
\begin{tabular}{cc}
\begin{minipage}{0.4\hsize}
\input{figure/gauge/A2/fund_fund_1.tikz}
\end{minipage}
\begin{minipage}{0.4\hsize}
\input{figure/gauge/A2/fund_fund_2.tikz}
\end{minipage}
\end{tabular}

\begin{tabular}{cc}
\begin{minipage}{0.4\hsize}
\input{figure/gauge/A2/fund_2sym_1.tikz}
\end{minipage}
\begin{minipage}{0.4\hsize}
\input{figure/gauge/A2/fund_2sym_2.tikz}
\end{minipage}
\end{tabular}

\caption{Some examples of the hexagonal tilings and the dyonic charges
that we have already identified \label{fig:hexex}}
\end{figure}

Now let us make  the dictionary between the  bipartite hexagon tilings and the dyonic charges.
Instead of thinking of filling a torus by hexagons, 
we can take the quotient of the bipartite hexagon tiling filling the entire plane, 
and then we define the vectors $e_1$ and $e_2$ there, see Fig.\ref{fig:hexagon_orientation}.
To specify a bipartite hexagon tiling, we choose the $\alpha$ and the $\beta$ cycles of the torus from $\bbZ e_1 \oplus \bbZ e_2$ so that they are linearly independent. 
In  Fig.~\ref{fig:hexex} we show the hexagonal tilings and the dyonic charges that already appeared in our analysis so far. 

From these examples, we can find the general map. 
We first naturally identify the $A_2$ weight lattice $\Lambda$ and the dual lattice of the hexagonal tiling. 
Then, the rule is \begin{equation}
(\lambda_e,\lambda_m) \mapsto (\alpha,\beta)=(\lambda_m,-\lambda_e).\label{eq:map}
\end{equation}
under the condition $q_1 p_2 -q_2 p_1 >0$ where $\lambda_e=q_1 \omega_1+q_2 \omega_2$ and $\lambda_m=p_1 \omega_1+p_2 \omega_2$.

\begin{figure}[ht]
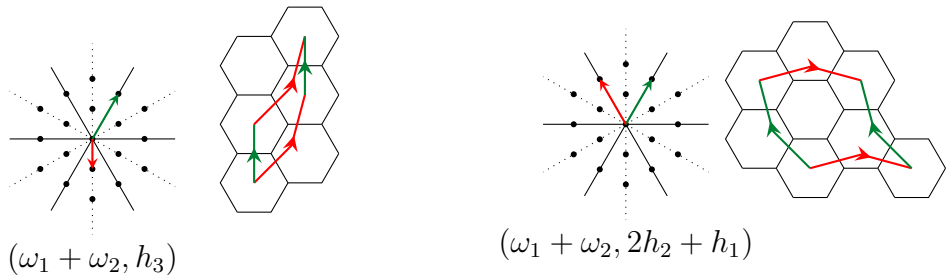

\centering

\begin{tabular}{cc}
\begin{minipage}{0.4\hsize}
\input{figure/gauge/A2/fund_adj_1.tikz}
\end{minipage}
\begin{minipage}{0.4\hsize}
\input{figure/gauge/A2/adj_adj_1.tikz}
\end{minipage}
\end{tabular}
\caption{Some more examples of the hexagonal tilings and the dyonic charges \label{fig:more}}
\end{figure}

It is clear that the action of the Weyl group is consistent.
Because the cycles $\alpha,\beta$ define the basis of charges, the action of $SL(2,\bbZ)$ on the  dyonic charges $(\lambda_e,\lambda_m)$  and that on the cycles $(\alpha,\beta)$ should be transpose of each other, and indeed the mapping \eqref{eq:map} satisfies this condition.
Let us end this section by exhibiting some more examples of the mapping between the dyonic charges and the hexagonal tilings, see Fig.~\ref{fig:more}.

In this section we only discussed the $A_2$ case. 
It would be interesting to find a general map from the charge of the dyonic loops to the networks for or general $A_{k>2}$ cases.

\section{Some examples in the superconformal index}
\label{sec:T3junction}

So far we have been studying the properties of the networks of the class S theories. 
In this section, we would like to study a few concrete manifestations of our analysis when the four-dimensional side is $S^1\times S^3$, or equivalently, in the setting of the superconformal index.

As has been well established, the superconformal index of a class S theory of type $\SU(N)$, defined by a Riemann surface $C$ of genus $g$ with $n$ full punctures, is given by the (zero-area limit of the) $(p,q,t)$-deformed Yang-Mills theory on the same Riemann surface \cite{Gadde:2009kb,Gadde:2011ik,Gadde:2011uv}.
Here we consider the case $q=t$ where the $p$ dependence automatically drops out,
and the superconformal index becomes the standard $q$-deformed Yang-Mills theory,
defined in \cite{Buffenoir:1994fh,Alekseev:1994pa,Aganagic:2004js}.
In this case, the supercharge defining the superconformal index is compatible with the presence of 
 a half-BPS line  wrapping the $S^1$ direction \cite{Dimofte:2011py,Gang:2012yr}.

Therefore, in this setup, a defect of a class S theory, specified by a network with labels  on the Riemann surface $C$  should be realized as a concrete object in the $q$-deformed Yang-Mills theory, defined by the same network with labels on $C$. 
Below, we show that they are just given by the Wilson lines\footnote{This statement was independently obtained by  W. Peelaers and L. Rastelli.  } and the Wilson junctions in the $q$-deformed Yang-Mills theory.
In particular, we compute the superconformal index of the nontrivial line operator of the $T_3$ theory, namely the 6d $\SUSYN{(2,0)}$ theory of type $\SU(3)$ compactified on a sphere with three full punctures. 
Reassuringly, we will find that the result shows an enhancement of the symmetry from $\SU(3)^3$ to $E_6$, as it should be, since the $T_3$ theory is the $E_6$ theory of Minahan and Nemeshcansky.

Below, we will first very briefly review the relation between the superconformal index and the $q$-deformed Yang-Mills in Sec.~\ref{sec:SCIwithout}. 
Then we check that the superconformal index with a Wilson line in the 4d gauge theory is given by the Wilson line in the 2d gauge theory in Sec.~\ref{sec:SCIwith}.
Finally, we will compute the superconformal index of a network in the $T_3$ theory in Sec.~\ref{sec:SCInetwork}.
Note that in this section we use the 2d Yang-Mills normalization of the junctions as discussed in Sec.~\ref{subsec:removal}.

\subsection{The superconformal indices without lines}\label{sec:SCIwithout}

Consider the class S theory $S_{g,n}$ for a genus $g$ surface with $n$ full punctures. 
Its superconformal index  is defined as \begin{equation}
\cI_g(a_1,\ldots, a_n)=\Tr (-1)^F q^{\Delta-I_3} a_1 \cdots a_n 
\end{equation} 
where the trace is over the Hilbert space of the system on $S^3$
or equivalently on the space of the operators, 
$\Delta$ is the scaling dimension,
$I_3$ is the Cartan generator of $\SU(2)_R$ symmetry, and 
$a_i\in \SU(N)_i$ is a group element of the flavor symmetry $\SU(N)_i$ of the theory associated to the $i$-th full puncture. 

This is known to be given by the following explicit expression: \begin{equation}
\cI_g(a_1,\ldots,a_n)=\sum_\lambda \frac{\prod_i K(a_i)\chi_\lambda(a_i) }{ K_0^{2g-2+n} \chi_\lambda(q^\rho)^{2g-2+n} }\label{SCIgn}
\end{equation}
where the sum is over all the irreducible representations $\lambda$ of $\SU(N)$,
$\chi_\lambda(a)$ is the character of $a$ in the representation $\lambda$,
$q^\rho$ is the element \begin{equation}
q^\rho=\textrm{diag}(q^{(N-1)/2},q^{(N-3)/2},\ldots, q^{(1-N)/2}),
\end{equation}
and $K(a)$, $K_0$ are given by 
\begin{equation}
K(a)=\PE[\frac{q}{1-q}(N-1+\sum_{i\neq j} a_i/a_j)],\qquad 
K_0=\PE\left[\frac{q^2+q^3+\cdots+q^N}{1-q}\right].
\end{equation}
Here we took $a$ to be diagonal, \begin{equation}
a=\textrm{diag}(a_1,\ldots,a_N)\in \SU(N),
\end{equation}
and $\PE$ is the plethystic exponential, defined by \begin{equation}
\PE \left[\sum_{n\geq 1} a_n t^n \right]_{t} = \prod_{n\ge 1}(1-t^n)^{-a_n}.
\end{equation}

Take two class S theories $S_{g,n}$ and $S_{g',n'}$,
pick one full puncture from each, say the last one from $S_{g,n}$ and the first one from $S_{g',n'}$.
Let us then couple  $\SU(N)$ gauge multiplets to the diagonal combination of the $\SU(N)$ symmetries carried by the two full punctures thus chosen.
The combined system has the superconformal index \begin{multline}
\cI_{g,g'}(a_1,\ldots,a_{n-1}; b_2,\ldots,b_{n'}) \\
=\int [dz]_\text{Haar} K(z)^{-2} \cI_{g}(a_1,\ldots,a_{n-1},z)
\cI_{g'}(z^{-1},b_1,\ldots,b_{n'})\label{SCIgauge}
\end{multline}
where $[dz]_\text{Haar}$ is the natural measure on the Cartan of $\SU(N)$  given by \begin{equation}
[dz]_\text{Haar}=\frac{1}{N!}\prod_{i=1}^{N-1} \frac{dz_i}{2\pi \sqrt{-1} z_i} \prod_{i\neq j} (1-z_i/z_j).
\end{equation}
Plugging in the expression \eqref{SCIgn} to \eqref{SCIgauge} and using the orthogonality of characters \begin{equation}
\int [dz]_\text{Haar} \chi_\lambda(z)\chi_\mu(z^{-1})=\delta_{\mu\lambda},
\end{equation} 
we see that the equation \eqref{SCIgauge} reduces to \eqref{SCIgn} for the genus $g+g'$ surface and $n+n'-2$ full punctures.
This is as it should be. 

When $g=0$ and $n=3$,   what we have is the $T_N$ theory.  
For $N=3$, the symmetry $\SU(N)^3$ enhances to $E_6$.
The decomposition is \begin{equation}
\cI_{T_3}(q,\alpha)=1+q\chi_{\Esix{000001}}
+q^2(\chi_{\Esix{000002}}+\chi_{\Esix{000001}})
+q^3(\chi_{\Esix{000003}}+\chi_{\Esix{001000}}+\chi_{\Esix{000002}}+2\chi_{\Esix{000001}}+2)q^3+\cdots
\end{equation} where $\alpha=(a,b,c)$ is the fugacity for $E_6$ flavor symmetry, and 
$\chi_\lambda=\chi_\lambda(\alpha)$ is the character of $E_6$ in the irreducible representation with Dynkin label $\lambda$.

\subsection{4d Wilson Lines}\label{sec:SCIwith}
Let us move on to the superconformal index in the presence of a loop operator wrapping $S^1$ \cite{Gang:2012yr,Gang:2012ff}. 
In this subsection, for simplicity, we only study a loop operator that is just a Wilson line with respect to a 4d gauge group. 
For class S theories, this covers every operator that is a genuine loop $L$ labeled by a representation $R$ on the Riemann surface (\ie all networks without junctions), since we can always cut the Riemann surface along $L$ to go to a duality frame where that particular loop wraps a tube once.  

To compute the superconformal index with a Wilson loop, let us first consider a more general situation.
Take a theory $X$ with $G$ flavor symmetry whose superconformal index is $\cI_X(a)$. Suppose we can couple it with the $G$ vector multiplet, such that the gauge coupling is exactly marginal. 
Let us insert  a BPS Wilson line in the representation $R$ of $G$. The resulting index is simply \begin{equation}
\cI_{R}(q)=\int[dz]_\text{Haar} K(z)^{-2} \chi_R(z)\cI_X(z).
\end{equation} 
This reduces to the formula \eqref{SCIgauge} when $R$ is a trivial representation, as it should be.

Now consider the case when $X$ consists of two copies of the $T_N$ theory, such that we gauge a diagonal combination of $G$ symmetries. We have \begin{align}
\cI_R(q)&=\int[dz]_\text{Haar}  \frac{K(a_1)K(a_2)K(z)}{K_0}
\left[\sum_{\lambda}\frac{\chi_{\lambda}(a_1)\chi_{\lambda}(a_2)\chi_{\lambda}(z)}{\chi_{\lambda}(q^\rho)} \right]  \chi_R(z) \nonumber \\
&\qquad \times K(z)^{-2}
\frac{K(z^{-1})K(a_3)K(a_4)}{K_0}
\left[\sum_{\lambda'}\frac{\chi_{\lambda'}(z^{-1})\chi_{\lambda'}(a_3)\chi_{\lambda'}(a_4)}{\chi_{\lambda'}(q^\rho)} \right]  \\
&= \frac{K(a_1)K(a_2)K(a_3)K(a_4)}{K_0{}^2}
\left[\sum_{\lambda,\lambda'}\frac{\chi_{\lambda}(a_1)\chi_{\lambda}(a_2)n_{\lambda R}^{\lambda'}\chi_{\lambda'}(a_3)\chi_{\lambda'}(a_4)}{\chi_{\lambda}(q^\rho)\chi_{\lambda'}(q^\rho)} \right]\label{WilsonSCI}
\end{align} where \begin{equation}
n_{\lambda R}^{\lambda'}=\int [da]_\text{Haar} \chi_R(a)\chi_{\lambda}(a)\chi_{\lambda'}(a^{-1})
\end{equation} counts how many times the irreducible representation $\lambda'$ appear in the tensor product $\lambda\otimes R$.

\begin{figure}
\[
\inc{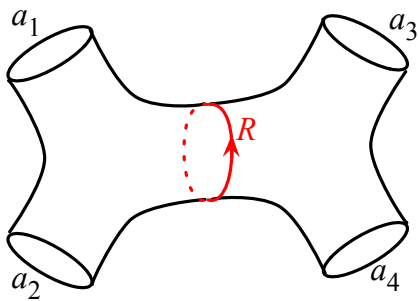}
\]
\caption{Wilson line around a tube corresponding to a gauge group.\label{TnWilson}}
\end{figure}

The result \eqref{WilsonSCI} is, up to the prefactor involving $K$, the unnormalized correlator of the $q$-deformed Yang-Mills theory with the Wilson line with the representation $R$, around the tube associated to the gauge group, see Fig.~\ref{TnWilson}.
There, we displayed full punctures as boundaries, as would be more common in the 2d Yang-Mills viewpoint. 

For non-deformed 2d Yang-Mills theory, this is an immediate consequence from the fact that the Hilbert space of the system on $S^1$ is the space of class functions on $G$, that is spanned by $\chi_\lambda(z)$, and the Wilson loop in the representation $R$ acts by a multiplication by $\chi_R(z)$ almost by definition.
For $q$-deformed 2d Yang-Mills theory, we need to use the fact that the structure of the tensor product decomposition of the representations of the quantum group $G_q$ is not deformed as long as $q$ is generic.

\subsection{A network in the $T_3$ theory}\label{sec:SCInetwork}
\begin{figure}[h]
\[
\inc{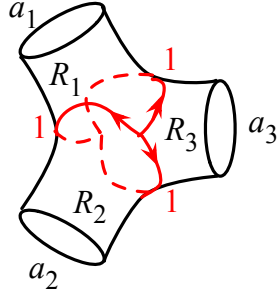}
\]
\caption{A line operator of $T_3$ theory, represented as a Wilson junction on the 2d side.\label{T3junction}}
\end{figure}

As an example of the network that is not just a loop, 
let us consider the $T_3$ theory and the 2d network  shown in Fig.~\ref{T3junction}.

The superconformal index is given by \begin{equation}
\cI_{T_3,\text{network}}=
\frac{K(a_1)K(a_2)K(a_3)}{K_0} \sum_{R_1,R_2,R_3} c_{R_1,R_2,R_3} \chi_{R_1}(a_1)\chi_{R_2}(a_2)\chi_{R_3}(a_3)
\label{T3line}
\end{equation} where $c_{R_1,R_2,R_3}$ is the amplitude of the $q$-deformed Yang-Mills, 
coupling three states $\chi_{R_{i}}(a_i)$ $(i=1,2,3)$ on the three boundaries. 

In the undeformed 2d Yang-Mills theory, $c_{R_1,R_2,R_3}$ is given by the integral
\begin{multline}
c_{R_1,R_2,R_3}=\int [dU_1][dU_2][dU_3]
\chi_{R_1}(U_2U_3^{-1})
\chi_{R_2}(U_3U_1^{-1})
\chi_{R_3}(U_1U_2^{-1}) \\
\times \epsilon_{ijk}\epsilon^{\bar i\bar j \bar k} (U_1)^i_{\bar i}(U_2)^j_{\bar j}(U_3)^k_{\bar k},
\end{multline}
where $U_{1,2,3}$ are holonomies of the 2d gauge field from one junction point to the other junction point, along three different segments. 

\begin{figure}
\[
\begin{tikzpicture}[tri/.style={{cm={-0.75,1.3,1.5,0,(0,0)}}}]
  \path[tri] (0,0) node  (00) {};
  \path[tri] (.66,.33) node {};
  \path[tri] (1,0) node  (10) {$\omega_2$};
  \path[tri] (1,1) node  (01) {$\omega_1$};
  \path[tri] (1.66,.33) node {};
  \path[tri] (1.66,1.33) node {};
  \path[tri] (1.33,.66) node {};
  \path[tri] (2,0) node  (20) {};
  \path[tri] (2,1) node  (11)    {};
  \path[tri] (2,2) node  (02)   {};
  \path[tri] (2.66,.33) node {};
  \path[tri] (2.66,1.33) node {};
  \path[tri] (2.66,2.33) node {};
  \path[tri] (2.33,.66) node {};
  \path[tri] (2.33,1.66) node {};
  \path[tri] (3,0) node  (30) {};
  \path[tri] (3,1) node  (21) {};
  \path[tri] (3,2) node  (12) {};
  \path[tri] (3,3) node  (03) {};
  \path[tri] (4,0) node  (50) {};
  \path[tri] (4,4) node  (05) {};
  \draw[->] (00)--(10);
  \draw[->] (10)--(20);
  \draw[->] (20)--(30);
  \draw[->] (01)--(11);
  \draw[->] (11)--(21);
  \draw[->] (02)--(12);
  \draw[<-] (00)--(01);
  \draw[<-] (01)--(02);
  \draw[<-] (02)--(03);
  \draw[<-] (10)--(11);
  \draw[<-] (11)--(12);
  \draw[<-] (20)--(21);
  \draw[->] (10)--(01);
  \draw[->] (20)--(11);
  \draw[->] (11)--(02);
  \draw[->] (30)--(21);
  \draw[->] (21)--(12);
  \draw[->] (12)--(03);
  \draw[dashed] (30)--(50);
  \draw[dashed] (03)--(05);
\end{tikzpicture}
\]
\caption{ An arrow $R\to R'$ connects two highest weights $R$, $R'$ of $\SU(3)$ if $R'\otimes V$ contains $R$. $c_{R_1,R_2,R_3}$ is nonzero only when the arrows form a triangle $R_{1}\to R_2\to R_3\to R_1$.   \label{su3reps}}
\end{figure}
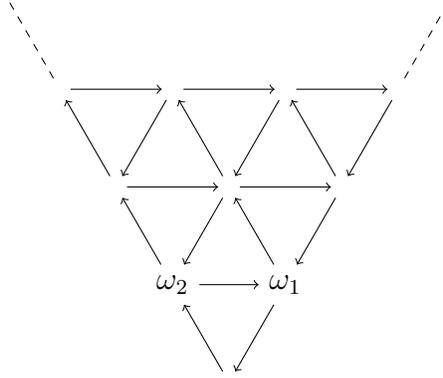

In the $q$-deformed Yang-Mills theory, we need to perform the integral above in the sense of the quantum group \cite{Alekseev:1994pa,Buffenoir:1994fh}. 
The $U_3$ integral gives a nonzero result only when  $R_1$ is contained in $R_2\otimes V$, where $V$ is the standard three-dimensional representation of $\SU(3)_q$.  
Since the rules of irreducible decompositions of tensor products are unchanged under the $q$ deformation,
we see that $c_{R_1,R_2,R_3}$ is nonzero only when the highest weight of $R_1$ is given by adding to the highest weight of $R_2$ one of the three weights of $V$. 
This means that there is an arrow connecting $R_1\to R_2$ in the diagram of irreducible representations as shown in Fig.~\ref{su3reps}.
We immediately see that there we should have arrows similarly for $R_2\to R_3$ and $R_3\to R_1$. 
Therefore $c_{R_1,R_2,R_3}$ is nonzero only when $R_1\to R_2\to R_3\to R_1$ forms a triangle in Fig.~\ref{su3reps}. 

\begin{figure}
\[
\inc{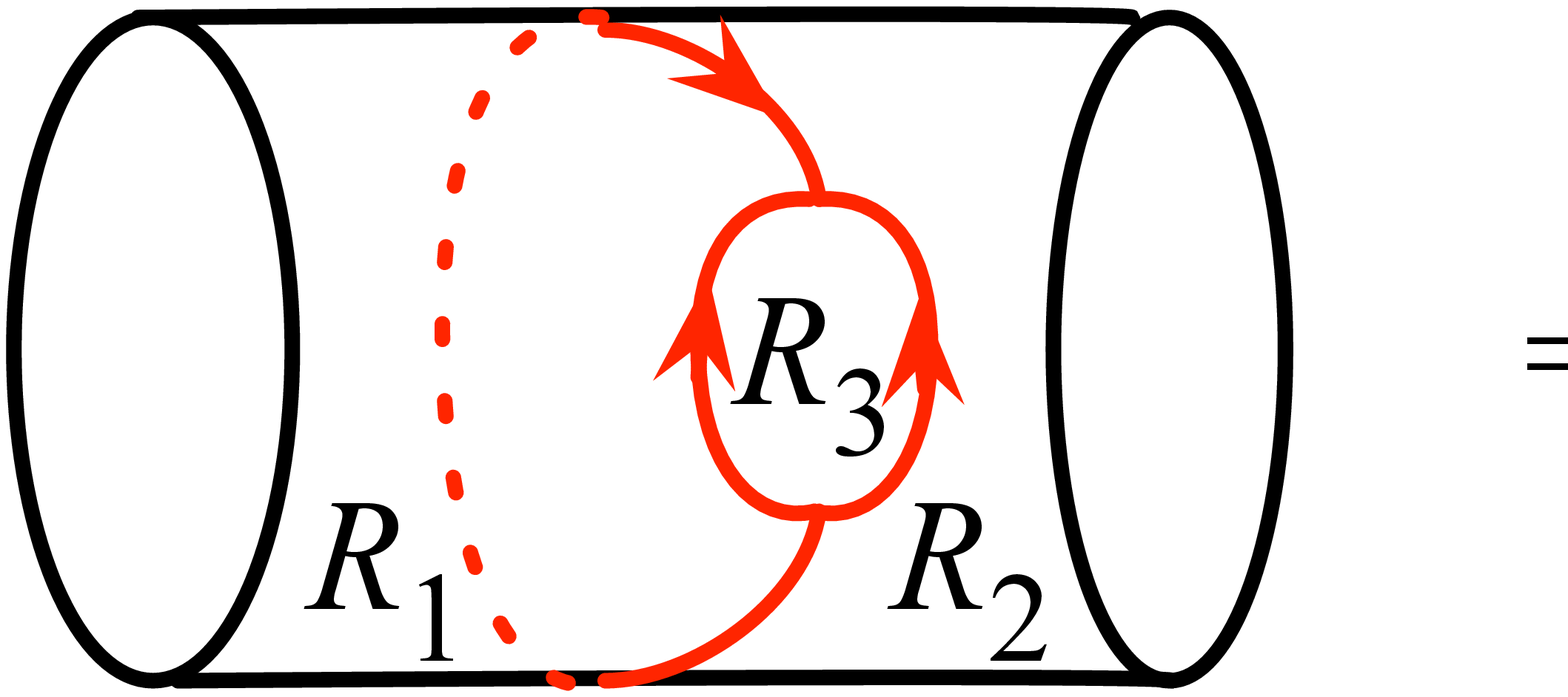}
\]
\caption{A trick\label{trick}}
\end{figure}

Next, consider what happens when the holonomy at a puncture $a_3$ is set to the special value $a_3=q^\rho=(q,1,q^{-1})$. 
This is equivalent to the absence of the puncture, and we have the situation in Fig.~\ref{trick}.
Using the skein relation, we see that the left hand side and the right hand side should be proportional by a factor of $[2]$. The right hand side is just $\sum_{R_1\to R_2} \chi_{R_1}(a_1)\chi_{R_2}(a_2)$.
Therefore, $c_{R_1,R_2,R_3}$ is determined by an overconstrained set of equations \begin{equation}
 \sum_{R_2\to R_3\to R_1\to R_2}  c_{R_1,R_2,R_3} \chi_{R_3}(q^\rho) = [2]_q\delta_{R_1\to R_2} 
\end{equation} where the sum on the right hand side is over $R_3$ that fit in the triangle. 
This equation can be recursively solved starting from the triangle closest to the origin. 
We can check that the following gives the general solution\footnote{The authors thank the twitter user \texttt{@LT\_shu} for kindly providing this general solution in \href{https://twitter.com/LT_shu/status/349136595632390144}{\texttt{https://twitter.com/LT\_shu/status/349136595632390144}}.}: \begin{align}
\dim_q (n\omega_1+m\omega_2)  =\chi_{n\omega_1+m\omega_2}(q^\rho)  &=\frac{[n+1][m+1][n+m+2]}{[2]} ,\\[2ex]
c_{
\begin{tikzpicture}[baseline=(base)]
  \path (0,.75) node  (base) {};
  \path (-1.2,.5) node  (10) {$\scriptstyle (n-1)\omega_1+m\omega_2$};
  \path (1.2,.5) node  (01) {$\scriptstyle  n\omega_1+(m-1)\omega_2$};
  \path (0,1) node  (11) {$\scriptstyle  n\omega_1+m\omega_2$};
  \draw[<-] (01)--(10);
  \draw[<-] (10)--(11.west);
  \draw[<-] (11.east)--(01);
\end{tikzpicture}
} &=\frac{[2]}{[n+1][m+1][n+m+1]},\\
c_{
\begin{tikzpicture}[baseline=(base)]
  \path (0,0) node  (00) {$\scriptstyle  n\omega_1+m\omega_2$};
  \path (-1.2,.5) node  (10) {$\scriptstyle  n\omega_1+(m+1)\omega_2$};
  \path (1.2,.5) node  (01) {$\scriptstyle (n+1)\omega_1+m\omega_2$};
  \path (0,.25) node  (base) {};
  \draw[<-] (01)--(10);
  \draw[<-] (10)--(00.west);
  \draw[<-] (00.east)--(01);
\end{tikzpicture}
} 
 &=\frac{[2]}{[n+1][m+1][n+m+3]}.
\end{align}

Plugging them into \eqref{T3line}, we find it nicely becomes a sum of representations of $E_6$:
\begin{multline}
\cI_{T_3,\text{Line}}(q,\alpha)=q^{1/2}\chi_{\Esix{100000}} 
+q^{3/2}(\chi_{\Esix{100001}})\\
+q^{5/2}(\chi_{\Esix{100000}}+\chi_{\Esix{000100}}+\chi_{\Esix{100001}}+\chi_{\Esix{100002}})+\cdots.
\end{multline}

\paragraph{Acknowledgments}
The authors would like to thank Dongmin Gang, Takuya Okuda and Masahito Yamazaki for helpful discussions.
YT's work is supported in part by JSPS Grant-in-Aid for Scientific Research No. 25870159.
NW is supported by the Advanced Leading Graduate Course for Photon Science, one of the Program for Leading Graduate Schools lead by Japan Society for the Promotion of Science, MEXT, and YT are also supported in part by WPI Initiative, MEXT, Japan at IPMU, the University of Tokyo.

\appendix

\section{Examples of OPE and charge/network dictionary}\label{app:Examples}

Here we present some more examples of the OPE and the correspondence of the dyonic charge and the networks for the $\SUSYN{4}$ theory, or equivalently for the torus.
The results of the OPE are for $A_{N-1>2}$ unless otherwise stated.
For the $A_2$ case, the edges with $3$ should be removed and those with $2$ can be  replaced by the reversed ones with $1$.
We use the convention that the magnetic weight lies in the fundamental Weyl chamber $\Lambda/\calW$ but the electric one is unrestricted.

\subsection{Pure dyonic loops}
Dyonic loops can be roughly classified into two, which we call pure and complex.
The pure ones are those that can be mapped to a Wilson loop in a duality frame,
and the complex ones are those without any such duality frame.
Let us first discuss the representation of the pure ones as loops on the torus.

We abbreviate a bundle of arrows with charges $s_1$, $s_2$, \ldots, $s_{r-1}$ and $s_r$ as an single arrow without any label:
\begin{align}
\raisebox{-0.45\height}{\tikzsetnextfilename{-figure-app-pure-arrow}
\begin{tikzpicture}
\def\vl{1.8}

\coordinate (O) at (0,0);
\coordinate (V) at (90:\vl);

\draw[white,line width=1cm] (O) -- (V);
\draw[mid arrow] (O) -- (V);
\end{tikzpicture}
}\;
:=
\quad\raisebox{-0.4\height}{\tikzsetnextfilename{-figure-app-pure-bundle_of_arrows}
\begin{tikzpicture}
\def\vl{1.8}
\def\wl{0.4}

\coordinate (V) at (90:\vl);

\coordinate (D1) at (0,0);
\coordinate (U1) at ($(D1)+(V)$);
\coordinate (D2) at (\wl,0);
\coordinate (U2) at ($(D2)+(V)$);
\coordinate (D3) at (2*\wl,0);
\coordinate (U3) at ($(D3)+(V)$);
\coordinate (D4) at (5*\wl,0);
\coordinate (U4) at ($(D4)+(V)$);
\coordinate (D5) at (6*\wl,0);
\coordinate (U5) at ($(D5)+(V)$);

\draw[mid arrow] (D1) -- (U1) node[above]{{\footnotesize $s_1$}};
\draw[mid arrow] (D2) -- (U2) node[above]{{\footnotesize $s_2$}};
\draw[mid arrow] (D3) -- (U3) node[above]{{\footnotesize $s_3$}};
\draw[dotted,ultra thick] ($(D3)!0.5!(U3)+(0.2,0)$) -- ($(D4)!0.5!(U4)-(0.2,0)$);
\draw[mid arrow] (D4) -- (U4) node[above]{{\footnotesize $s_{{\tiny r\!-\!1}}\;\;$}};
\draw[mid arrow] (D5) -- (U5) node[above]{{\footnotesize $s_{r}$}};

\end{tikzpicture}
}.
\end{align}
Let  $\displaystyle \omega_I = \sum_{i=1}^{r} \omega_{s_i}$.
Then there are four types as follows:
\begin{align}
\nonumber&
 (\mu_m,\lambda_e)=(p\omega_I,q\omega_I),
&&
 (\mu_m,\lambda_e)=(p\omega_I,-q\omega_I)
\\
q \ge p:\qquad
&
\raisebox{-0.35\height}{\tikzsetnextfilename{-figure-app-pure-pure1}
\begin{tikzpicture}
\def\vl{1.8}
\def\hl{2}
\coordinate (H) at (0:\hl);
\coordinate (V) at (90:\vl);
\coordinate (DL) at (0,0);
\coordinate (DR) at ($(DL)+(H)$);
\coordinate (UL) at ($(DL)+(V)$);
\coordinate (UR) at ($(UL)+(H)$);

\coordinate (L1) at ($(UL)!0.2!(DL)$);
\coordinate (L2) at ($(UL)!0.3!(DL)$);
\coordinate (L3) at ($(UL)!0.4!(DL)$);
\coordinate (L4) at ($(UL)!0.5!(DL)$);
\coordinate (L5) at ($(UL)!0.6!(DL)$);
\coordinate (R1) at ($(UR)!0.4!(DR)$);
\coordinate (R2) at ($(UR)!0.5!(DR)$);
\coordinate (R3) at ($(UR)!0.6!(DR)$);
\coordinate (R4) at ($(UR)!0.7!(DR)$);
\coordinate (R5) at ($(UR)!0.8!(DR)$);
\coordinate (U1) at ($(UL)!0.2!(UR)$);
\coordinate (U2) at ($(UL)!0.3!(UR)$);
\coordinate (U3) at ($(UL)!0.4!(UR)$);
\coordinate (D1) at ($(DL)!0.6!(DR)$);
\coordinate (D2) at ($(DL)!0.7!(DR)$);
\coordinate (D3) at ($(DL)!0.8!(DR)$);

\draw (UL) -- (UR) -- (DR) -- (DL) -- (UL);

\def\a{35}
\def\b{0.5}

\newcommand{\curve}[4]{\draw[mid arrow] (#1) .. controls ($(#1)!#4!#3:(#2)$) and ($(#2)!#4!-#3:(#1)$) .. (#2);}

\curve{L1}{U1}{-\a}{\b}
\curve{L2}{U2}{-\a}{\b}
\curve{L3}{U3}{-\a}{\b}
\draw[mid arrow] (L4) -- (R1);
\draw[mid arrow] (L5) -- (R2);
\curve{D1}{R3}{\a}{\b}
\curve{D2}{R4}{\a}{\b}
\curve{D3}{R5}{\a}{\b}

\draw [decorate,decoration={brace,amplitude=5pt},xshift=-4pt,yshift=0pt] ($(UR)+(0.1,0)$) -- ($(DR)+(0.1,0)$) node [black,midway,xshift=+0.4cm]  {\footnotesize $q$};
\draw [decorate,decoration={brace,amplitude=5pt},xshift=-4pt,yshift=0pt] ($(UL)+(0,0.1)$) -- ($(UR)+(0,0.1)$) node [black,midway,yshift=+0.4cm]  {\footnotesize $p$};

\end{tikzpicture}
},
&&
\raisebox{-0.35\height}{\tikzsetnextfilename{-figure-app-pure-pure3}
\begin{tikzpicture}
\def\vl{1.8}
\def\hl{2}
\coordinate (H) at (0:\hl);
\coordinate (V) at (90:\vl);
\coordinate (DL) at (0,0);
\coordinate (DR) at ($(DL)+(H)$);
\coordinate (UL) at ($(DL)+(V)$);
\coordinate (UR) at ($(UL)+(H)$);

\coordinate (L1) at ($(UL)!0.4!(DL)$);
\coordinate (L2) at ($(UL)!0.5!(DL)$);
\coordinate (L3) at ($(UL)!0.6!(DL)$);
\coordinate (L4) at ($(UL)!0.7!(DL)$);
\coordinate (L5) at ($(UL)!0.8!(DL)$);
\coordinate (R1) at ($(UR)!0.2!(DR)$);
\coordinate (R2) at ($(UR)!0.3!(DR)$);
\coordinate (R3) at ($(UR)!0.4!(DR)$);
\coordinate (R4) at ($(UR)!0.5!(DR)$);
\coordinate (R5) at ($(UR)!0.6!(DR)$);
\coordinate (U1) at ($(UL)!0.6!(UR)$);
\coordinate (U2) at ($(UL)!0.7!(UR)$);
\coordinate (U3) at ($(UL)!0.8!(UR)$);
\coordinate (D1) at ($(DL)!0.2!(DR)$);
\coordinate (D2) at ($(DL)!0.3!(DR)$);
\coordinate (D3) at ($(DL)!0.4!(DR)$);

\draw (UL) -- (UR) -- (DR) -- (DL) -- (UL);

\def\a{35}
\def\b{0.5}

\newcommand{\curve}[4]{\draw[mid arrow] (#1) .. controls ($(#1)!#4!#3:(#2)$) and ($(#2)!#4!-#3:(#1)$) .. (#2);}

\curve{R1}{U3}{\a}{\b}
\curve{R2}{U2}{\a}{\b}
\curve{R3}{U1}{\a}{\b}
\draw[mid arrow] (R4) -- (L1);
\draw[mid arrow] (R5) -- (L2);
\curve{D3}{L3}{-\a}{\b}
\curve{D2}{L4}{-\a}{\b}
\curve{D1}{L5}{-\a}{\b}

\draw [decorate,decoration={brace,amplitude=5pt},xshift=-4pt,yshift=0pt] ($(UR)+(0.1,0)$) -- ($(DR)+(0.1,0)$) node [black,midway,xshift=+0.4cm]  {\footnotesize $q$};
\draw [decorate,decoration={brace,amplitude=5pt},xshift=-4pt,yshift=0pt] ($(UL)+(0,0.1)$) -- ($(UR)+(0,0.1)$) node [black,midway,yshift=+0.4cm]  {\footnotesize $p$};

\end{tikzpicture}
},
\\
p \ge q:\qquad
&
\raisebox{-0.35\height}{\tikzsetnextfilename{-figure-app-pure-pure4}
\begin{tikzpicture}
\def\vl{1.8}
\def\hl{2}
\coordinate (H) at (0:\hl);
\coordinate (V) at (90:\vl);
\coordinate (DL) at (0,0);
\coordinate (DR) at ($(DL)+(H)$);
\coordinate (UL) at ($(DL)+(V)$);
\coordinate (UR) at ($(UL)+(H)$);

\coordinate (L1) at ($(UL)!0.2!(DL)$);
\coordinate (L2) at ($(UL)!0.3!(DL)$);
\coordinate (L3) at ($(UL)!0.4!(DL)$);
\coordinate (R1) at ($(UR)!0.6!(DR)$);
\coordinate (R2) at ($(UR)!0.7!(DR)$);
\coordinate (R3) at ($(UR)!0.8!(DR)$);
\coordinate (U1) at ($(UL)!0.2!(UR)$);
\coordinate (U2) at ($(UL)!0.3!(UR)$);
\coordinate (U3) at ($(UL)!0.4!(UR)$);
\coordinate (U4) at ($(UL)!0.5!(UR)$);
\coordinate (U5) at ($(UL)!0.6!(UR)$);
\coordinate (D1) at ($(DL)!0.4!(DR)$);
\coordinate (D2) at ($(DL)!0.5!(DR)$);
\coordinate (D3) at ($(DL)!0.6!(DR)$);
\coordinate (D4) at ($(DL)!0.7!(DR)$);
\coordinate (D5) at ($(DL)!0.8!(DR)$);

\draw (UL) -- (UR) -- (DR) -- (DL) -- (UL);

\def\a{35}
\def\b{0.5}

\newcommand{\curve}[4]{\draw[mid arrow] (#1) .. controls ($(#1)!#4!#3:(#2)$) and ($(#2)!#4!-#3:(#1)$) .. (#2);}

\curve{L1}{U1}{-\a}{\b}
\curve{L2}{U2}{-\a}{\b}
\curve{L3}{U3}{-\a}{\b}
\draw[mid arrow] (D1) -- (U4);
\draw[mid arrow] (D2) -- (U5);
\curve{D3}{R1}{\a}{\b}
\curve{D4}{R2}{\a}{\b}
\curve{D5}{R3}{\a}{\b}

\draw [decorate,decoration={brace,amplitude=5pt},xshift=-4pt,yshift=0pt] ($(UR)+(0.1,0)$) -- ($(DR)+(0.1,0)$) node [black,midway,xshift=+0.4cm]  {\footnotesize $q$};
\draw [decorate,decoration={brace,amplitude=5pt},xshift=-4pt,yshift=0pt] ($(UL)+(0,0.1)$) -- ($(UR)+(0,0.1)$) node [black,midway,yshift=+0.4cm]  {\footnotesize $p$};

\end{tikzpicture}
},
&&
\raisebox{-0.35\height}{\tikzsetnextfilename{-figure-app-pure-pure2}
\begin{tikzpicture}
\def\vl{1.8}
\def\hl{2}
\coordinate (H) at (0:\hl);
\coordinate (V) at (90:\vl);
\coordinate (DL) at (0,0);
\coordinate (DR) at ($(DL)+(H)$);
\coordinate (UL) at ($(DL)+(V)$);
\coordinate (UR) at ($(UL)+(H)$);

\coordinate (L1) at ($(UL)!0.6!(DL)$);
\coordinate (L2) at ($(UL)!0.7!(DL)$);
\coordinate (L3) at ($(UL)!0.8!(DL)$);
\coordinate (R1) at ($(UR)!0.2!(DR)$);
\coordinate (R2) at ($(UR)!0.3!(DR)$);
\coordinate (R3) at ($(UR)!0.4!(DR)$);
\coordinate (U1) at ($(UL)!0.4!(UR)$);
\coordinate (U2) at ($(UL)!0.5!(UR)$);
\coordinate (U3) at ($(UL)!0.6!(UR)$);
\coordinate (U4) at ($(UL)!0.7!(UR)$);
\coordinate (U5) at ($(UL)!0.8!(UR)$);
\coordinate (D1) at ($(DL)!0.2!(DR)$);
\coordinate (D2) at ($(DL)!0.3!(DR)$);
\coordinate (D3) at ($(DL)!0.4!(DR)$);
\coordinate (D4) at ($(DL)!0.5!(DR)$);
\coordinate (D5) at ($(DL)!0.6!(DR)$);

\draw (UL) -- (UR) -- (DR) -- (DL) -- (UL);

\def\a{35}
\def\b{0.5}

\newcommand{\curve}[4]{\draw[mid arrow] (#1) .. controls ($(#1)!#4!#3:(#2)$) and ($(#2)!#4!-#3:(#1)$) .. (#2);}

\curve{D1}{L3}{-\a}{\b}
\curve{D2}{L2}{-\a}{\b}
\curve{D3}{L1}{-\a}{\b}
\draw[mid arrow] (D4) -- (U1);
\draw[mid arrow] (D5) -- (U2);
\curve{R3}{U3}{\a}{\b}
\curve{R2}{U4}{\a}{\b}
\curve{R1}{U5}{\a}{\b}

\draw [decorate,decoration={brace,amplitude=5pt},xshift=-4pt,yshift=0pt] ($(UR)+(0.1,0)$) -- ($(DR)+(0.1,0)$) node [black,midway,xshift=+0.4cm]  {\footnotesize $q$};
\draw [decorate,decoration={brace,amplitude=5pt},xshift=-4pt,yshift=0pt] ($(UL)+(0,0.1)$) -- ($(UR)+(0,0.1)$) node [black,midway,yshift=+0.4cm]  {\footnotesize $p$};

\end{tikzpicture}
}.
\end{align}
This is essentially the same as the discussion in \cite{DMO09}.

\subsection{$W_{\Box\!\Box} \times T_{\Box}$}
Let us compute the skein relation of $W_{\Box\!\Box} $ and $T_{\Box}$.  
Comparing with what we expect from the gauge theory, we can then identify various networks with complex dyonic loops.
From the skein relation, we have
\begin{align}
\nonumber \raisebox{-0.35\height}{\tikzsetnextfilename{-figure-app-W2S_Tf-OPE}
\begin{tikzpicture}
\def\vl{2.3}
\def\hl{2.5}
\coordinate (H) at (0:\hl);
\coordinate (V) at (90:\vl);
\coordinate (DL) at (0,0);
\coordinate (DR) at ($(DL)+(H)$);
\coordinate (UL) at ($(DL)+(V)$);
\coordinate (UR) at ($(UL)+(H)$);

\coordinate (L1) at ($(DL)!0.5!(UL)$);
\coordinate (R1) at ($(DR)!0.5!(UR)$);
\coordinate (U1) at ($(UL)!0.5!(UR)$);
\coordinate (D1) at ($(DL)!0.5!(DR)$);

\draw (UL) -- (UR) -- (DR) -- (DL) -- (UL);

\draw[green,mid-end arrow,ultra thick] (D1) -- (U1) node[black,above]{$\Box$};
\draw[white,line width=3pt] ($(L1)+0.1*(H)$) -- ($(R1)-0.1*(H)$);
\draw[red,mid-end arrow,ultra thick] (L1) -- (R1) node[black,right]{$\Box\!\Box$};
\end{tikzpicture}
}\;
=\;\frq^{2/N}\; &\left[\quad\raisebox{-0.45\height}{\tikzsetnextfilename{-figure-app-W2S_Tf-0A}
\begin{tikzpicture}
\def\vl{2.3}
\def\hl{2.5}
\coordinate (H) at (0:\hl);
\coordinate (V) at (90:\vl);
\coordinate (DL) at (0,0);
\coordinate (DR) at ($(DL)+(H)$);
\coordinate (UL) at ($(DL)+(V)$);
\coordinate (UR) at ($(UL)+(H)$);

\coordinate (L1) at ($(DL)!0.7!(UL)$);
\coordinate (L2) at ($(DL)!0.3!(UL)$);
\coordinate (R1) at ($(DR)!0.7!(UR)$);
\coordinate (R2) at ($(DR)!0.3!(UR)$);
\coordinate (U1) at ($(UL)!0.5!(UR)$);
\coordinate (D1) at ($(DL)!0.5!(DR)$);

\coordinate (M1) at ($(U1)!0.2!(D1)$);
\coordinate (M2) at ($(U1)!0.4!(D1)$);
\coordinate (M3) at ($(D1)!0.4!(U1)$);
\coordinate (M4) at ($(D1)!0.2!(U1)$);

\draw (UL) -- (UR) -- (DR) -- (DL) -- (UL);

\draw[mid arrow] (L1) -- (M2);
\draw[mid arrow] (L2) -- (M4);
\draw[mid arrow] (M1) -- (R1);
\draw[mid arrow] (M3) -- (R2);
\draw[mid arrow] (D1) -- (M4);
\draw[mid arrow] (M1) -- (U1);
\draw[mid arrow] (M3) -- (M2);

\draw[blue,mid arrow] (M4) -- node[left]{2} (M3);
\draw[blue,mid arrow] (M2) -- node[left]{2} (M1);
\end{tikzpicture}
}\;
-\quad\raisebox{-0.45\height}{\tikzsetnextfilename{-figure-app-W2S_Tf-0B}
\begin{tikzpicture}
\def\vl{2.3}
\def\hl{2.5}
\coordinate (H) at (0:\hl);
\coordinate (V) at (90:\vl);
\coordinate (DL) at (0,0);
\coordinate (DR) at ($(DL)+(H)$);
\coordinate (UL) at ($(DL)+(V)$);
\coordinate (UR) at ($(UL)+(H)$);

\coordinate (L1) at ($(DL)!0.5!(UL)$);
\coordinate (R1) at ($(DR)!0.5!(UR)$);
\coordinate (U1) at ($(UL)!0.5!(UR)$);
\coordinate (D1) at ($(DL)!0.5!(DR)$);

\coordinate (M1) at ($(UR)!0.3!(DL)$);
\coordinate (M2) at ($(DL)!0.3!(UR)$);

\draw (UL) -- (UR) -- (DR) -- (DL) -- (UL);

\draw[blue,mid arrow] (L1) -- node[above right]{2} (M2);
\draw[mid arrow] (D1) -- (M2);
\draw[blue,mid arrow] (M1) -- node[below]{2} (R1);
\draw[mid arrow] (M1) -- (U1);
\draw[red,mid arrow] (M2) --  node[below right]{3} (M1);
\end{tikzpicture}
} \right. \\
&\left. +\; \frq^{-1} \quad\raisebox{-0.45\height}{\tikzsetnextfilename{-figure-app-W2S_Tf-1A}
\begin{tikzpicture}
\def\vl{1.8}
\def\hl{2}
\coordinate (H) at (0:\hl);
\coordinate (V) at (90:\vl);
\coordinate (DL) at (0,0);
\coordinate (DR) at ($(DL)+(H)$);
\coordinate (UL) at ($(DL)+(V)$);
\coordinate (UR) at ($(UL)+(H)$);

\coordinate (L1) at ($(DL)!0.5!(UL)$);
\coordinate (R1) at ($(DR)!0.5!(UR)$);
\coordinate (U1) at ($(UL)!0.5!(UR)$);
\coordinate (D1) at ($(DL)!0.5!(DR)$);

\coordinate (M1) at ($(UL)!0.3!(DR)$);
\coordinate (M2) at ($(DR)!0.3!(UL)$);

\draw (UL) -- (UR) -- (DR) -- (DL) -- (UL);

\draw[blue,mid arrow] (L1) -- node[below]{2} (M1);
\draw[mid arrow] (D1) -- (M2);
\draw[blue,mid arrow] (M2) -- node[above]{2} (R1);
\draw[mid arrow] (M1) -- (U1);
\draw[mid arrow] (M1) -- (M2);
\end{tikzpicture}
}\;
+\; \frq^{-2} \quad\raisebox{-0.45\height}{\tikzsetnextfilename{-figure-app-W2S_Tf-2A}
\begin{tikzpicture}
\def\vl{2.3}
\def\hl{2.5}
\coordinate (H) at (0:\hl);
\coordinate (V) at (90:\vl);
\coordinate (DL) at (0,0);
\coordinate (DR) at ($(DL)+(H)$);
\coordinate (UL) at ($(DL)+(V)$);
\coordinate (UR) at ($(UL)+(H)$);

\coordinate (L1) at ($(DL)!0.7!(UL)$);
\coordinate (L2) at ($(DL)!0.3!(UL)$);
\coordinate (R1) at ($(DR)!0.7!(UR)$);
\coordinate (R2) at ($(DR)!0.3!(UR)$);
\coordinate (U1) at ($(UL)!0.5!(UR)$);
\coordinate (D1) at ($(DL)!0.5!(DR)$);

\draw (UL) -- (UR) -- (DR) -- (DL) -- (UL);

\draw[mid arrow] (L2) -- (R1);
\draw[mid arrow] (D1) -- (R2);
\draw[mid arrow] (L1) -- (U1);
\end{tikzpicture}
}\quad  \right]
\end{align}
The first term is a new one, the second and third ones are elementary and the final one is pure.
Then we identify:
\begin{align}
 D_{[(\omega_1,2h_{i(\neq 1)})]} &\sim
\raisebox{-0.45\height}{\tikzsetnextfilename{-figure-app-W2S_Tf-0A}
\begin{tikzpicture}
\def\vl{2.3}
\def\hl{2.5}
\coordinate (H) at (0:\hl);
\coordinate (V) at (90:\vl);
\coordinate (DL) at (0,0);
\coordinate (DR) at ($(DL)+(H)$);
\coordinate (UL) at ($(DL)+(V)$);
\coordinate (UR) at ($(UL)+(H)$);

\coordinate (L1) at ($(DL)!0.7!(UL)$);
\coordinate (L2) at ($(DL)!0.3!(UL)$);
\coordinate (R1) at ($(DR)!0.7!(UR)$);
\coordinate (R2) at ($(DR)!0.3!(UR)$);
\coordinate (U1) at ($(UL)!0.5!(UR)$);
\coordinate (D1) at ($(DL)!0.5!(DR)$);

\coordinate (M1) at ($(U1)!0.2!(D1)$);
\coordinate (M2) at ($(U1)!0.4!(D1)$);
\coordinate (M3) at ($(D1)!0.4!(U1)$);
\coordinate (M4) at ($(D1)!0.2!(U1)$);

\draw (UL) -- (UR) -- (DR) -- (DL) -- (UL);

\draw[mid arrow] (L1) -- (M2);
\draw[mid arrow] (L2) -- (M4);
\draw[mid arrow] (M1) -- (R1);
\draw[mid arrow] (M3) -- (R2);
\draw[mid arrow] (D1) -- (M4);
\draw[mid arrow] (M1) -- (U1);
\draw[mid arrow] (M3) -- (M2);

\draw[blue,mid arrow] (M4) -- node[left]{2} (M3);
\draw[blue,mid arrow] (M2) -- node[left]{2} (M1);
\end{tikzpicture}
},
& D_{[(\omega_1,h_{i(\neq 1)}+h_{j(>i)})]} &\sim
\raisebox{-0.45\height}{\tikzsetnextfilename{-figure-app-W2S_Tf-0B}
\begin{tikzpicture}
\def\vl{2.3}
\def\hl{2.5}
\coordinate (H) at (0:\hl);
\coordinate (V) at (90:\vl);
\coordinate (DL) at (0,0);
\coordinate (DR) at ($(DL)+(H)$);
\coordinate (UL) at ($(DL)+(V)$);
\coordinate (UR) at ($(UL)+(H)$);

\coordinate (L1) at ($(DL)!0.5!(UL)$);
\coordinate (R1) at ($(DR)!0.5!(UR)$);
\coordinate (U1) at ($(UL)!0.5!(UR)$);
\coordinate (D1) at ($(DL)!0.5!(DR)$);

\coordinate (M1) at ($(UR)!0.3!(DL)$);
\coordinate (M2) at ($(DL)!0.3!(UR)$);

\draw (UL) -- (UR) -- (DR) -- (DL) -- (UL);

\draw[blue,mid arrow] (L1) -- node[above right]{2} (M2);
\draw[mid arrow] (D1) -- (M2);
\draw[blue,mid arrow] (M1) -- node[below]{2} (R1);
\draw[mid arrow] (M1) -- (U1);
\draw[red,mid arrow] (M2) --  node[below right]{3} (M1);
\end{tikzpicture}
}, \\
 D_{[(\omega_1,h_1+h_{i(\neq 1)})]} &\sim
\raisebox{-0.45\height}{\tikzsetnextfilename{-figure-app-W2S_Tf-1A}
\begin{tikzpicture}
\def\vl{1.8}
\def\hl{2}
\coordinate (H) at (0:\hl);
\coordinate (V) at (90:\vl);
\coordinate (DL) at (0,0);
\coordinate (DR) at ($(DL)+(H)$);
\coordinate (UL) at ($(DL)+(V)$);
\coordinate (UR) at ($(UL)+(H)$);

\coordinate (L1) at ($(DL)!0.5!(UL)$);
\coordinate (R1) at ($(DR)!0.5!(UR)$);
\coordinate (U1) at ($(UL)!0.5!(UR)$);
\coordinate (D1) at ($(DL)!0.5!(DR)$);

\coordinate (M1) at ($(UL)!0.3!(DR)$);
\coordinate (M2) at ($(DR)!0.3!(UL)$);

\draw (UL) -- (UR) -- (DR) -- (DL) -- (UL);

\draw[blue,mid arrow] (L1) -- node[below]{2} (M1);
\draw[mid arrow] (D1) -- (M2);
\draw[blue,mid arrow] (M2) -- node[above]{2} (R1);
\draw[mid arrow] (M1) -- (U1);
\draw[mid arrow] (M1) -- (M2);
\end{tikzpicture}
},
& D_{[(\omega_1,2\omega_1)]}  &\sim
\raisebox{-0.45\height}{\tikzsetnextfilename{-figure-app-W2S_Tf-2A}
\begin{tikzpicture}
\def\vl{2.3}
\def\hl{2.5}
\coordinate (H) at (0:\hl);
\coordinate (V) at (90:\vl);
\coordinate (DL) at (0,0);
\coordinate (DR) at ($(DL)+(H)$);
\coordinate (UL) at ($(DL)+(V)$);
\coordinate (UR) at ($(UL)+(H)$);

\coordinate (L1) at ($(DL)!0.7!(UL)$);
\coordinate (L2) at ($(DL)!0.3!(UL)$);
\coordinate (R1) at ($(DR)!0.7!(UR)$);
\coordinate (R2) at ($(DR)!0.3!(UR)$);
\coordinate (U1) at ($(UL)!0.5!(UR)$);
\coordinate (D1) at ($(DL)!0.5!(DR)$);

\draw (UL) -- (UR) -- (DR) -- (DL) -- (UL);

\draw[mid arrow] (L2) -- (R1);
\draw[mid arrow] (D1) -- (R2);
\draw[mid arrow] (L1) -- (U1);
\end{tikzpicture}
}.
\end{align}

\subsection{$W_{(2,1)} \times T_{\Box}$}
Let us next consider $W_{(2,1)} \times T_{\Box}$:
\begin{multline}
 \raisebox{-0.35\height}{\tikzsetnextfilename{-figure-app-WAS_Tf-OPE}
\begin{tikzpicture}
\def\vl{2.3}
\def\hl{2.5}
\coordinate (H) at (0:\hl);
\coordinate (V) at (90:\vl);
\coordinate (DL) at (0,0);
\coordinate (DR) at ($(DL)+(H)$);
\coordinate (UL) at ($(DL)+(V)$);
\coordinate (UR) at ($(UL)+(H)$);

\coordinate (L1) at ($(DL)!0.5!(UL)$);
\coordinate (R1) at ($(DR)!0.5!(UR)$);
\coordinate (U1) at ($(UL)!0.5!(UR)$);
\coordinate (D1) at ($(DL)!0.5!(DR)$);

\draw (UL) -- (UR) -- (DR) -- (DL) -- (UL);

\draw[green,mid-end arrow,ultra thick] (D1) -- (U1) node[black,above]{$\Box$};
\draw[white,line width=3pt] ($(L1)+0.1*(H)$) -- ($(R1)-0.1*(H)$);
\draw[red,mid-end arrow,ultra thick] (L1) -- (R1) node[black,right]{\tiny $\yng(2,1)$};
\end{tikzpicture}
}\;
=\;\frq^{3/N}\; \left[\quad\raisebox{-0.45\height}{\tikzsetnextfilename{-figure-app-WAS_Tf-0A}
\begin{tikzpicture}
\def\vl{2.3}
\def\hl{2.5}
\coordinate (H) at (0:\hl);
\coordinate (V) at (90:\vl);
\coordinate (DL) at (0,0);
\coordinate (DR) at ($(DL)+(H)$);
\coordinate (UL) at ($(DL)+(V)$);
\coordinate (UR) at ($(UL)+(H)$);

\coordinate (L1) at ($(DL)!0.7!(UL)$);
\coordinate (L2) at ($(DL)!0.3!(UL)$);
\coordinate (R1) at ($(DR)!0.7!(UR)$);
\coordinate (R2) at ($(DR)!0.3!(UR)$);
\coordinate (U1) at ($(UL)!0.5!(UR)$);
\coordinate (D1) at ($(DL)!0.5!(DR)$);

\coordinate (M1) at ($(U1)!0.2!(D1)$);
\coordinate (M2) at ($(U1)!0.4!(D1)$);
\coordinate (M3) at ($(D1)!0.4!(U1)$);
\coordinate (M4) at ($(D1)!0.2!(U1)$);

\draw (UL) -- (UR) -- (DR) -- (DL) -- (UL);

\draw[mid arrow] (L1) -- (M2);
\draw[blue,mid arrow] (L2) -- node[above]{2} (M4);
\draw[mid arrow] (M1) -- (R1);
\draw[blue,mid arrow] (M3) -- node[below]{2} (R2);
\draw[mid arrow] (D1) -- (M4);
\draw[mid arrow] (M1) -- (U1);
\draw[mid arrow] (M3) -- (M2);

\draw[red,mid arrow] (M4) -- node[left]{3} (M3);
\draw[blue,mid arrow] (M2) -- node[left]{2} (M1);
\end{tikzpicture}
}\;
-\quad\raisebox{-0.45\height}{\tikzsetnextfilename{-figure-app-WAS_Tf-0B}
\begin{tikzpicture}
\def\vl{2.3}
\def\hl{2.5}
\coordinate (H) at (0:\hl);
\coordinate (V) at (90:\vl);
\coordinate (DL) at (0,0);
\coordinate (DR) at ($(DL)+(H)$);
\coordinate (UL) at ($(DL)+(V)$);
\coordinate (UR) at ($(UL)+(H)$);

\coordinate (L1) at ($(DL)!0.5!(UL)$);
\coordinate (R1) at ($(DR)!0.5!(UR)$);
\coordinate (U1) at ($(UL)!0.5!(UR)$);
\coordinate (D1) at ($(DL)!0.5!(DR)$);

\coordinate (M1) at ($(UR)!0.3!(DL)$);
\coordinate (M2) at ($(DL)!0.3!(UR)$);

\draw (UL) -- (UR) -- (DR) -- (DL) -- (UL);

\draw[red,mid arrow] (L1) -- node[above right]{3} (M2);
\draw[mid arrow] (D1) -- (M2);
\draw[red,mid arrow] (M1) -- node[below]{3} (R1);
\draw[mid arrow] (M1) -- (U1);
\draw[green,mid arrow] (M2) --  node[below right]{4} (M1);
\end{tikzpicture}
} \right. \\
\left. +\; \frq^{-1} \quad\raisebox{-0.45\height}{\tikzsetnextfilename{-figure-app-WAS_Tf-1A}
\begin{tikzpicture}
\def\vl{2.3}
\def\hl{2.5}
\coordinate (H) at (0:\hl);
\coordinate (V) at (90:\vl);
\coordinate (DL) at (0,0);
\coordinate (DR) at ($(DL)+(H)$);
\coordinate (UL) at ($(DL)+(V)$);
\coordinate (UR) at ($(UL)+(H)$);

\coordinate (L1) at ($(DL)!0.7!(UL)$);
\coordinate (L2) at ($(DL)!0.3!(UL)$);
\coordinate (R1) at ($(DR)!0.7!(UR)$);
\coordinate (R2) at ($(DR)!0.3!(UR)$);
\coordinate (U1) at ($(UL)!0.3!(UR)$);
\coordinate (D1) at ($(DL)!0.3!(DR)$);

\coordinate (M1) at ($(R1)!0.3!(L1)$);
\coordinate (M2) at ($(R1)!0.7!(L1)$);
\coordinate (M3) at ($(R2)!0.7!(L2)$);
\coordinate (M4) at ($(R2)!0.3!(L2)$);

\draw (UL) -- (UR) -- (DR) -- (DL) -- (UL);

\draw[mid arrow] (M1) -- (U1);
\draw[mid arrow] (M1) -- (R1);
\draw[blue,mid arrow] (M2) -- node[below]{2} (M1);
\draw[mid arrow] (L1) -- (M2);
\draw[mid arrow] (M3) -- (M2);
\draw[blue,mid arrow] (L2) -- node[below]{2} (M3);
\draw[mid arrow] (M3) -- (M4);
\draw[blue,mid arrow] (M4) -- node[above]{2} (R2);
\draw[mid arrow] (D1) -- (M4);
\end{tikzpicture}
}\;
+\; \frq^{-2} \quad\raisebox{-0.45\height}{\tikzsetnextfilename{-figure-app-WAS_Tf-2A}
\begin{tikzpicture}
\def\vl{2.3}
\def\hl{2.5}
\coordinate (H) at (0:\hl);
\coordinate (V) at (90:\vl);
\coordinate (DL) at (0,0);
\coordinate (DR) at ($(DL)+(H)$);
\coordinate (UL) at ($(DL)+(V)$);
\coordinate (UR) at ($(UL)+(H)$);

\coordinate (L1) at ($(DL)!0.7!(UL)$);
\coordinate (L2) at ($(DL)!0.3!(UL)$);
\coordinate (R1) at ($(DR)!0.7!(UR)$);
\coordinate (R2) at ($(DR)!0.3!(UR)$);
\coordinate (U1) at ($(UL)!0.3!(UR)$);
\coordinate (D1) at ($(DL)!0.3!(DR)$);

\coordinate (M1) at ($(DR)!0.3!(L1)$);
\coordinate (M2) at ($(DR)!0.7!(L1)$);

\draw (UL) -- (UR) -- (DR) -- (DL) -- (UL);

\draw[blue,mid arrow] (L2) -- node[below]{2} (M2);
\draw[mid arrow] (D1) -- (M1);
\draw[blue,mid arrow] (M1) -- node[below]{2} (R2);
\draw[mid arrow] (M2) -- (R1);
\draw[mid arrow] (L1) -- (U1);
\draw[mid arrow] (M2) -- (M1);
\end{tikzpicture}
}\quad  \right]
\end{multline}
The first and the third terms are new, the second one is elementary and the fourth one is obtained by a T action of some elementary one.
When $N=3$, the second one does not appear and indeed this case is the same as Example ~\ref{subsec:WAdj_Tf_OPE}. 

We can therefore identify:
\begin{align}
 D_{[(\omega_1,2h_{i(\neq 1)}+h_{j(>i)})]}&  \sim
\raisebox{-0.45\height}{\tikzsetnextfilename{-figure-app-WAS_Tf-0A}
\begin{tikzpicture}
\def\vl{2.3}
\def\hl{2.5}
\coordinate (H) at (0:\hl);
\coordinate (V) at (90:\vl);
\coordinate (DL) at (0,0);
\coordinate (DR) at ($(DL)+(H)$);
\coordinate (UL) at ($(DL)+(V)$);
\coordinate (UR) at ($(UL)+(H)$);

\coordinate (L1) at ($(DL)!0.7!(UL)$);
\coordinate (L2) at ($(DL)!0.3!(UL)$);
\coordinate (R1) at ($(DR)!0.7!(UR)$);
\coordinate (R2) at ($(DR)!0.3!(UR)$);
\coordinate (U1) at ($(UL)!0.5!(UR)$);
\coordinate (D1) at ($(DL)!0.5!(DR)$);

\coordinate (M1) at ($(U1)!0.2!(D1)$);
\coordinate (M2) at ($(U1)!0.4!(D1)$);
\coordinate (M3) at ($(D1)!0.4!(U1)$);
\coordinate (M4) at ($(D1)!0.2!(U1)$);

\draw (UL) -- (UR) -- (DR) -- (DL) -- (UL);

\draw[mid arrow] (L1) -- (M2);
\draw[blue,mid arrow] (L2) -- node[above]{2} (M4);
\draw[mid arrow] (M1) -- (R1);
\draw[blue,mid arrow] (M3) -- node[below]{2} (R2);
\draw[mid arrow] (D1) -- (M4);
\draw[mid arrow] (M1) -- (U1);
\draw[mid arrow] (M3) -- (M2);

\draw[red,mid arrow] (M4) -- node[left]{3} (M3);
\draw[blue,mid arrow] (M2) -- node[left]{2} (M1);
\end{tikzpicture}
},
& D_{[(\omega_1,h_{i(\neq 1)}+h_{j(>i)}+h_{k(>j)})]} & \sim
\raisebox{-0.45\height}{\tikzsetnextfilename{-figure-app-WAS_Tf-0B}
\begin{tikzpicture}
\def\vl{2.3}
\def\hl{2.5}
\coordinate (H) at (0:\hl);
\coordinate (V) at (90:\vl);
\coordinate (DL) at (0,0);
\coordinate (DR) at ($(DL)+(H)$);
\coordinate (UL) at ($(DL)+(V)$);
\coordinate (UR) at ($(UL)+(H)$);

\coordinate (L1) at ($(DL)!0.5!(UL)$);
\coordinate (R1) at ($(DR)!0.5!(UR)$);
\coordinate (U1) at ($(UL)!0.5!(UR)$);
\coordinate (D1) at ($(DL)!0.5!(DR)$);

\coordinate (M1) at ($(UR)!0.3!(DL)$);
\coordinate (M2) at ($(DL)!0.3!(UR)$);

\draw (UL) -- (UR) -- (DR) -- (DL) -- (UL);

\draw[red,mid arrow] (L1) -- node[above right]{3} (M2);
\draw[mid arrow] (D1) -- (M2);
\draw[red,mid arrow] (M1) -- node[below]{3} (R1);
\draw[mid arrow] (M1) -- (U1);
\draw[green,mid arrow] (M2) --  node[below right]{4} (M1);
\end{tikzpicture}
}, \\
 D_{[(\omega_1,h_1+2h_{i(\neq 1)})]} &\sim
\raisebox{-0.45\height}{\tikzsetnextfilename{-figure-app-WAS_Tf-1A}
\begin{tikzpicture}
\def\vl{2.3}
\def\hl{2.5}
\coordinate (H) at (0:\hl);
\coordinate (V) at (90:\vl);
\coordinate (DL) at (0,0);
\coordinate (DR) at ($(DL)+(H)$);
\coordinate (UL) at ($(DL)+(V)$);
\coordinate (UR) at ($(UL)+(H)$);

\coordinate (L1) at ($(DL)!0.7!(UL)$);
\coordinate (L2) at ($(DL)!0.3!(UL)$);
\coordinate (R1) at ($(DR)!0.7!(UR)$);
\coordinate (R2) at ($(DR)!0.3!(UR)$);
\coordinate (U1) at ($(UL)!0.3!(UR)$);
\coordinate (D1) at ($(DL)!0.3!(DR)$);

\coordinate (M1) at ($(R1)!0.3!(L1)$);
\coordinate (M2) at ($(R1)!0.7!(L1)$);
\coordinate (M3) at ($(R2)!0.7!(L2)$);
\coordinate (M4) at ($(R2)!0.3!(L2)$);

\draw (UL) -- (UR) -- (DR) -- (DL) -- (UL);

\draw[mid arrow] (M1) -- (U1);
\draw[mid arrow] (M1) -- (R1);
\draw[blue,mid arrow] (M2) -- node[below]{2} (M1);
\draw[mid arrow] (L1) -- (M2);
\draw[mid arrow] (M3) -- (M2);
\draw[blue,mid arrow] (L2) -- node[below]{2} (M3);
\draw[mid arrow] (M3) -- (M4);
\draw[blue,mid arrow] (M4) -- node[above]{2} (R2);
\draw[mid arrow] (D1) -- (M4);
\end{tikzpicture}
},
& D_{[(\omega_1,2h_1+h_{i(\neq 1)}]} &\sim
\raisebox{-0.45\height}{\tikzsetnextfilename{-figure-app-WAS_Tf-2A}
\begin{tikzpicture}
\def\vl{2.3}
\def\hl{2.5}
\coordinate (H) at (0:\hl);
\coordinate (V) at (90:\vl);
\coordinate (DL) at (0,0);
\coordinate (DR) at ($(DL)+(H)$);
\coordinate (UL) at ($(DL)+(V)$);
\coordinate (UR) at ($(UL)+(H)$);

\coordinate (L1) at ($(DL)!0.7!(UL)$);
\coordinate (L2) at ($(DL)!0.3!(UL)$);
\coordinate (R1) at ($(DR)!0.7!(UR)$);
\coordinate (R2) at ($(DR)!0.3!(UR)$);
\coordinate (U1) at ($(UL)!0.3!(UR)$);
\coordinate (D1) at ($(DL)!0.3!(DR)$);

\coordinate (M1) at ($(DR)!0.3!(L1)$);
\coordinate (M2) at ($(DR)!0.7!(L1)$);

\draw (UL) -- (UR) -- (DR) -- (DL) -- (UL);

\draw[blue,mid arrow] (L2) -- node[below]{2} (M2);
\draw[mid arrow] (D1) -- (M1);
\draw[blue,mid arrow] (M1) -- node[below]{2} (R2);
\draw[mid arrow] (M2) -- (R1);
\draw[mid arrow] (L1) -- (U1);
\draw[mid arrow] (M2) -- (M1);
\end{tikzpicture}
}.
\end{align}

\subsection{$W_{\Box\!\Box\!\Box} \times T_{\Box}$}
Our next example is $W_{\Box\!\Box\!\Box} \times T_{\Box}$:
\begin{multline}
 \raisebox{-0.35\height}{\input{figure/app/W3S_Tf/OPE.tikz}}\;
=\;\frq^{3/N}\; \left[\quad\raisebox{-0.45\height}{\input{figure/app/W3S_Tf/0A.tikz}}\; \right. \\
 -2\quad\raisebox{-0.45\height}{\input{figure/app/W3S_Tf/0B.tikz}}\;
+\quad\raisebox{-0.45\height}{\input{figure/app/W3S_Tf/0C.tikz}}
  \\
 +\; \frq^{-1} \left(\quad\raisebox{-0.45\height}{\input{figure/app/W3S_Tf/1A.tikz}}\;
-\quad\raisebox{-0.45\height}{\input{figure/app/W3S_Tf/1B.tikz}}\right) \\
\left. \qquad +\; \frq^{-2} \quad\raisebox{-0.45\height}{\input{figure/app/W3S_Tf/2A.tikz}}\;
+\; \frq^{-3} \quad\raisebox{-0.45\height}{\input{figure/app/W3S_Tf/3A.tikz}}\quad  \right]
\end{multline}
The first term is new, the fifth one is elementary, the seventh one is pure and others have appeared in the previous case.
When $N=3$, the third one does not appear.
We therefore identify:
\begin{align}
 D_{[(\omega_1,3h_{i(\neq 1)})]} &\sim
\raisebox{-0.45\height}{\input{figure/app/W3S_Tf/0A.tikz}},
& D_{[(\omega_1,h_1+h_{i(\neq 1)}+h_{j(>i)})]} &\sim
\raisebox{-0.45\height}{\input{figure/app/W3S_Tf/1B.tikz}}, \\
 D_{[(\omega_1,3\omega_1)]} & \sim
\raisebox{-0.45\height}{\input{figure/app/W3S_Tf/3A.tikz}}.
\end{align}

\subsection{$W_\text{Adj} \times T_{\Box}$}
\label{subsec:WAdj_Tf_OPE}
As a further example, let us consider $W_\text{Adj} \times T_{\Box}$:
\begin{multline}
 \raisebox{-0.35\height}{\tikzsetnextfilename{-figure-app-WAdj_Tf-OPE}
\begin{tikzpicture}
\def\vl{2.3}
\def\hl{2.5}
\coordinate (H) at (0:\hl);
\coordinate (V) at (90:\vl);
\coordinate (DL) at (0,0);
\coordinate (DR) at ($(DL)+(H)$);
\coordinate (UL) at ($(DL)+(V)$);
\coordinate (UR) at ($(UL)+(H)$);

\coordinate (L1) at ($(DL)!0.5!(UL)$);
\coordinate (R1) at ($(DR)!0.5!(UR)$);
\coordinate (U1) at ($(UL)!0.5!(UR)$);
\coordinate (D1) at ($(DL)!0.5!(DR)$);

\draw (UL) -- (UR) -- (DR) -- (DL) -- (UL);

\draw[green,mid-end arrow,ultra thick] (D1) -- (U1) node[black,above]{$\Box$};
\draw[white,line width=3pt] ($(L1)+0.1*(H)$) -- ($(R1)-0.1*(H)$);
\draw[red,mid-end arrow,ultra thick] (L1) -- (R1) node[black,right]{Adj};
\end{tikzpicture}
}\; 
=\\
\;\frq \quad\raisebox{-0.45\height}{\tikzsetnextfilename{-figure-app-WAdj_Tf-0A}
\begin{tikzpicture}
\def\vl{2.3}
\def\hl{2.5}
\coordinate (H) at (0:\hl);
\coordinate (V) at (90:\vl);
\coordinate (DL) at (0,0);
\coordinate (DR) at ($(DL)+(H)$);
\coordinate (UL) at ($(DL)+(V)$);
\coordinate (UR) at ($(UL)+(H)$);

\coordinate (L1) at ($(DL)!0.7!(UL)$);
\coordinate (L2) at ($(DL)!0.3!(UL)$);
\coordinate (R1) at ($(DR)!0.7!(UR)$);
\coordinate (R2) at ($(DR)!0.3!(UR)$);
\coordinate (U1) at ($(UL)!0.3!(UR)$);
\coordinate (D1) at ($(DL)!0.3!(DR)$);

\coordinate (M1) at ($(UR)!0.3!(L2)$);
\coordinate (M2) at ($(UR)!0.7!(L2)$);

\draw (UL) -- (UR) -- (DR) -- (DL) -- (UL);

\draw[blue,mid arrow] (M2) -- node[above left]{2} (M1);
\draw[mid arrow] (L1) -- (M2);
\draw[mid arrow] (R2) -- (M2);
\draw[mid arrow] (M1) -- (R1);
\draw[mid arrow] (D1) -- (L2);
\draw[mid arrow] (M1) -- (U1);
\end{tikzpicture}
}\;
+\quad\raisebox{-0.45\height}{\tikzsetnextfilename{-figure-app-WAdj_Tf-1A}
\begin{tikzpicture}
\def\vl{2.3}
\def\hl{2.5}
\coordinate (H) at (0:\hl);
\coordinate (V) at (90:\vl);
\coordinate (DL) at (0,0);
\coordinate (DR) at ($(DL)+(H)$);
\coordinate (UL) at ($(DL)+(V)$);
\coordinate (UR) at ($(UL)+(H)$);

\coordinate (L1) at ($(DL)!0.7!(UL)$);
\coordinate (L2) at ($(DL)!0.3!(UL)$);
\coordinate (R1) at ($(DR)!0.7!(UR)$);
\coordinate (R2) at ($(DR)!0.3!(UR)$);
\coordinate (U1) at ($(UL)!0.7!(UR)$);
\coordinate (D1) at ($(DL)!0.7!(DR)$);

\coordinate (M1) at ($(R1)!0.3!(L1)$);
\coordinate (M2) at ($(R1)!0.7!(L1)$);
\coordinate (M3) at ($(R2)!0.7!(L2)$);
\coordinate (M4) at ($(R2)!0.3!(L2)$);

\draw (UL) -- (UR) -- (DR) -- (DL) -- (UL);

\draw[mid arrow] (M1) -- (U1);
\draw[mid arrow] (M1) -- (R1);
\draw[blue,mid arrow] (M2) -- node[below]{2} (M1);
\draw[mid arrow] (L1) -- (M2);
\draw[mid arrow] (M3) -- (M2);
\draw[blue,mid arrow] (M4) -- node[below]{2} (M3);
\draw[mid arrow] (M3) -- (L2);
\draw[mid arrow] (R2) -- (M4);
\draw[mid arrow] (D1) -- (M4);
\end{tikzpicture}
}\;
+\; \frq^{-1} \quad\raisebox{-0.45\height}{\tikzsetnextfilename{-figure-app-WAdj_Tf-2A}
\begin{tikzpicture}
\def\vl{2.3}
\def\hl{2.5}
\coordinate (H) at (0:\hl);
\coordinate (V) at (90:\vl);
\coordinate (DL) at (0,0);
\coordinate (DR) at ($(DL)+(H)$);
\coordinate (UL) at ($(DL)+(V)$);
\coordinate (UR) at ($(UL)+(H)$);

\coordinate (L1) at ($(DL)!0.7!(UL)$);
\coordinate (L2) at ($(DL)!0.3!(UL)$);
\coordinate (R1) at ($(DR)!0.7!(UR)$);
\coordinate (R2) at ($(DR)!0.3!(UR)$);
\coordinate (U1) at ($(UL)!0.3!(UR)$);
\coordinate (D1) at ($(DL)!0.3!(DR)$);

\coordinate (M1) at ($(DR)!0.3!(L1)$);
\coordinate (M2) at ($(DR)!0.7!(L1)$);

\draw (UL) -- (UR) -- (DR) -- (DL) -- (UL);

\draw[blue,mid arrow] (M1) -- node[below left]{2} (M2);
\draw[mid arrow] (D1) -- (M1);
\draw[mid arrow] (R2) -- (M1);
\draw[mid arrow] (M2) -- (R1);
\draw[mid arrow] (L1) -- (U1);
\draw[mid arrow] (M2) -- (L2);
\end{tikzpicture}
}
\end{multline}
The second term is new, and the first and the third ones are obtained by some duality actions of some elementary one. Our identifications are:
\begin{align}
& D_{[(\omega_1,h_{i(\neq 1)}-h_1)]} \sim
\raisebox{-0.45\height}{\tikzsetnextfilename{-figure-app-WAdj_Tf-0A}
\begin{tikzpicture}
\def\vl{2.3}
\def\hl{2.5}
\coordinate (H) at (0:\hl);
\coordinate (V) at (90:\vl);
\coordinate (DL) at (0,0);
\coordinate (DR) at ($(DL)+(H)$);
\coordinate (UL) at ($(DL)+(V)$);
\coordinate (UR) at ($(UL)+(H)$);

\coordinate (L1) at ($(DL)!0.7!(UL)$);
\coordinate (L2) at ($(DL)!0.3!(UL)$);
\coordinate (R1) at ($(DR)!0.7!(UR)$);
\coordinate (R2) at ($(DR)!0.3!(UR)$);
\coordinate (U1) at ($(UL)!0.3!(UR)$);
\coordinate (D1) at ($(DL)!0.3!(DR)$);

\coordinate (M1) at ($(UR)!0.3!(L2)$);
\coordinate (M2) at ($(UR)!0.7!(L2)$);

\draw (UL) -- (UR) -- (DR) -- (DL) -- (UL);

\draw[blue,mid arrow] (M2) -- node[above left]{2} (M1);
\draw[mid arrow] (L1) -- (M2);
\draw[mid arrow] (R2) -- (M2);
\draw[mid arrow] (M1) -- (R1);
\draw[mid arrow] (D1) -- (L2);
\draw[mid arrow] (M1) -- (U1);
\end{tikzpicture}
},
& D_{[(\omega_1,h_{i(\neq 1)}-h_{j(\neq i)}+h_{j(>i)})]} \sim
\raisebox{-0.45\height}{\tikzsetnextfilename{-figure-app-WAdj_Tf-1A}
\begin{tikzpicture}
\def\vl{2.3}
\def\hl{2.5}
\coordinate (H) at (0:\hl);
\coordinate (V) at (90:\vl);
\coordinate (DL) at (0,0);
\coordinate (DR) at ($(DL)+(H)$);
\coordinate (UL) at ($(DL)+(V)$);
\coordinate (UR) at ($(UL)+(H)$);

\coordinate (L1) at ($(DL)!0.7!(UL)$);
\coordinate (L2) at ($(DL)!0.3!(UL)$);
\coordinate (R1) at ($(DR)!0.7!(UR)$);
\coordinate (R2) at ($(DR)!0.3!(UR)$);
\coordinate (U1) at ($(UL)!0.7!(UR)$);
\coordinate (D1) at ($(DL)!0.7!(DR)$);

\coordinate (M1) at ($(R1)!0.3!(L1)$);
\coordinate (M2) at ($(R1)!0.7!(L1)$);
\coordinate (M3) at ($(R2)!0.7!(L2)$);
\coordinate (M4) at ($(R2)!0.3!(L2)$);

\draw (UL) -- (UR) -- (DR) -- (DL) -- (UL);

\draw[mid arrow] (M1) -- (U1);
\draw[mid arrow] (M1) -- (R1);
\draw[blue,mid arrow] (M2) -- node[below]{2} (M1);
\draw[mid arrow] (L1) -- (M2);
\draw[mid arrow] (M3) -- (M2);
\draw[blue,mid arrow] (M4) -- node[below]{2} (M3);
\draw[mid arrow] (M3) -- (L2);
\draw[mid arrow] (R2) -- (M4);
\draw[mid arrow] (D1) -- (M4);
\end{tikzpicture}
}, \\
& D_{[(\omega_1,h_1-h_{i(\neq 1)})]} \sim
\raisebox{-0.45\height}{\tikzsetnextfilename{-figure-app-WAdj_Tf-2A}
\begin{tikzpicture}
\def\vl{2.3}
\def\hl{2.5}
\coordinate (H) at (0:\hl);
\coordinate (V) at (90:\vl);
\coordinate (DL) at (0,0);
\coordinate (DR) at ($(DL)+(H)$);
\coordinate (UL) at ($(DL)+(V)$);
\coordinate (UR) at ($(UL)+(H)$);

\coordinate (L1) at ($(DL)!0.7!(UL)$);
\coordinate (L2) at ($(DL)!0.3!(UL)$);
\coordinate (R1) at ($(DR)!0.7!(UR)$);
\coordinate (R2) at ($(DR)!0.3!(UR)$);
\coordinate (U1) at ($(UL)!0.3!(UR)$);
\coordinate (D1) at ($(DL)!0.3!(DR)$);

\coordinate (M1) at ($(DR)!0.3!(L1)$);
\coordinate (M2) at ($(DR)!0.7!(L1)$);

\draw (UL) -- (UR) -- (DR) -- (DL) -- (UL);

\draw[blue,mid arrow] (M1) -- node[below left]{2} (M2);
\draw[mid arrow] (D1) -- (M1);
\draw[mid arrow] (R2) -- (M1);
\draw[mid arrow] (M2) -- (R1);
\draw[mid arrow] (L1) -- (U1);
\draw[mid arrow] (M2) -- (L2);
\end{tikzpicture}
}.
&
\end{align}

\subsection{$W_\text{Adj} \times T_\text{Adj}$}
We now move on to the example $W_\text{Adj} \times T_\text{Adj}$:
\begin{multline}
\raisebox{-0.35\height}{\input{figure/app/WAdj_TAdj/OPE.tikz}}\;
=
\quad\raisebox{-0.45\height}{\input{figure/app/WAdj_TAdj/2B.tikz}}\;
+\quad\raisebox{-0.45\height}{\input{figure/app/WAdj_TAdj/WAdj.tikz}}\;
+\quad\raisebox{-0.45\height}{\input{figure/app/WAdj_TAdj/TAdj.tikz}} \\
+\; \frq^{-2} \quad\raisebox{-0.45\height}{\input{figure/app/WAdj_TAdj/DAdj1.tikz}}\;
+\; \frq^{2} \quad\raisebox{-0.45\height}{\input{figure/app/WAdj_TAdj/DAdj2.tikz}} \\
+\; (2[N-2]+[N-4])\quad\raisebox{-0.45\height}{\input{figure/app/WAdj_TAdj/flavor.tikz}}\;
+\; (3+\frq^{2}+\frq^{-2})\quad\raisebox{-0.45\height}{\tikzsetnextfilename{-figure-app-WAdj_TAdj-empty}
\begin{tikzpicture}
\def\vl{2.3}
\def\hl{2.5}
\coordinate (H) at (0:\hl);
\coordinate (V) at (90:\vl);
\coordinate (DL) at (0,0);
\coordinate (DR) at ($(DL)+(H)$);
\coordinate (UL) at ($(DL)+(V)$);
\coordinate (UR) at ($(UL)+(H)$);

\draw (UL) -- (UR) -- (DR) -- (DL) -- (UL);
\end{tikzpicture}
} \\
+\frq^{-1}\; \left( \quad\raisebox{-0.45\height}{\input{figure/app/WAdj_TAdj/3A.tikz}}\;
+\quad\raisebox{-0.45\height}{\input{figure/app/WAdj_TAdj/3B.tikz}}\right) \\
+\frq\; \left( \quad\raisebox{-0.45\height}{\input{figure/app/WAdj_TAdj/1A.tikz}}\; 
+\quad\raisebox{-0.45\height}{\input{figure/app/WAdj_TAdj/1B.tikz}}\right)
\end{multline}
The first one does not appear when $N=3$ but reduces to a sum of simpler ones.

Then the  charge/network dictionary for the new ones is:
\begin{align}
 D_{[(\lambda_{\text{Adj}},h_{i(\neq 1,N)}-h_1)]} &\sim
\quad\raisebox{-0.45\height}{\input{figure/app/WAdj_TAdj/2B.tikz}},\;  \\
 D_{[(\lambda_{\text{Adj}},h_{i(\neq 1)}-h_{N})]} &\sim
\raisebox{-0.45\height}{\input{figure/app/WAdj_TAdj/3A.tikz}},
& D_{[(\lambda_{\text{Adj}},h_1-h_{i(\neq 1,N)})]} &\sim
\raisebox{-0.45\height}{\input{figure/app/WAdj_TAdj/3B.tikz}}, \\
 D_{[(\lambda_{\text{Adj}},-h_1+h_{i(\neq 1,N)})]}& \sim
\raisebox{-0.45\height}{\input{figure/app/WAdj_TAdj/1A.tikz}},
& D_{[(\lambda_{\text{Adj}},h_N-h_{i(\neq 1,N)})]} &\sim
\raisebox{-0.45\height}{\input{figure/app/WAdj_TAdj/1B.tikz}}
\end{align}
where $\lambda_{\text{Adj}}=h_1-h_{N}$ is the highest weight of the adjoint representation.

\subsection{An OPE of complex dyonic loops}
Finally we give an example of the OPE of  two elementary dyonic loops $D^{1,1}_{(0)}$ and $D^{\ell,k}_{(0)}$ for $\ell,k \le N/2$ in $\SUSYN{4}$ SYM:
\begin{multline}
 \raisebox{-0.35\height}{\tikzsetnextfilename{-figure-app-D110_Dlk0-OPE}
\begin{tikzpicture}
\def\vl{2.3}
\def\hl{2.5}
\coordinate (H) at (0:\hl);
\coordinate (V) at (90:\vl);
\coordinate (DL) at (0,0);
\coordinate (DR) at ($(DL)+(H)$);
\coordinate (UL) at ($(DL)+(V)$);
\coordinate (UR) at ($(UL)+(H)$);

\def\x{0.3}
\coordinate (L1) at ($(UL)!\x!(DL)$);
\coordinate (L2) at ($(DL)!\x!(UL)$);
\coordinate (R1) at ($(UR)!\x!(DR)$);
\coordinate (R2) at ($(DR)!\x!(UR)$);
\coordinate (U1) at ($(UL)!\x!(UR)$);
\coordinate (U2) at ($(UR)!\x!(UL)$);
\coordinate (D1) at ($(DL)!\x!(DR)$);
\coordinate (D2) at ($(DR)!\x!(DL)$);

\coordinate (M1) at ($(DL)!0.2!(UR)$);
\coordinate (M2) at ($(DL)!0.4!(UR)$);
\coordinate (M3) at ($(DL)!0.6!(UR)$);
\coordinate (M4) at ($(DL)!0.8!(UR)$);

\draw (UL) -- (UR) -- (DR) -- (DL) -- (UL);

\draw[mid arrow] (L2) node[left]{$1$} -- (M1);
\draw[mid arrow] (D1) node[below]{$1$} -- (M1);
\draw[mid arrow] (M1) -- node[below]{$2$} (M2);
\draw[mid arrow] (M2)-- (U1) node[above]{$1$};
\draw[mid arrow] (M2) -- (R2) node[right]{$1$};

\draw[line width=1.5mm,white] (L1) -- (M3);
\draw[line width=1.5mm,white] (D2) -- (M3);
\draw[mid arrow,blue] (L1) node[left]{$k$} -- (M3);
\draw[mid arrow,blue] (D2) node[below]{$\ell$} -- (M3);
\draw[mid arrow,blue] (M3) -- node[below right]{$k+\ell$} (M4);
\draw[mid arrow,blue] (M4) -- (U2) node[above]{$\ell$};
\draw[mid arrow,blue] (M4) -- (R1) node[right]{$k$};

\end{tikzpicture}
}\; 
 = \;\frq^{k-\ell/N}\; \left[\quad
\raisebox{-0.45\height}{\tikzsetnextfilename{-figure-app-D110_Dlk0-A}
\begin{tikzpicture}
\def\vl{2.3}
\def\hl{2.5}
\coordinate (H) at (0:\hl);
\coordinate (V) at (90:\vl);
\coordinate (DL) at (0,0);
\coordinate (DR) at ($(DL)+(H)$);
\coordinate (UL) at ($(DL)+(V)$);
\coordinate (UR) at ($(UL)+(H)$);

\def\x{0.25}
\coordinate (L1) at ($(UL)!\x!(DL)$);
\coordinate (L2) at ($(DL)!\x!(UL)$);
\coordinate (R1) at ($(UR)!\x!(DR)$);
\coordinate (R2) at ($(DR)!\x!(UR)$);
\coordinate (U1) at ($(UL)!\x!(UR)$);
\coordinate (U2) at ($(UR)!\x!(UL)$);
\coordinate (D1) at ($(DL)!\x!(DR)$);
\coordinate (D2) at ($(DR)!\x!(DL)$);

\coordinate (LM) at ($(UL)!0.5!(DL)$);
\coordinate (RM) at ($(UR)!0.5!(DR)$);
\coordinate (UM) at ($(UR)!0.5!(UL)$);
\coordinate (DM) at ($(DL)!0.5!(DR)$);

\coordinate (M1) at ($(DL)!0.2!(UR)$);
\coordinate (M2) at ($(DL)!0.4!(UR)$);
\coordinate (M3) at ($(DL)!0.6!(UR)$);
\coordinate (M4) at ($(DL)!0.8!(UR)$);

\def\y{0.35}
\coordinate (M5) at ($(UM)!\y!(LM)$);
\coordinate (M6) at ($(LM)!\y!(UM)$);
\coordinate (M7) at ($(RM)!\y!(DM)$);
\coordinate (M8) at ($(DM)!\y!(RM)$);

\draw (UL) -- (UR) -- (DR) -- (DL) -- (UL);

\draw[mid arrow] (L2) node[left]{$1$} -- (M1);
\draw[mid arrow] (D1) node[below]{$1$} -- (M1);
\draw[mid arrow,blue] (M1) -- node[below]{$2$} (M2);
\draw[mid arrow] (M2)-- (M6);
\draw[mid arrow] (M2) -- (M8);

\draw[mid arrow,red] (L1) node[left]{$k$} -- (M6);
\draw[mid arrow,green] (D2) node[below]{$\ell$} -- (M8);
\draw[mid arrow,blue] (M6) node[above left]{$k+1$} -- (M5);
\draw[mid arrow,blue] (M8) -- node[below right]{$\ell+1$} (M7);
\draw[mid arrow] (M5)-- (U1);
\draw[mid arrow] (M7) -- (R2);

\draw[mid arrow,red] (M5) -- (M3);
\draw[mid arrow,green] (M7) -- (M3);
\draw[mid arrow,ultra thick,blue] (M3) node[right]{$k+\ell$} -- (M4);
\draw[mid arrow,green] (M4) -- (U2) node[above]{$\ell$};
\draw[mid arrow,red] (M4) -- (R1) node[right]{$k$};

\end{tikzpicture}
}\; \right.  \\
\left. 
+\; \frq^{-1} \raisebox{-0.55\height}{\tikzsetnextfilename{-figure-app-D110_Dlk0-B}
\begin{tikzpicture}
\def\vl{2.3}
\def\hl{2.5}
\coordinate (H) at (0:\hl);
\coordinate (V) at (90:\vl);
\coordinate (DL) at (0,0);
\coordinate (DR) at ($(DL)+(H)$);
\coordinate (UL) at ($(DL)+(V)$);
\coordinate (UR) at ($(UL)+(H)$);

\coordinate (L1) at ($(UL)!0.2!(DL)$);
\coordinate (L2) at ($(DL)!0.35!(UL)$);
\coordinate (R1) at ($(UR)!0.2!(DR)$);
\coordinate (R2) at ($(DR)!0.35!(UR)$);
\coordinate (U1) at ($(UL)!0.5!(UR)$);
\coordinate (D1) at ($(DL)!0.5!(DR)$);

\coordinate (M1) at ($(L2)!0.3!(R2)$);
\coordinate (M2) at ($(D1)!0.15!(U1)$);
\coordinate (M3) at ($(R2)!0.3!(L2)$);
\coordinate (M4) at ($(D1)!0.55!(U1)$);

\coordinate (M5) at ($(UL)!0.6!(M4)$);
\coordinate (M6) at ($(L2)!0.7!(UR)$);

\draw (UL) -- (UR) -- (DR) -- (DL) -- (UL);

\draw[mid arrow,red] (L2) node[left]{$k$} -- (M1);
\draw[mid arrow,green] (D1) node[below]{$\ell+1$} -- (M2);
\draw[mid arrow,blue] (M1) -- (M2);
\draw[mid arrow,blue] (M2) -- (M3);
\draw[mid arrow,blue] (M3) -- (M4);
\draw[mid arrow,blue] (M1) -- node[above left]{$1$} (M4);
\draw[mid arrow,green] (M4) -- (M5);
\draw[mid arrow,red] (M3) -- (R2);

\draw[mid arrow] (L1) node[left]{$1$} -- (M5);
\draw[mid arrow,blue] (M5) -- (M6) node[below]{$\ell+2$};
\draw[mid arrow,green] (M6) -- (U1);
\draw[mid arrow] (M6) -- (R1);

\end{tikzpicture}
}\;
+\; \frq \raisebox{-0.55\height}{\tikzsetnextfilename{-figure-app-D110_Dlk0-C}
\begin{tikzpicture}
\def\vl{2.3}
\def\hl{2.5}
\coordinate (H) at (0:\hl);
\coordinate (V) at (90:\vl);
\coordinate (DL) at (0,0);
\coordinate (DR) at ($(DL)+(H)$);
\coordinate (UL) at ($(DL)+(V)$);
\coordinate (UR) at ($(UL)+(H)$);

\coordinate (U1) at ($(UL)!0.35!(UR)$);
\coordinate (U2) at ($(UR)!0.2!(UL)$);
\coordinate (D1) at ($(DL)!0.35!(DR)$);
\coordinate (D2) at ($(DR)!0.2!(DL)$);
\coordinate (L1) at ($(UL)!0.5!(DL)$);
\coordinate (R1) at ($(UR)!0.5!(DR)$);

\coordinate (M1) at ($(U1)!0.3!(D1)$);
\coordinate (M2) at ($(L1)!0.15!(R1)$);
\coordinate (M3) at ($(D1)!0.3!(U1)$);
\coordinate (M4) at ($(L1)!0.55!(R1)$);

\coordinate (M5) at ($(DR)!0.6!(M4)$);
\coordinate (M6) at ($(D1)!0.7!(UR)$);

\draw (UL) -- (UR) -- (DR) -- (DL) -- (UL);

\draw[mid arrow,green] (D1) node[below]{$\ell$} -- (M3);
\draw[mid arrow,red] (L1) node[left]{$k+1$} -- (M2);
\draw[mid arrow,blue] (M2) -- node[above left]{$1$} (M1);
\draw[mid arrow,blue] (M2) -- (M3);
\draw[mid arrow,blue] (M4) -- (M1);
\draw[mid arrow,blue] (M3) -- (M4);
\draw[mid arrow,red] (M4) -- (M5);
\draw[mid arrow,green] (M1) -- (U1);

\draw[mid arrow] (D2) node[below]{$1$} -- (M5);
\draw[mid arrow,blue] (M5) node[right]{$k+2$} -- (M6);
\draw[mid arrow] (M6) -- (U2);
\draw[mid arrow,red] (M6) -- (R1);

\end{tikzpicture}
}
\quad \right]
\end{multline}
where 
\begin{itemize}
\item the first term corresponds to $D_{[(\omega_\ell+\omega_1,2h_{j_{s=1}}+\sum_{s=2}^{k} h_{j_s})]}$ where $j_s > \ell$, 
\item the second one to $D_{[(\omega_{\ell+1},h_{i(\le \ell+1)}+2h_{j_{s=1}}+\sum_{s=2}^{k-1}h_{j_s})]}$ where $j_s > \ell+1$,
\item the third one to $D_{[(\omega_{\ell}+\omega_1,h_{(1<)i(\le \ell)}+\sum_{s=1}^{k}h_{j_s})]}$ where $j_s > \ell$.
\end{itemize}
Here we require $j_s$ differ for each $s$.
It would be interesting  to apply the diagrammatic approach to more complicated OPEs and read off the charge information from networks in general.


\bibliographystyle{JHEP}
\bibliography{yuji,qYM,defect,defectAGT,books,classS,AGT,math}

\end{document}